\documentclass[reqno,12pt]{amsart}

\usepackage[footnotesize,normal]{caption}
\usepackage{amssymb, amsmath, amsthm, amscd, graphicx, amsfonts, color}
\usepackage{vmargin}

\newtheorem{theo}{Theorem}[section]
\newtheorem{cor}{Corollary}[section]
\newtheorem{lem}{Lemma}[section]
\newtheorem{pro}{Proposition}[section]
\newtheorem{defin}{Definition}[section]

\newcommand{\bp}{\begin{proof}}
\newcommand{\ep}{\end{proof}}
\newcommand{\blem}{\begin{lem}}
\newcommand{\elem}{\end{lem}}

\newcommand{\nouvellepage}{}

\newcommand{\optionel}[1]{#1}

\newcommand{\grasA}[1]{\textsf{#1}}
\newcommand{\grasB}[1]{\textsf{#1}}

\numberwithin{equation}{section}

\newcommand{\eq}[1]{\begin{equation}#1\end{equation}}
\newcommand{\al}[1]{\begin{align}#1\end{align}}
\newcommand{\gras}[1]{\textsf{#1}}

\newcommand{\ov}[1]{\overline{#1}}

\newcommand{\cb}{\eta_{\Lambda^c}}

\newcommand{\real}{\mathrm{Re}\,}
\newcommand{\imag}{\mathrm{Im}\,}

\newcommand{\epsr}{\epsilon_r}

\newcommand{\monphi}{{\varphi_\Lambda^+(\Ga)}}

\newcommand{\pardef}{:=}

\newcommand{\newplus}{{+}}

\newcommand{\bigdot}{{{\bullet}}}

\newcommand{\hatP}{{\hat{P}}}
\newcommand{\hatcP}{{\hat{\cP}}}
\newcommand{\hatcC}{{\hat{\cC}}}

\newcommand{\UN}{{1}}

\newcommand{\tildelta}{{\tilde{\delta}}}

\newcommand{\tildomega}{{\widetilde{\omega}}}
\newcommand{\ovsigma}{\ov{\sigma}}

\newcommand{\ga}{\gamma}
\newcommand{\Ga}{\Gamma}
\newcommand{\hatga}{{\hat{\gamma}}}

\newcommand{\uga}{\underline{\gamma}}

\newcommand{\ppetit}{\preceq}

\newcommand{\Gamax}{\Ga^{\mathrm{max}}}

\newcommand{\animal}{{animal}}
\newcommand{\animals}{{animals}}
\newcommand{\hatcD}{{\hat{\cD}}}

\newcommand{\rZ}{{Z_r}}
\newcommand{\rcZ}{{{\cZ}_r}}

\newcommand{\eGa}{{X}}

\newcommand{\hateGa}{{\hat{\eGa}}}
\newcommand{\hatcX}{{\hat{\cX}}}

\newcommand{\eZ}{{\Xi}}

\newcommand{\inte}{\mathrm{int}}

\newcommand{\ext}{\mathrm{ext}}

\newcommand{\usd}{\frac{1}{d}}
\newcommand{\usdmu}{\frac{1}{d-1}}

\newcommand{\ddmu}{\frac{d}{d-1}}
\newcommand{\dmusd}{\frac{d-1}{d}}
\newcommand{\supp}{\mathrm{supp}\,}

\newcommand{\diff}{\mathrm{d}}
\newcommand{\deriv}[1]{\frac{\diff}{\diff #1}}
\newcommand{\derivp}[1]{\frac{\partial}{\partial #1}}
\newcommand{\derivk}[2]{\frac{\diff^{#2}}{\diff {#1}^{#2}}}

\newcommand{\ttt}{{\mathtt{t}}}

\newcommand{\cA}{\mathcal{A}}
\newcommand{\cB}{\mathcal{B}}
\newcommand{\cC}{\mathcal{C}}
\newcommand{\cD}{\mathcal{D}}
\newcommand{\cE}{\mathcal{E}}
\newcommand{\cF}{\mathcal{F}}
\newcommand{\cG}{\mathcal{G}}

\newcommand{\cL}{\mathcal{L}}

\newcommand{\cP}{\mathcal{P}}

\newcommand{\cT}{\mathcal{T}}

\newcommand{\cX}{\mathcal{X}}

\newcommand{\cZ}{\mathcal{Z}}

\newcommand{\bC}{\mathbb{C}}
\newcommand{\bR}{\mathbb{R}}

\newcommand{\bN}{\mathbb{N}}

\newcommand{\bZ}{\mathbf{Z}}

\def\data{\the\day\space\ifcase\month\or January \or February \or March \or
April \or May \or June \or July \or August \or September
\or October \or November \or December \fi\space\the\year}

\def\data{\the\day\space\ifcase\month\or January \or February \or March \or
April \or May \or June \or July \or August \or September
\or October \or November \or December \fi\the\year}
\begin{document}
\centerline{\bf \sf NON-ANALYTICITY AND THE VAN DER WAALS LIMIT$~^{(*)}$}
\vspace*{.5cm}
\centerline{S. Friedli$~^{(a)}$ and C.-E. Pfister$~^{(b)}$}
\vspace*{0.5cm}

\textbf{Abstract:}
\noindent 
We study the analyticity properties of the free energy $f_\ga(m)$
of the 
Kac model at points of first order phase transition,
in the van der Waals limit $\ga\searrow 0$. 
We show that there exists an inverse temperature 
$\beta_0$ and $\ga_0>0$ such that for all $\beta\geq
\beta_0$ and for all $\ga\in(0,\ga_0)$,
$f_\ga(m)$ has no analytic continuation along the path $m\searrow m^*$
($m^*$ denotes spontaneous magnetization).
The proof consists in studying high order derivatives of the pressure
$p_\ga(h)$, which is related to the free energy $f_\ga(m)$ by a 
Legendre transform.
\tableofcontents
\noindent 
Date: \data.\\
a) {\em Institute for 
Theoretical Physics}, EPFL, CH-1015  Lausanne, Switzerland.\\
b) {\em Institute of Analysis and Scientific Computing}, School 
of Mathematics,
EPFL, CH-1015  Lausanne, Switzerland.\\
(*) Supported by the Swiss National Foundation for Scientific Research.
\nouvellepage
\section{Introduction}
The first equation of state giving precise predictions on
the liquid-vapor equilibrium at low
temperature was given by van der Waals \cite{vdW}:
\eq{\label{K0.0}
\Big(p+\frac{a}{v^2}\Big)\Big(v-b\Big)=RT\,.}
This equation follows from the hypothesis that the molecules
interact via 
1) a short range hard core repulsion, due to the assumption that
molecules are extended in space, 2) an attractive potential,
whose range is assumed to be comparable to the size of the
system. Nowadays, such an approximation is called a \emph{mean field}
approximation. As well known, there exists a critical temperature
$T_c=T_c(a,b)$ such that for $T<T_c$, $\derivp{v}p\geq 0$ for some values of
$v$, which implies thermodynamic instability.
On physical and geometrical grounds, the graph of the pressure
was modified by Maxwell who replaced $p(v)$, on a suitably chosen interval
$[v_l,v_g]$, by a
flat horizontal segment (the ``equal area rule''). 
\begin{figure}[htbp]
\begin{center}
\input{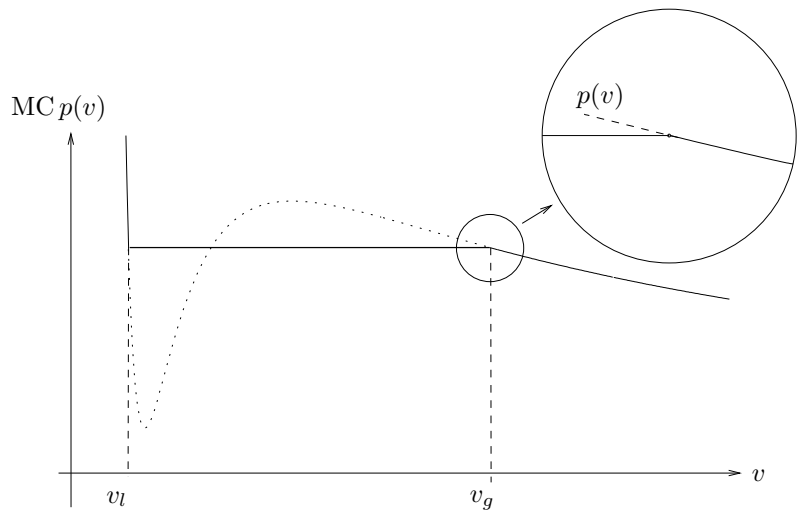}
\end{center}
\caption{The equation of state modified by Maxwell and the analytic
continuation at the condensation point.}
\label{KF0.0} 
\end{figure}
The new function obtained, written
$\mathrm{MC}\,{p}(v)$, describes precisely what is observed in the laboratory: $v_l$
is called the evaporation point and $v_g$ is the condensation point (see Figure
\ref{KF0.0}).\\
A particularity of this scenario is that $\mathrm{MC}\,{p}$ can be continued
analytically
along the paths $v\nearrow v_l$ and $v\searrow v_g$: the liquid and
gas branches can be joined analytically by a \emph{single} function, which is
nothing but the original isotherm $p$ given in
\eqref{K0.0}. The pressure obtained by analytic continuation was
originally considered as the pressure of a meta-stable state (see 
Figure \ref{KF0.0}). For instance, the meta-stable state obtained by analytic
continuation along the path $v\searrow v_g$ is
called a \emph{super-saturated vapor}.

Much later, Kac, Uhlenbeck and Hemmer \cite{KUH}
showed how the Maxwell construction
could be rigorously justified for a one
dimensional model, from first principles of statistical mechanics, using a
double limiting process: if the range of interaction diverges \emph{after} the
thermodynamic limit, then convexity is preserved and
the free energy converges to the convex
envelope of mean field theory.
Later this was generalized and 
extended to higher dimensions by Lebowitz and
Penrose \cite{LP}. From the
point of view of analyticity, these results
imply, as in the theory of van der
Waals, that the free energy can be continued
analytically across condensation/evaporation points.\\
In the mean time, arguments were given, saying that
when the range of interaction is finite, the free energy
might have some singularities that forbid analytic
continuation across the transition points. In
\cite{F} and \cite{L1}, Fisher and Langer 
analyzed in details simple models
to illustrate this phenomenon, but it  was
not until the seminal work of Isakov
\cite{I1} that this was shown for the Ising model.

An important issue is thus to understand how the
breakdown of analyticity at a first order phase transition point
relates to the range of interaction.
Since Kac potentials give a way of interpolating finite range systems
and mean field,
it seems an interesting problem to study the dependence on the scaling
parameter $\ga$ of the analyticity properties of
the Kac model at low temperature. The
aim of this work is to show that as
long as this parameter is strictly positive, i.e. as long as 
the range of interaction is \emph{finite}, the free energy has no analytic
continuation at first order phase transition points. Only \emph{after} the van
der Waals limit ($\ga\searrow 0$)
does the free energy have analytic continuation.
This result answers a question raised
by Joel Lebowitz at a con\-ference devoted to Kac potentials,
\emph{In-homogeneous Random Systems}, held in Paris, January 2001.\\

In Section \ref{KSSmeanfield} we remind the main properties of the free energy
for mean field
and Kac potentials in the case of Ising spins. In Section \ref{KSSresults}
we state our main results and give the strategy of the proof.

\subsection{Mean Field and Kac Potentials}\label{KSSmeanfield}
\begin{figure}[htb]
\vspace{0.5cm}
\begin{center}
\input{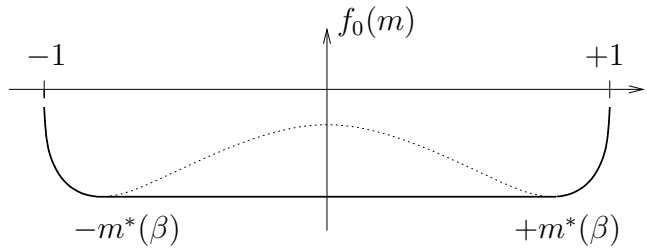}
\end{center}
\caption{{The free energy $f_0(m)$ when $\beta>1$. The dotted line is the
analytic continuation provided by $f_{MF}(m)$.}}
\label{KF0b}
\end{figure}

We consider the lattice $\bZ^d$, $d\geq 2$, with a distance $d(x,y)=\|x-y\|$,
where
\eq{\label{K1}\|x\|\pardef \max_{i=1,\dots,d}|x_i|\,.}
This distance will also be used for points of $\bR^d$.
The letter $\Lambda$ will always denote a finite subset of $\bZ^d$.
At each site $i\in \bZ^d$ lives a spin $\sigma_i\in\{\pm 1\}$. 
The configuration
space is $\Omega=\{\pm 1\}^{\bZ^d}$. For any set $\Lambda$,
$\Omega_\Lambda=\{\pm 1\}^\Lambda$. Our notations are often
inspired by those of Presutti \cite{Pr}.\\
\textbf{Mean Field.}
In a mean field model, the interactions ignore the spatial positions of the
spins, and the hamiltonian in a volume $\Lambda$ containing $N$ sites is
($\sigma\in \Omega_\Lambda$)
\eq{H_\Lambda^{MF}
(\sigma)\pardef-\frac{1}{N}\sum_{\substack{\{i,j\}
\subset\Lambda\\i\neq j}}\sigma_i\sigma_j\,.}
As is well known, the free energy can be easily computed. For $m\in[-1,+1]$,
\eq{f_{MF}(m)=-\frac{1}{2}m^2-\frac{1}{\beta}I(m)\,,}
where
\eq{\label{K13}I(m):=-\frac{1-m}{2}\log \frac{1-m}{2}-
\frac{1+m}{2}\log \frac{1+m}{2}\,.}
When $\beta\leq 1$ $f_{MF}$ is strictly convex, but when $\beta >1$, $f_{MF}$
has two minima at $\pm m^*(\beta)$, where 
$m^*(\beta)$ is the positive non-trivial solution of $m=\tanh(\beta m)$.
$\beta_c\pardef 1$ is the critical temperature of mean field theory.
As in van der Waals theory, $f_{MF}$ is non convex when $\beta>\beta_c$, in 
contradiction with thermodynamic stability.\\
\textbf{Kac Potentials.} Kac potentials are defined as follows.
Consider $J:\bR^d\to\bR^+$ supported
by the unit cube $\{y\in\bR^d:\|y\|\leq 1\}=[-1,+1]^d$ such that 
the \grasA{overall strength} equals unity, i.e.
\eq{\label{K2.1}\int_{\bR^d}J(x)\diff x=1\,.}
Let $\ga\in (0,1)$ be the \grasA{scaling parameter}. Define
$J_\ga:\bZ^d\to \bR^+$ as follows:
\eq{\label{K3}J_\ga(x)\pardef c_\ga\ga^d J(\ga x)\,,}
where $c_\ga$ is defined so that 
\eq{\label{K4}\sum_{x\neq 0}J_\ga(x)= 1\,.}
It is easy to see that \eqref{K2.1} implies $\lim_{\ga\searrow 0}c_\ga=1$.
Since $J_\ga(x)=0$ if $\|x\|>\ga^{-1}$, we call $R\pardef \ga^{-1}$ the
\grasB{range of the interaction}. Unless stated explicitly, $R$ will always
denote the range of interaction, i.e. $\ga^{-1}$.
For a finite $\Lambda$, $\sigma\in \Omega_\Lambda$,
the \grasB{Kac hamiltonian} is defined by
\eq{\label{K8}
H_\Lambda^h(\sigma)=-
\sum_{\substack{\{i,j\}\subset \Lambda\\i\neq j}}J_\ga(i-j)
\sigma_i\sigma_j-h\sum_{i\in\Lambda}\sigma_i\,,}
where $h\in\bR$ is the magnetic field.
The \grasB{magnetization in} $\Lambda$ is
\eq{\label{K9}m_\Lambda(\sigma)=\frac{1}{|\Lambda|}
\sum_{i\in \Lambda}\sigma_i\,} and takes values
in a set $\chi_\Lambda\subset [-1,+1]$. 
The \grasB{canonical partition function} 
is defined by
($\beta>0$ is the inverse temperature, $m\in \chi_\Lambda$):
\eq{\label{K10}
Z(\Lambda,m)=\sum_{\substack{\sigma_\Lambda\in 
\Omega_\Lambda\\ m_\Lambda(\sigma_\Lambda)=m}}
\exp\big(-\beta H_\Lambda^0(\sigma)\big)\,.}
The \gras{free energy density} is, for $m\in [-1,+1]$,
\eq{\label{K12}
f_\ga(m)=-\lim_{\Lambda\nearrow \bZ^d}\frac{1}{\beta|\Lambda|}\log
Z(\Lambda,m(\Lambda))\,,} where
the thermodynamic limit $\Lambda\nearrow \bZ^d$ is 
along a sequence of cubes, and the sequence $m(\Lambda)$ is
such that $m(\Lambda)\to m$. The function $f_\ga$ exists and is convex.
The 
Theorem of Lebowitz-Penrose
\cite{LP} gives a closed form for the free energy in the
\grasB{van der Waals limit} $\ga\searrow 0$. For a function
$f(x)$, let $\mathrm{CE}\,f(x)$ denote its convex envelope.
\begin{theo}{\bf\cite{LP}}\label{KT0}
For any $\beta>0$, $m\in [-1,+1]$,
\eq{\label{K14}f_0(m)\pardef \lim_{\ga\searrow 0}f_\ga(m)=\mathrm{CE}\,
f_{MF}(m)\,.}
\end{theo}
\noindent When
$\beta>1$, the graph of $f_0(m)$ is thus horizontal 
between $-m^*(\beta)$ and $+m^*(\beta)$, giving a rigorous justification of the
Maxwell construction (see Figure \ref{KF0b}).

\noindent From the point of view of analyticity, we have
\begin{cor}\label{KC1}
When $\beta>1$,
$f_0$ is analytic everywhere except at $\pm m^*(\beta)$, and 
has analytic continuations along the (real) paths $m\nearrow -m^*(\beta)$,
$m\searrow +m^*(\beta)$. The unique analytic 
continuation is given by the mean
field free energy $f_{MF}$.
\end{cor}
\noindent
That is: \emph{after} 
the van der Waals limit, all the analyticity properties of the free energy are
known explicitely. There exists
no formula for $f_\ga$ when $\ga>0$, and
it was not shown, until the papers of Cassandro and
Presutti \cite{CP} and Bovier and Zahradn\'\i k \cite{BZ1}, that the system
exhibits a first order phase transition \emph{before} reaching the mean
field regime: for all $\beta>1$, the graph of $f_\ga(m)$ already has a plateau
$[-m^*(\beta,\ga),+m^*(\beta,\ga)]$ when
$\ga$ is small enough. In this sense, one can say that mean field, together with
the Maxwell construction, is a good approximation
to long but finite range interactions (and vice versa). 
Our purpose is to show that
from the point of view of analyticity, the situation is very different.
\subsection{Obstruction for $\ga>0$; Main Results}\label{KSSresults}
Our results hold for Kac potentials for which Lemmas \ref{KL1} and \ref{KL7}
hold, but we believe them to be true for any ferromagnetic
potential satisfying \eqref{K2.1}. For the sake of simplicity, we focus
on a particular potential, i.e. on the \grasA{step function}
\eq{\label{K0.00}J(x)\pardef2^{-d}\UN_{\{\|y\|\leq 1\}}(x)\,,}
In this setting, our main result for the free energy density is the following:
\begin{theo}\label{KT1}
There exists $\beta_0$ and $\ga_0>0$ 
such that for all
$\beta\geq \beta_0$, $\ga\in (0,\ga_0)$, $f_\ga$ is analytic
everywhere except at $\pm m^*(\beta,\ga)$, but has
no analytic continuation along the paths $m\nearrow -m^*(\beta,\ga)$, 
$m\searrow +m^*(\beta,\ga)$.
\end{theo}
This result is in favor of the original ideas of Fisher and Langer, saying that
\emph{finiteness of the range of interaction is responsible for absence of
analytic continuation}. In particular it excludes the possibility
of obtaining the free energy by a  Maxwell construction: when $\ga>0$ the phases
$+$ and $-$ cannot be joined analytically.\\

\begin{figure}[Htb]
\begin{center}
\input{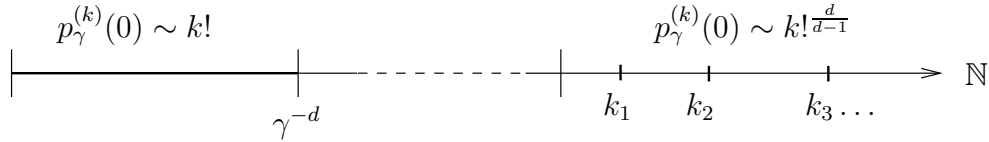}
\end{center}
\caption{The derivatives of the pressure at $h=0$, when $\ga>0$.
The first ones ($k\leq \ga^{-d}$) behave like those of an analytic
function, but non-analyticity always dominates for large $k$.}
\label{KF00.01}
\end{figure}

The proof of Theorem \ref{KT1}
will be done by working in the more appropriate grand
canonical ensemble (in the lattice gas terminology), in which 
the constraint on the magnetization is replaced by a magnetic field. Let
\eq{\label{K15}Z(\Lambda)=\sum_{\substack{\sigma_\Lambda\in \Omega_\Lambda}}
\exp\big(-\beta H_\Lambda^h(\sigma_\Lambda)\big)\,.}
Define the \grasA{pressure density} by
\eq{\label{K17}p_{\ga}(h)\pardef
\lim_{\Lambda\nearrow \bZ^d}p_{\ga,\Lambda}(h)\,,
\text{  where  }p_{\ga,\Lambda}(h)=\frac{1}{\beta|\Lambda|}\log Z(\Lambda)\,.}
The free energy and pressure densities are related by a
Legendre transform:
\eq{\label{K18}f_\ga(m)=\sup_{h\in\bR}(hm-p_\ga(h))\,.}
See for instance \cite{Pr} for a proof of this property.
The analytic properties of $f_\ga$ at $\pm m^*(\beta,\ga)$ will 
be obtained from those of
$p_\ga$ at $h=0$. By the Theorem of Lee and
Yang \cite{LY}, $p_\ga$ is analytic outside the imaginary axis.
The main result of the paper is the following characterization of the
analyticity properties of the pressure at $h=0$.
\begin{theo}\label{KT2}
There exists $\beta_0$, $\ga_0>0$ and a constant $C_r>0$
such that for all $\beta\geq \beta_0$, $\ga\in
(0,\ga_0)$, the following holds:\\

\noindent 1) The directional derivatives 
$p_\ga^{(k),\leftarrow}(0)$ exists for all $k\in \bN$, i.e. $p_\ga$ is
$C^\infty$ at $h=0$. Moreover,
there exists a constant $C_+>0$ such that
for all $k\in \bN$,
\eq{\label{K19}\sup_{0\leq \real h\leq \epsilon}
|p_\ga^{(k),\leftarrow}(h)|
\leq \big(C_+\ga^\ddmu\beta^{-\frac{1}{d-1}}\big)^kk!^{\ddmu}+C_r^kk!\,.}
2) The pressure has no analytic continuation at $h=0$. More precisely,
there exists $C_->0$ and an unbounded increasing sequence of integers
$k_1,k_2,\dots$ such that for all $k\in \{k_1,k_2,\dots\}$,
\eq{\label{K20}|p_\ga^{(k),\leftarrow}(0)|\geq \big(C_-
\ga^{\ddmu}\beta^{-\frac{1}{d-1}}\big)^kk!^{\ddmu}-C_r^kk!\,.}
\end{theo}
The lower bound \eqref{K20} becomes irrelevant when $\ga\searrow 0$. Moreover,
we should mention that each integer $k_i$ depends on $\ga$ and $\beta$, 
with $\lim_{\ga\searrow 0}k_i=+\infty$:
information about non-analyticity is lost in the van der Waals limit.
Since we know from the Lebowitz-Penrose Theorem that $p_\ga$ converges, when
$\ga\searrow 0$, to a function that is
is analytic at $h=0$, it is worthwhile
considering the low order derivatives of $p_\ga$. Considering the upper bound
\eqref{K19}, it easy to show the
\begin{cor}
There exists $C=C(\beta)$ such that for small values of $k$,
i.e. for $k\leq \ga^{-d}$, we have the upper bound
\eq{\sup_{0\leq\real h\leq\epsilon}|p_\ga^{(k),\leftarrow}(h)|\leq C^kk!\,,}
\end{cor}
\noindent This 
shows that a close inspection of the derivatives of the pressure allows to
detect how analyticity starts to manifest when $\ga$ approaches
$0$. These different behaviours are illustrated on Figure \ref{KF00.01}.\\

To show Theorem \ref{KT2}, we first construct the
phase diagram of the Kac model with a complex magnetic field, at low
temperatures, $\ga$ small. Then, we adapt the technique of Isakov to
obtain lower bounds on the derivatives of the pressure in a
finite volume. These two essential steps deserve a few comments.
\begin{enumerate}
\item Phase diagrams of lattice systems can be studied in
the general framework of Pirogov-Sinai Theory (\cite{PS}, \cite{Z1}), which 
applies when the system under consideration has a finite number of ground 
states,
and for which the unperturbed hamiltonian satisfies the Peierls condition.
In our case, the Kac potential has two ground states 
which are the pure $+$ and pure $-$ configurations, but the Peierls constant
(computed with respect to these two ground states)
goes to zero when $\ga\searrow 0$ since 
in the van der Waals limit, the interaction
between two arbitrary spins vanishes. Therefore, a direct application of
Pirogov-Sinai Theory would lead to a range of temperature shrinking to
zero in the van der Waals limit.\\
Recently, Bovier and Zahradn\'\i k \cite{BZ2}
proposed a systematic method to study spin systems with long but finite range
interactions. Their technique allows to study, for instance, the 
Kac model with a complex
magnetic field, in a range of temperature that is 
\emph{uniform in} $\ga$. In their approach, the 
ground states of Pirogov-Sinai Theory 
are replaced by  \emph{restricted phases}, i.e. by \emph{sets} 
of configurations. In the 
$+$-restricted phase, for example, all the points are $+$-correct, i.e. their
$\ga^{-1}$-neighbourhood contains a majority of spins $+$. 
When a point is in neither of the
restricted phases, it is in the
support of a contour $\Ga$, and it can then be
shown that the contours defined in this way
satisfy the Peierls condition with a Peierls constant $\rho$ that is 
\emph{uniform} in $\ga$: $\|\Gamma\|\geq \rho |\Ga|$ where $\|\Ga\|$ is the
surface energy of $\Ga$.
In Section \ref{KSrestricted} we show that a polymer representation can be
obtained for the restricted phases, and that their corresponding free energies
behave analytically at $h=0$. The full phase diagram is then completed in
Section \ref{KSextended}: we give precise domains in which the
partition function can be exponentiated. These domains are
made optimal by introducing special isoperimetric constants associated to
contours (see the discussion hereafter, and \eqref{K38}).
Complications arise from the fact that polymers of the restricted phases
induce interactions among
contours. Besides the definition of the restricted ensembles, our
analysis of the phase diagram is independent of the
paper \cite{BZ2}. In a different setting, restricted 
ensembles were also studied in \cite{BS}, \cite{DS} and 
\cite{BKL}.
\item To implement the mechanism used by Isakov, we consider the pressure 
$p_{\ga,\Lambda}^+$ in a finite box $\Lambda$, with a pure 
$+$-boundary condition. By introducing an order among the
contours inside $\Lambda$, the pressure can be written as a \emph{finite} sum:
\eq{\label{K20.1}p_{\ga,\Lambda}^+=
\frac{1}{\beta|\Lambda|}\log Z^+_r(\Lambda)+\frac{1}{\beta|\Lambda|}
\sum_{\Ga\in \cC^+(\Lambda)}u_\Lambda^+(\Ga)\,,}
where $Z^+_r(\Lambda)$ is the restricted partition function and $\cC^+(\Lambda)$
is the family of all contours of type $+$ in $\Lambda$.
With the analysis of Sections \ref{KSrestricted} and \ref{KSextended},
the derivatives of the functions $u_\Lambda^+(\Ga)$ can be
estimated using a stationary phase
analysis.
When $\Lambda$ is sufficiently large, the contributions to
$p_{\ga,\Lambda}^{(k)}(0)$ are the following: because it is analytic,
the restricted phase contributes a
factor $C_r^kk!$. Then, a class of contours called $k$-large gives a
contribution of order $k!^\ddmu$. The rest of the contours is shown to have a
negligible contribution in comparison of the $k$-large ones. 
This gives a lower bound
\eq{\label{K20b} |p_{\ga,\Lambda}^{+(k)}(0)|\geq \big(C_-
\ga^{\ddmu}\beta^{-\frac{1}{d-1}}\big)^kk!^{\ddmu}-C_r^kk!\,.}
In the last step of
the proof we show that $\lim_{\Lambda}p_{\ga,\Lambda}^{+,(k)}(0)
=p_{\ga}^{(k),\leftarrow}(0)$, and so
\eqref{K20b} extends to the thermodynamic limit
$\Lambda\nearrow \bZ^d$, which gives \eqref{K20}.
\end{enumerate}

Before going further, we make an important remark.
In \cite{I1}, Isakov proved Theorem \ref{KT2} for the Ising model. An attempt
was then made, in a second paper \cite{I2}, to extend the method to any two
phase model for which the Peierls condition holds. Unfortunately, this extension
could only be done under two additional assumptions which we briefly describe.
Associate to each phase
a discrete isoperimetric problem of the following type: let $V(\Ga)$ denote the
volume of the contour $\Ga$ (of a given type) 
and $\|\Ga\|$ its surface energy. For $N\in\bN$, consider the problem:
\al{\begin{cases}
\text{Find}\text{ the best constant }C(N)
\text{ such that }
\frac{V(\Ga)}{\|\Ga\|}\leq C(N)V(\Ga)^{\usd}\nonumber\\
\text{for all contour $\Ga$ with }V(\Ga)\leq N\,.\nonumber
\end{cases}} 
The assumptions of Isakov are then that in the limit $N\to\infty$, 
1) the asymptotic behaviour of the constant $C(N)$ is the \emph{same}
for the two phases, 2) there exist maximizers of arbitrary large size.\\
Clearly, these assumptions are satisfied by the Ising model, for which
$\|\Ga\|=|\Ga|$ (the number of dual bonds on the dual lattice) 
and the maximizers are always given by cubes, i.e. $C(N)=(2d)^{-1}$ for all $N$.
But for a model with
no symmetry or with interactions that are more complicated than nearest
neighbours, these assumptions can be very hard to check. The
problem comes
from the fact that the surface energy $\|\Ga\|$ depends on the
detailed structure of the hamiltonian.
In our case, symmetry reduces the difficulty to the existence of large
maximizers. We will see that the construction of the phase diagram can be done
when the isoperimetric problem is formulated
as follows:
\al{\begin{cases}
\text{Find}\text{ the best constant }K(N)\text{ such that }
\frac{V(\Ga)}{\|\Ga\|}\leq K(N)V(\Ga)^{\usd}\nonumber\\
\text{for all contour $\Ga$ with }V(\Ga)\geq N\,.\nonumber
\end{cases}} 
By formulating the problem in this way, the existence of large maximizers is
immediate, and we avoid the necessity of
solving the isoperimetric problem explicitly.\\
It was actually shown in \cite{FP} that the two assumptions of Isakov
can be swept out, and that the result of \cite{I2} can be extended to
the whole class of two phase models treated generally in Pirogov-Sinai theory,
the only necessary ingredient for non-analyticity being the Peierls condition.
The general theorem of \cite{FP} applies to the Kac model but with some
restriction $\beta\geq \beta_0(\ga)$ where $\beta_0(\ga)$ diverges when
$\ga\searrow 0$.  In the present paper we
study the van der Waals limit at fixed $\beta$.\\

The description of the model in terms of contours
and the verification of the Peierls condition for $\|\Ga\|$
will be done in Section \ref{KScontours}. 
Section \ref{KSrestricted} is entirely devoted to the study of restricted 
phases and to their analyticity properties, adapting the technique of
\cite{BZ2}. Section
\ref{KSextended} is the construction of the phase diagram in the complex plane
of the magnetic field. Section \ref{KSderivatives} contains the proofs
of our main results. In Appendix \ref{KSLSP} we give 
the details of the stationary
phase analysis for the study of the derivatives of the functions
$u_\Lambda^+(\Ga)$, and Appendix \ref{KSnewCLEXP} contains basic definitions
for the cluster expansion technique.\\

\noindent \emph{Conventions:} we will 
often use the norm $\|f\|_D\pardef \sup_{z\in D}|f(z)|$. When $G$ is a
graph we denote by $V(G)$ its set of vertices and by $E(G)$ its set of edges.\\

\noindent \emph{Acknowledgments:} We wish to thank
Anton Bovier and Milo\v s Zahradn\'\i k for many useful
discussions concerning \cite{BZ2}.

\section{Contour Description}\label{KScontours}
For the description of configurations in terms of contours, we use the notion of
correct/incorrect point introduced by Bovier and Zahradn\'\i k in \cite{BZ2}.
There are two major requirements for the way in which contours should be
defined.
\begin{enumerate}
\item They are defined on a coarse-grained scale, and
a Peierls condition must hold for 
the surface energy of each contour, with a Peierls
constant that is \emph{uniform} in $\ga$. See Proposition \ref{KP1}.
\item Outside contours, a partial re-summation over configurations 
will lead to restricted phases. To obtain convergent expansions
for these phases, care must be taken in the definition of contours. See the
parameter $\tildelta$ in \eqref{K27}.
\end{enumerate}
\subsection{Definition of Contours}\label{KSScorrect}
We introduce some more notations. We have
$d(x,\Lambda)=\inf\{d(x,y):y\in \Lambda\}$. For $N\geq 1$, define the 
box $B_N(x)\pardef \{y\in\bZ^d:d(x,y)\leq N\}$, and
$B_N^\bigdot(x)\pardef 
B_N(x)\backslash \{x\}$. The $N$-neighbourhood of $\Lambda$ is
\eq{\label{K22.11}
[\Lambda]_N\pardef\bigcup_{x\in\Lambda}B_N(x)\,,}
and the boundaries
\al{\partial_N^+\Lambda&=\{x\in \Lambda^c:d(x,\Lambda)\leq N\}\,,\\
\partial_N^-\Lambda&=\{x\in \Lambda:d(x,\Lambda^c)\leq N\}\,.}
A set
$\Lambda$ is $N$-connected if for all $x,y\in\Lambda$ there exists a sequence
$x_1,x_2,\dots,x_{n-1},x_n$ with $x_1=x$, $x_n=y$, 
$x_i\in \Lambda$, and $d(x_i,x_{i+1})\leq N$.
If $\sigma_\Lambda\in \Omega_\Lambda$,
$\eta_{\Lambda^c}\in \Omega_{\Lambda^c}$, we define
the concatenation $\sigma_\Lambda
\eta_{\Lambda^c}\in\Omega$ in the usual way:
\eq{
(\sigma_\Lambda\eta_{\Lambda^c})_i=
\begin{cases}
(\sigma_\Lambda)_i&\text{ if }i\in\Lambda\,,\\
(\eta_{\Lambda^c})_i&\text{ if }i\in\Lambda^c\,.
\end{cases}}
We often use the symbol $\#$ to denote either of the symbols $+$ or $-$, or the
constant configuration taking the value $\#$ at each site of $\bZ^d$.
We define
\eq{\label{K23}\phi_{ij}(\sigma_i,\sigma_j)\pardef-\frac{1}{2}J_\ga(i-j)
(\sigma_i\sigma_j-1)\,,}
Let $\phi_{ij}\pardef \phi_{ij}(+,-)$. The \grasA{overall
interaction strength} is the upper bound on the energy of interaction 
of a single spin with the rest of the system, and equals
\eq{\label{K7}\sum_{j:j\neq i}\phi_{ij}=\sum_{j:j\neq i}J_\ga(i-j)=1\,.}
Relevant functions for the study of nearly constant spin regions
are the following (they will appear naturally
 later when reformulating the
hamiltonian):
\eq{\label{K24}w_{ij}^\#(\sigma_i,\sigma_j)\pardef 
\phi_{ij}(\sigma_i,\sigma_j)-\phi_{ij}(\#,\sigma_j)-\phi_{ij}(\sigma_i,\#)\,.}
Notice that $w_{ij}^\#(\#,\sigma_j)=w_{ij}^\#(\sigma_i,\#)=0$.
Let $\delta\in (0,1)$, $\sigma\in\Omega$. With regard to the
step function $J$ defined in \eqref{K0.00}, we define a point
$i$ to be $(\delta,+)$-\grasA{correct for} $\sigma$ if 
\eq{\label{K24.1}|B_R^\bigdot(i)\cap\{j:\sigma_j=-1\}|\leq 
{\textstyle\frac{\delta}{2}}|B_R(i)|\,.}
That is, the $R$-neighbourhood of a $(\delta,+)$-correct point 
contains a majority of $+$ spins.
Although we will always consider the
step function, it is often easier to formulate proofs with the help of the
functions $w_{ij}^\#$, since they will appear naturally later 
in the re-formulation of
the hamiltonian. We thus define the notion of correct/incorrect point in the
general case.
\begin{defin}\label{KDEF2}
Let $\delta\in(0,1)$, $\sigma\in \Omega$, $i\in \bZ^d$. 
\begin{enumerate}
\item $i$ is \emph{\grasA{$(\delta,+)$-correct for }}$\sigma$ 
if $\sum_{j:j\neq
i}|w_{ij}^+(-,\sigma_j)|\leq \delta$.
\item $i$ is \emph{\grasA{$(\delta,-)$-correct for }}$\sigma$ if $\sum_{j:j\neq
i}|w_{ij}^-(+,\sigma_j)|\leq \delta$.
\item $i$ is \emph{\grasA{$\delta$-correct for }}$\sigma$ if 
it is either $(\delta,+)$-
or $(\delta,-)$-correct for $\sigma.$
\item $i$ is \emph{\grasA{$\delta$-incorrect for }}$\sigma$ if 
it is not $\delta$-correct.
\end{enumerate}
\end{defin}
\noindent 
It is easy to see that this
definition coincides with \eqref{K24.1} when
$J$ is the step function.\\
The notion of correctness for a point $i$ depends on the spins in the
$R$-neighbour\-hood of $i$ but neither
on the value of $\sigma_i$, nor on the magnetic field. 
Notice that if $\delta=0$ this notion of correct point essentially 
coincides with the one of Zahradn\'\i k in \cite{Z1}. 
We first show that when $\delta$ is small, regions of
$(\delta,+)$- and $(\delta,-)$-correct points are distant.
In particular, a point
$i$ cannot be at the same time $(\delta,+)$- and $(\delta,-)$-correct.
\begin{lem}\label{KL1}
Let $\delta\in(0,2^{-d})$, $\sigma\in \Omega$. Then\\
1) If $i$ is $(\delta,+)$-correct, the box $B_R(i)$ contains either
$(\delta,+)$-correct, or $\delta$-incorrect points (but no
$(\delta,-)$-correct points).\\
2) If $i$ is $(\delta,-)$-correct, the box $B_R(i)$ contains either
$(\delta,-)$-correct, or $\delta$-incorrect points (but no
$(\delta,+)$-correct points).
\end{lem}
\bp 
Suppose $i$ is $(\delta,+)$-correct for $\sigma$.
Consider $j\in B_R(i)$ and compute
\al{\sum_{k:k\neq j}|w_{jk}^-(+,\sigma_k)|&=\sum_{\substack{k\in
B_R^\bigdot(j)\\ \sigma_k=+1}}2\phi_{jk}
\geq \sum_{\substack{k\in
B_R^\bigdot(j)\cap B_R^\bigdot(i)\\ \sigma_k=+1}}2\phi_{jk}}
Using the properties of the function $J$~\footnote{At this point we use the
particularity of the step function: $\phi_{jk}$ is constant on the intersection
$B_R^\bigdot(j)\cap B_R^\bigdot(i)$.},
we can exchange $j$ and $i$ and write
\al{\sum_{\substack{k\in
B_R^\bigdot(j)\cap B_R^\bigdot(i)\\ \sigma_k=+1}}2\phi_{jk}&=
\sum_{\substack{k\in
B_R^\bigdot(j)\cap B_R^\bigdot(i)\\ \sigma_k=+1}}2\phi_{ik}
&=\sum_{\substack{k\neq i
\\ \sigma_k=+1}}2\phi_{ik}-\sum_{\substack{k\not\in
B_R^\bigdot(j)\cap B_R^\bigdot(i)\\ \sigma_k=+1}}2\phi_{ik}\label{K24.2}}
Using \eqref{K7} and $|B_R(j)\cap B_R(i)|\geq 2^{-d}|B_R(i)|$,
this last sum can be bounded by
\eq{\sum_{\substack{k\not\in
B_R^\bigdot(j)\cap B_R^\bigdot(i)\\ \sigma_k=+1}}\phi_{ik}\leq
\frac{2^d-1}{2^d}\,.}
Then, since $i$ is $(\delta,+)$-correct for $\sigma$,
\al{\sum_{\substack{k\neq i
\\ \sigma_k=+1}}2\phi_{ik}&=2-\sum_{\substack{k\neq i
\\ \sigma_k=-1}}2\phi_{ik}
=2-\sum_{k:k\neq i}|w_{ik}^+(-,\sigma_k)|\geq 2-\delta\,.}
We thus have the lower bound 
\eq{\sum_{k:k\neq j}|w_{jk}^-(+,\sigma_k)|\geq
2-\delta-2\frac{2^d-1}{2^d}>\delta\,,}
i.e. $j$ cannot be $(\delta,-)$-correct for $\sigma$,
which finishes the proof.
\ep
\noindent
In the sequel we will always assume that $\delta\in(0,2^{-d})$ is fixed.
The \grasA{cleaned configuration} $\ovsigma\in \Omega$ is defined as follows:
\eq{\label{K25}
\ovsigma_i\pardef
\begin{cases}
+1&\text{ if }i\text{ is }(\delta,+)\text{-correct for }\sigma\,,\\
-1&\text{ if }i\text{ is }(\delta,-)\text{-correct for }\sigma\,,\\
\sigma_i&\text{ if }i\text{ is }\delta\text{-incorrect for }\sigma\,.
\end{cases}}
For any set $M\subset \bZ^d$, we can always consider the \grasA{partial 
cleaning}
$\sigma_M\ovsigma_{M^c}$ which coincides with $\sigma$ on $M$ and with 
$\ovsigma$
on $M^c$.
In the sequel, the cleaning and partial cleaning are 
always done according to the \emph{original}
configuration $\sigma$, with a fixed $\delta$. 
Notice that if a point $i$ is, say, 
$(\delta,+)$-correct for
$\sigma$, then the cleaning of $\sigma$ has the only effect, in the box
$B_R(i)$, of changing $-$ spins into $+$ spins (and not $+$ spins into $-$
spins). This is a consequence of Lemma \ref{KL1}.
We denote by $I_\delta(\sigma)$ the set of $\delta$-incorrect points of the
configuration $\sigma$.
The important property of the cleaning operation is stated in the
following lemma.
\begin{lem}\label{KL2}
Let $M_1\subset M_2$, $\delta'\in (0,\delta]$. 
Then $I_{\delta'}(\sigma_{M_1}\ovsigma_{M_1^c})\subset
I_{\delta'}(\sigma_{M_2}\ovsigma_{M_2^c})$.
\end{lem}
\bp

Let $i$ be a $(\delta',+)$-correct point of
$\sigma_{M_2}\ovsigma_{M_2^c}$.
Using the fact that $\sigma_{M_1}\ovsigma_{M_1^c}$ and
$\sigma_{M_2}\ovsigma_{M_2^c}$ coincide on $M_1$ and $M_2^c$, we decompose
\eq{\sum_{k:k\neq i}|w_{ik}^+(-,(\sigma_{M_1}\ovsigma_{M_1^c})_k)|=
\sum_{\substack{k:k\neq i\\k\in M_1\cup M_2^c}}
|w_{ik}^+(-,(\sigma_{M_2}\ovsigma_{M_2^c})_k)|
+\sum_{\substack{k:k\neq i\\k\in M_2\backslash M_1}}
|w_{ik}^+(-,\ovsigma_k)|\nonumber}
There are at most
three possibilities for a point $k$ of the last sum. 1) If $k$ is
$(\delta,+)$-correct for $\sigma$ then $\ovsigma_k=+1$ and so
$|w_{ik}^+(-,\ovsigma_k)|=0$. 2) If $k$ is
$\delta$-incorrect for $\sigma$ then $\ovsigma_k=\sigma_k
=(\sigma_{M_2}\ovsigma_{M_2^c})_k$. 3) If $k$ is
$(\delta,-)$-correct for $\sigma$ then it is also $(\delta,-)$-correct for 
$\sigma_{M_2}\ovsigma_{M_2^c}$. By Lemma \ref{KL1}, $i$ is not
$(\delta,+)$-correct for $\sigma_{M_2}\ovsigma_{M_2^c}$. This is a contradiction
with the fact that $i$ is $(\delta',+)$-correct 
for $\sigma_{M_2}\ovsigma_{M_2^c}$, so there are no such $k$. \\
We can then bound the whole sum by
$\delta'$. This shows that $i$ is $(\delta',+)$-correct for 
$\sigma_{M_1}\ovsigma_{M_1^c}$, and finishes the proof.
\ep
Contours are defined on a coarse-grained scale.
Consider the partition of $\bZ^d$ into disjoint
cubes $C^{(l)}$ of side length $l\in \bN$, $l> 2R$, whose centers lie
on the sites of a square sub-lattice of $\bZ^d$. We denote by $C_i^{(l)}$ the
unique box of this partition containing the site $i\in \bZ^d$. $\cC^{(l)}$
will denote the family of all subsets of $\bZ^d$ that are unions of boxes
$C^{(l)}$. For any set $A\subset\bZ^d$, consider the thickening (compare with 
\eqref{K22.11})
\eq{\label{K26}
\{A\}_{l}\pardef \bigcup_{i\in A}C_i^{(l)}\,.}
In the sequel we
always consider $l$ such that $l=\nu R$, with $\nu>2$.\\
We will need to decouple contours from the rest of the system. Since
interactions are of arbitrary large finite range, we follow \cite{BZ2} and
introduce a second parameter
$\tildelta\in(0,\delta)$. This new parameter is crucial; its 
importance will be seen later, for instance in the proof of the analyticity of
the restricted phases. For each
$\sigma\in\Omega$ with $|I_\tildelta(\sigma)|<\infty$, consider the following
set:
\eq{\label{K27}\cE(\sigma)\pardef\big\{M\in\cC^{(l)}:M\supset
[I_\delta(\sigma)]_R,\,M\supset 
[I_\tildelta(\sigma_M\ovsigma_{M^c})]_R\big\}\,.}
First we show that $\cE(\sigma)$ is not empty. Consider
$M_0\pardef\{[I_\tildelta(\sigma)]_R\}_{l}$. If $M_0=\emptyset$ then
$I_{\tildelta}(\sigma)=I_{\delta}(\sigma)=\emptyset$ 
and any subset of $\bZ^d$ is in $\cE(\sigma)$.
So we assume $M_0\neq \emptyset$. 
This gives $\cE(\sigma)\neq \emptyset$ since $M_0\in\cC^{(l)}$,
$M_0\supset [I_\tildelta(\sigma)]_R\supset[I_\delta(\sigma)]_R$ and
$M_0\supset [I_\tildelta(\sigma)]_R\supset [I_\tildelta(\sigma_{M_0}
\ovsigma_{M_0^c})]_R$
by Lemma \ref{KL2}.
We then show that $\cE(\sigma)$ is stable by intersection. Suppose $A,B\in
\cE(\sigma)$. Then clearly $A\cap B\supset [I_\delta(\sigma)]_R$ and
using again Lemma \ref{KL2},
\al{A&\supset [I_\tildelta(\sigma_A\ovsigma_{A^c})]_R\supset
[I_\tildelta(\sigma_{A\cap B}\ovsigma_{(A\cap B)^c})]_R\,,\\
B&\supset [I_\tildelta(\sigma_B\ovsigma_{B^c})]_R\supset
[I_\tildelta(\sigma_{A\cap B}\ovsigma_{(A\cap B)^c})]_R\,,}
which implies $A\cap B\in \cE(\sigma)$.
The following set is thus well defined, and is the candidate for describing the
contours of the configuration $\sigma$:
\eq{\label{K28} I^*(\sigma)\pardef \bigcap_{M\in \cE(\sigma)}M\,.}
By construction, $I^*(\sigma)$ is the smallest element of $\cE(\sigma)$.
A first important property of $I^*(\sigma)$ is the following, which will be
essential to obtain the Peierls bound on the surface energy of contours.
\begin{lem}\label{KL3}
There exists, in the $2R$-neighbourhood of each box $C^{(l)}\subset
I^*(\sigma)$, a point $j\in I^*(\sigma)$ 
which is $\tildelta$-incorrect for the configuration
$\sigma_{I^*}\ovsigma_{{I^*}^c}$. 
\end{lem}
\bp  Let $C^{(l)}\subset I^*(\sigma)$. 
First, suppose $I_\delta(\sigma)\cap [C^{(l)}]_{2R}\neq
\emptyset$. Then each $j\in I_\delta(\sigma)\cap [C^{(l)}]_{2R}$ is
$\delta$-incorrect for $\sigma$, and hence $\tildelta$-incorrect for 
$\sigma_{I^*}\ovsigma_{{I^*}^c}$, since $\tildelta<\delta$ and
$\sigma$ and $\sigma_{I^*}\ovsigma_{{I^*}^c}$ coincide on $B_R(j)$.\\
Suppose there exists a box $C^{(l)}$ such
that~\footnote{Here we use the fact that $A\cap[B]_{2R}=\emptyset$ if and only
if $[A]_R\cap[B]_R=\emptyset$.}
$[I_\delta(\sigma)]_R\cap [C^{(l)}]_R= \emptyset$.
If $I_\tildelta(\sigma_{I^*}\ovsigma_{{I^*}^c})\cap
[C^{(l)}]_{2R}=\emptyset$, i.e. 
$[I_\tildelta(\sigma_{I^*}\ovsigma_{{I^*}^c})]_R\cap
[C^{(l)}]_{R}=\emptyset$, then we define
$I'\pardef I^*\backslash C^{(l)}$ and show that $I'\in \cE(\sigma)$, a
contradiction with the definition of $I^*$. First, $I'\supset
[I_\delta(\sigma)]_R$. Using Lemma \ref{KL2}, 
$I^*\supset [I_\tildelta(\sigma_{I^*}\ovsigma_{{I^*}^c})]_R
\supset [I_\tildelta
(\sigma_{I'}\ovsigma_{{I'}^c})]_R$.
Since $[I_\tildelta(\sigma_{I^*}\ovsigma_{{I^*}^c})]_R\cap
[C^{(l)}]_R=\emptyset$, this implies $I'\supset
[I_\tildelta(\sigma_{I'}\ovsigma_{{I'}^c})]_R$, i.e. $I'\in\cE(\sigma)$.
\ep
When studying restricted phases, we will need to re-sum over the set of
configurations that have the same set of contours, that is to consider, for a
fixed $\sigma$ (we assume $I^*(\sigma)\neq \emptyset$),
\eq{\cA(\sigma)\pardef\big\{\sigma':\sigma'_{I^*(\sigma)}=
\sigma_{I^*(\sigma)},\,I^*(\sigma')=I^*(\sigma)\big\}\,.}
It is important to have an
\emph{explicit} characterization of the set $\cA(\sigma)$.
Let $\Lambda^\#(\sigma)$ denote the set of
points of $I^*(\sigma)^c$ that are $(\delta,\#)$-correct for $\sigma$. By 
Lemma
\ref{KL1} we have $d(\Lambda^+(\sigma),\Lambda^-(\sigma))>l$, and we
have the partition
\eq{\bZ^d=I^*(\sigma)\cup \Lambda^+(\sigma)\cup\Lambda^-(\sigma)\,.}
We now show that the set $\cA(\sigma)$ can be characterized explicitly
by
\al{\cD(\sigma)\pardef\big\{\sigma':&\,\sigma'_{I^*(\sigma)}=
\sigma_{I^*(\sigma)},\text{ each $i\in [\Lambda^\#(\sigma)]_R$}\nonumber
\text{ is $(\delta,\#)$-correct for $\sigma'$}\big\}\,.}
\begin{pro}\label{KP0.1}
If $I^*(\sigma)\neq \emptyset$, then $\cA(\sigma)=\cD(\sigma)$.
\end{pro}
\bp
1) Assume $\sigma'\in\cA(\sigma)$. Then $I^*\equiv I^*(\sigma)=I^*(\sigma')
\supset [I_\delta(\sigma')]_R$, so that each $i\in [{I^*}^c]_R$ is
$\delta$-correct for $\sigma'$. 
Let $A$ be a maximal connected component of $[{I^*}^c]_R$. There exists $i\in A$
such that $i\in I^*$, since we assumed $I^*\neq \emptyset$.
By Lemma \ref{KL1}, it suffices to show that $i$ is $(\delta,+)$-correct for
$\sigma$ if and only if it is $(\delta,+)$-correct for $\sigma'$. Assume this is
not the case, i.e. suppose $i$ is $(\delta,+)$-correct for $\sigma$ and
$(\delta,-)$-correct for $\sigma'$. That is,
\al{\sum_{j\neq i}|\omega^+_{ij}(-,(\sigma_{I^*}\ovsigma_{{I^*}^c})_j)|
&=\sum_{j\in B_R^\bigdot(i)\cap I^*}|w^+_{ij}(-,\sigma_j)|\leq \tildelta\,,\\
\sum_{j\neq i}|\omega^-_{ij}(+,(\sigma'_{I^*}\ovsigma'_{{I^*}^c})_j)|
&=\sum_{j\in B_R^\bigdot(i)\cap I^*}|w^-_{ij}(+,\sigma_j)|\leq \tildelta\,.}
Since $i\in I^*$ we have~\footnote{Here we use a property of the step
function, but this can be done for any Kac potential whose function $J$ has 
the symmetry  $J(x)=J(y)$ when $\|x\|=\|y\|$.}
\eq{\nonumber\sum_{j\in B_R^\bigdot(i)\cap {I^*}^c}|w^-_{ij}(+,(\sigma_{I^*}
\ovsigma_{{I^*}^c})_j)|\leq
\sum_{j\in B_R^\bigdot(i)\cap {I^*}^c}|w^-_{ij}(+,+)|
\leq 2(1-2^{-d})\,.}
Therefore we get a contradiction, since,
\al{2&=
\sum_{j\neq i}|w_{ij}^+(-,(\sigma_{I^*}\ovsigma_{{I^*}^c})_j)|
+|w_{ij}^-(+,(\sigma_{I^*}\ovsigma_{{I^*}^c})_j)|\nonumber\\
&\leq 2\tildelta+2 \sum_{\substack{j\in B_R^\bigdot(i)\cap {I^*}^c}}
|w^-_{ij}(+,(\sigma_{I^*}\ovsigma_{{I^*}^c})_j)|\leq 2\tildelta+2(1-2^{-d})<2\,,}
where we used the fact that $\tildelta<\delta<2^{-d}$.\\
2) Suppose $\sigma'\in\cD(\sigma)$. Since $\sigma'$ coincides with $\sigma$ on
$I^*(\sigma)$ and all points of $[I^*(\sigma)^c]_R$ are $\delta$-correct for
$\sigma'$, we have $I_\delta(\sigma')=I_\delta(\sigma)$. This gives
$I^*(\sigma)\supset [I_\delta(\sigma)]_R=[I_\delta(\sigma')]_R$. Then, since
${\sigma}_{I^*(\sigma)}
{\ovsigma}_{{I^*(\sigma)^c}}={\sigma'}_{I^*(\sigma)}
{\ovsigma'}_{{I^*(\sigma)^c}}$, we have 
$I^*(\sigma)\supset [I_\tildelta({\sigma}_{I^*(\sigma)}
{\ovsigma}_{{I^*(\sigma)^c}})]_R=
[I_\tildelta({\sigma'}_{I^*(\sigma)}
{\ovsigma'}_{{I^*(\sigma)^c}})]_R$.
This implies $I^*(\sigma)\in\cE(\sigma')$, i.e.
$I^*(\sigma')\subset I^*(\sigma)$. Assume 
$I^*(\sigma)\backslash I^*(\sigma')\neq \emptyset$. Using the fact that
$\sigma$ and $\sigma'$ coincide on $I^*(\sigma)\backslash I^*(\sigma')$, we 
have $\sigma_{I^*(\sigma')}\ovsigma_{I^*(\sigma')^c}=
\sigma'_{I^*(\sigma')}\ovsigma'_{I^*(\sigma')^c}$. This gives, like before, 
$I^*(\sigma')\supset [I_\tildelta({\sigma'}_{I^*(\sigma')}
{\ovsigma'}_{{I^*(\sigma')^c}})]_R=
[I_\tildelta({\sigma}_{I^*(\sigma')}
{\ovsigma}_{{I^*(\sigma')^c}})]_R$. With $I^*(\sigma')\supset
[I_\delta(\sigma')]_R=[I_\delta(\sigma)]_R$, this implies
$I^*(\sigma')\in\cE(\sigma)$, i.e. $I^*(\sigma')\supset I^*(\sigma)$. So
$\sigma'\in\cA(\sigma)$.
\ep
In particular, Proposition \ref{KP0.1} implies that the cleaned configuration
$\sigma_{I^*(\sigma)}\ovsigma_{I^*(\sigma)^c}$ is an element of $\cA(\sigma)$. 
\begin{defin}\label{KDEF3}. 
The connected components of $I^*(\sigma)$ form the support of the
\grasA{contours} of the configuration $\sigma$, and are written
$\supp\Ga_1,\dots,\supp\Ga_n$. A
\emph{\grasA{contour}} is 
thus a couple $\Ga=(\supp \Ga,\sigma_\Ga)$, 
where $\sigma_\Ga$ is the restriction of $\sigma$ to $\Ga$.\\
A family of contours $\{\Ga_1,\dots,\Ga_n\}$ is \emph{\grasA{admissible}} if 
there exists a configuration $\sigma$ such that $\{\Ga_1,\dots,\Ga_n\}$
are the contours of $\sigma$~\footnote{Note that this configuration is not
unique, unlike in the usual situation treated in Pirogov-Sinai Theory.}.
\end{defin}
The fact that the contours are defined on a coarse-grained scale will be crucial
when dealing with the entropy of contours, which we must control uniformly in
$\ga$. Notice that two (distinct)
contours are at distance at least $l$ from each other.
We will usually denote $\supp \Ga$ also by $\Ga$.
Contours should always be considered together with 
their type and labels, which we are about do define. The following
topological property is needed for the definition of labels.
\begin{lem}\label{KL4} Fix $R\geq 1$.
Let $B\subset \bZ^d$ be $R$-connected and bounded. Then $\partial_R^+A$ and
$\partial_R^-A$ are $R$-connected, where $A$ is any maximal
$R$-connected component
of $B^c=\bZ^d\backslash B$.
\end{lem}
\optionel{
\bp
Let $A$ be any maximal $R$-connected component of $B^c$. Then $A^c$ is
$R$-connected. Indeed, let $x,y\in A^c$, and consider a path
$x_1=x,x_2,\dots,x_n=y$, $d(x_i,x_{i+1})\leq R$. If $x_i\in A^c$ for all $i$
there is nothing to show. So suppose there exists $1\leq i_-\leq i_+\leq n$ such
that $\{x_1,\dots,x_{i_--1},x_{i_-}\}\subset A^c$, $x_{i_-+1}\in A$, 
$x_{i_+-1}\in A$, $\{x_{i_+},x_{i_++1},\dots,x_n\}\subset A^c$. Since $A$ is
maximal, we have $x_{i_-}\in B$, $x_{i_+}\in B$, and we can find a path from
$x_{i_-}$ to $x_{i_+}$ entirely contained in $B$, i.e. in $A^c$.\\
We then show that $\partial_1^+A$ is $R$-connected. Fix $\epsilon>0$
and consider the sets 
\al{X&=\{x\in \bR^d:d(x,A)\leq \frac{R}{2}+\epsilon\}\,,\\
Y&=\{y\in \bR^d:d(y,A^c)\leq \frac{R}{2}+\epsilon\}\,.}
Then $X,Y$ are closed arc-wise connected subsets of $\bR^d$, and $X\cup
Y=\bR^d$. By a Theorem of Kuratowski, $X\cap Y$ is 
arc-wise connected~\footnote{This property of $\bR^d$
is called \emph{unicoherence}. See \cite{Ku}, vol. $2$, 
Theorem $9$ of Chapter $57$.I,
and Theorem $2$ of Chapter $57$.II.}.
Let $\epsilon'>0$ and consider $x,y\in \partial_1^+A$, together with
$\tilde{x},\tilde{y}\in X\cap Y$ such that $d(x,\tilde{x})<\frac{1}{2}$,
$d(y,\tilde{y})<\frac{1}{2}$. Then consider any
sequence $\tilde{x}_1=\tilde{x},\dots,\tilde{x}_n=\tilde{y}$, 
$\tilde{x}_i\in X\cap Y$,
$d(\tilde{x}_i,\tilde{x}_{i+1})\leq \epsilon'$. For each $i$ we have
$d(\tilde{x}_i,A)\leq \frac{R}{2}+\epsilon$, 
$d(\tilde{x}_i,A^c)\leq \frac{R}{2}+\epsilon$. This implies that each box
$B_{\frac{R}{2}+\epsilon}(\tilde{x}_i)$ contains at least one element
$x_i'\in\partial_1^+A$, i.e. $d(\tilde{x}_i,x_i')\leq \frac{R}{2}+\epsilon$.
We have
\eq{d(x_i',x_{i+1}')\leq d(x_i',\tilde{x}_i)+
d(\tilde{x}_i,\tilde{x}_{i+1})+ d(\tilde{x}_{i+1},x_{i+1}')\leq
R+2\epsilon+\epsilon'\,.}
If $2\epsilon+\epsilon'<1$, this 
shows that $\partial_1^+A$ is $R$-connected, which implies that
$\partial_R^+A$ is $R$-connected. The same proof holds when $\partial_R^+A$ is
replaced by $\partial_R^-A$.
\ep}
Let $\Ga$ be a contour of $\sigma$, $A$ a maximal $R$-connected component of
$(\supp\Ga)^c$. Let $i\in \partial_R^-A$. By definition, $i$ is
$(\delta,\#)$-correct for $\sigma$ for
some $\#\in\{\pm 1\}$. By Lemmas \ref{KL4}
and \ref{KL1}, each $i'\in\partial_R^-A$ is 
$(\delta,\#)$-correct for $\sigma$ for
the \emph{same} value $\#$.  We call $\#$ the \grasA{label} of the component
$A$. There exists a unique unbounded component of $\Ga^c$. The label of this
component is called the \grasA{type} of the contour $\Ga$.
Let $\Ga$ be of type $+$ (resp. $-$). The union of all components of $\Ga^c$
with label $-$
(resp. $+$) is called the \grasA{interior} of $\Ga$, and is denoted
$\inte\Ga$. Notice that there is only one type of interior.
We define $V(\Ga)\pardef |\inte\Ga|$. The union
of the remaining components is called the \grasA{exterior} of $\Ga$, and is
denoted by $\ext \Ga$. A contour is 
\grasA{external} if it is not contained in the interior of another
contour.\\
Let $\Ga$ be a contour of some configuration
$\sigma$. Assume $\Ga$ is of type $+$.
Consider the configuration $\sigma[\Ga]$, which coincides with
$\sigma_\Ga$ on the support of $\Ga$, and which equals $+1$ on $\ext\Ga$,
$-1$ on $\inte\Ga$.
Using Proposition \ref{KP0.1}, it is easy to see that
$\sigma[\Ga]$ has a single contour, which is exactly $\Ga$. This can be
generalized to a family of external contours of the same type, as in the second
part of the following lemma.
\begin{lem}\label{KL4.1}
External contours have the following properties:\\
1) External contours of an admissible family have the same type.\\
2) Let $\{\Ga_1,\dots,\Ga_n\}$ be a family of external contours, all of the same
type. Then $\{\Ga_1,\dots,\Ga_n\}$ is admissible if and only if
$d(\Ga_i,\Ga_j)>l$ for all $i\neq j$.
\end{lem}
\bp
The first statement follows easily from Lemma \ref{KL4}. For the second, we can
assume that the contours are of type $+$. If $\{\Ga_1,\dots,\Ga_n\}$ is
admissible, then by construction the $\Ga_i$ 
are at distance at least $l$. Then,
assume $d(\Ga_i,\Ga_j)>l$ for all $i\neq j$.
Consider the 
configuration $\sigma[\Ga_1,\dots,\Ga_n]$, which coincides with
$\sigma_{\Ga_i}$ on the support of $\Ga_i$, which equals $+1$ on
$\bigcap_{i}\ext\Ga_i$ and $-1$ on $\bigcup_i\inte\Ga_i$. Then  the contours
of $\sigma[\Ga_1,\dots,\Ga_n]$ are given by $\{\Ga_1,\dots,\Ga_n\}$.
\ep
\subsection{Re-formulation of the Hamiltonian}
Consider a finite volume $\Lambda\in\cC^{(l)}$ with the pure $+$-boundary
condition $+_{\Lambda^c}\in \Omega_{\Lambda^c}$.
Let $\sigma_\Lambda\in\Omega_\Lambda$.
We set $\sigma\pardef \sigma_\Lambda +_{\Lambda^c}$.
The \grasA{hamiltonian with boundary condition} $+_{\Lambda^c}$ is defined by
\eq{\label{K28.55}
H_\Lambda(\sigma)=H_\Lambda(\sigma_\Lambda+_{\Lambda^c})
=\sum_{\substack{\{i,j\}\cap
\Lambda\neq \emptyset\\i\neq j}}
\phi_{ij}(\sigma_i,\sigma_j)+\sum_{i\in \Lambda}u(\sigma_i)\,,}
where $u(\sigma_i)=-h\sigma_i$, $h\in \bR$.
Since we work in a finite volume, we will from now on identify
$I^*(\sigma)$ with
$I^*(\sigma)\cap \Lambda$ and $\Lambda^\pm(\sigma)$ 
with $\Lambda^\pm(\sigma)\cap\Lambda$.
The following lemma shows how the
hamiltonian can be written in such a way that spins in correct regions interact
via the functions $w_{ij}^\#$ and are subject to an effective external field
$U^\#$.
\begin{lem}\label{KL5}
Define the potential $U^\#(\sigma_i)\pardef u(\sigma_i)+\sum_{j:j\neq
i}\phi_{ij}(\sigma_i,\#)$. Suppose
$\sigma_\Lambda$ is such that $I^*(\sigma)\cap
\partial_R^-\Lambda=\emptyset$. Then
\eq{\label{K28.6}H_\Lambda(\sigma)=
H_{I^*}(\sigma_{I^*}\ovsigma_{{I^*}^c})+\sum_{\#}
\Big(\sum_{\substack{\{i,j\}\cap \Lambda^\#\neq
\emptyset\\i\neq j}}w_{ij}^\#(\sigma_i,\sigma_j)
+\sum_{i\in \Lambda^\#}U^\#(\sigma_i)\Big)\,.}
\end{lem}
\bp
The proof is a simple rearrangement of the terms. Consider a pair $\{i,j\}$ 
appearing in $H_\Lambda(\sigma)$. 
Since $d(\Lambda^+,\Lambda^-)>R$ we have a certain number of 
cases to consider: 1)
$\{i,j\}\subset \Lambda^+$. In this case, write
\eq{\phi_{ij}(\sigma_i,\sigma_j)
=w^+_{ij}(\sigma_i,\sigma_j)+\phi_{ij}(\sigma_i,+)
+\phi_{ij}(+,\sigma_j)\,.}
The second term contributes to $U^+(\sigma_i)$, the third to $U^+(\sigma_j)$.
2) $i\in\Lambda^+$, $j\in I^*$.
In this case the third term contributes to
$H_{I^*}(\sigma_{I^*}\ovsigma_{{I^*}^c})$.
3) $i\in\Lambda^+$, $j\in\Lambda^c$; in this case, $\phi_{ij}(+,\sigma_j)=0$. 
The other cases are similar.
Notice that the case $i\in \Lambda^-$, $j\in \Lambda^c$ never occurs since
points of $\partial_R^-\Lambda$ can only be $(\delta,+)$-correct.
\ep
\subsection{Peierls Condition and Isoperimetric Constants}
We
take a closer look at the term $H_{I^*}$. Remember that contours are maximal 
$R$-connected components of $I^*$. For each contour $\Ga$,
$\sigma[\Ga]$ and $\sigma_{I^*}\ovsigma_{{I^*}^c}$ coincide on
$[I^*]_R$. Since
$d(\Ga,\Ga')>l$, we can decompose
\al{H_{I^*}(\sigma_{I^*}\ovsigma_{{I^*}^c})&=\sum_{\Ga}H_\Ga
(\sigma[\Ga])\label{K29}\\
&=\sum_{\Ga}\Big(\|\Ga\|+\sum_{i\in \Ga}u(\sigma[\Ga]_i)\Big)\,,\label{K29.1}}
where the sum is over contours of the configuration $\sigma$ (contained in
$\Lambda$), and  where the \grasA{surface energy} is defined as
\eq{\label{K30}\|\Ga\|\pardef \sum_{\substack{\{i,j\}\cap \Ga\neq
\emptyset\\i\neq j}}
\phi_{ij}(\sigma[\Ga]_i,\sigma[\Ga]_j)\,.}
The central result of this section is the following.
\begin{pro}\label{KP1}
The surface energy satisfies the \emph{\grasA{Peierls condition}}, i.e. 
there exists
$\rho=\rho(\tildelta,\nu)>0$ such that for all contour $\Ga$,
\eq{\|\Ga\|\geq \rho |\Ga|\,.}
The 
constant $\rho$ is independent of $\ga$ and is called the \emph{\grasA{Peierls
constant}}.
\end{pro}
\noindent We will need two lemmas. The first is purely geometric. 
\begin{lem}
For any finite set $A\subset \bZ^d$ and for all $R_0\in \bN$, there exists
$A_0\subset A$, called an \emph{\grasA{$R_0$-approximant of}} $A$, such that 
\begin{enumerate}
\item $A\subset [A_0]_{R_0}$\,,
\item $d(x,y)>R_0$ for all $x,y\in A_0$, $x\neq y$.
\end{enumerate}
\end{lem}

\noindent The 
second lemma is a property of the Kac potential. In
\cite{BZ2}, this property was called ``continuity'' for obvious reasons.
\begin{lem}\label{KL7}
Let $\sigma\in\Omega$, $i\in \bZ^d$, $\#\in\{\pm\}$. Define 
\eq{\label{K31.5}
V_\sigma(i;\#)\pardef \sum_{j:j\neq i}\phi_{ij}(\#,\sigma_j)\,.}
Then there exists $c_2>0$ such that for all $x,y$, $d(x,y)\leq R$,
\eq{\label{K32} |V_\sigma(x;\#)-V_\sigma(y;\#)|\leq c_2\frac{d(x,y)}{R}\,.}
\end{lem}
\bp
The difference $V_\sigma(x;\#)-V_\sigma(y;\#)$ can be expressed as follows:
\eq{\nonumber
\sum_{\substack{l\in B_R(x)\\ l\not\in B_R(y)}}\phi_{xl}(\#,\sigma_l)
+\sum_{\substack{l\in B_R(x)\cap B_R(y)}}\big(\phi_{xl}(\#,\sigma_l)-
\phi_{yl}(\#,\sigma_l)\big)
-\sum_{\substack{l\in B_R(y)\\l\not \in B_R(x)}}\phi_{yl}(\#,\sigma_l)}
The first and last sum can be estimated as follows:
\al{\sum_{\substack{l\in B_R(x)\\ l\not\in B_R(y)}}\phi_{xl}(\#,\sigma_l)&\leq
\big(|B_R(x)|-|B_R(x)\cap B_R(y)|\big)\sup \phi_{ij}\\
&\leq d c_\ga\big(\sup_t J(t)\big)
\Big(\frac{2R+1}{R}\Big)^{d-1}\frac{d(x,y)}{R}}
Since we are considering the step function, $\sup_t J(t)=2^{-d}$.
The middle sum vanishes~\footnote{Here we use for the second time
the fact that we are considering the step function \eqref{K0.00}.
Nevertheless, if $J$ is an arbitrary $K$-Lipshitz function:
\al{\sum_{\substack{l\in B_R(x)\cap B_R(y)}}\big|\phi_{xl}(\#,\sigma_l)-
\phi_{yl}(\#,\sigma_l)\big|&\leq
Kc_\ga\ga^d \sum_{\substack{l\in B_R(x)\cap B_R(y)}}d(\ga x,\ga y)\nonumber\\
&\leq Kc_\ga\ga^d|B_R(x)|\frac{d(x,y)}{R}\,.}\nonumber},
which finishes the proof.
\ep
\bp[Proof of Proposition \ref{KP1}:]
By Lemma \ref{KL3} there exists in the $2R$-neighbourhood of 
each $C^{(l)}\subset \Ga$ a point $j\in \Ga$ 
that is $\tildelta$-incorrect for $\sigma[\Ga]$. Let $A$ be
the set of all such points. We have $\Ga\subset [A]_{l+2R}$. Let $A_0$
be any $4R$-approximant of $A$.
We have $A\subset [A_0]_{4R}$, i.e. $\Ga\subset
[A_0]_{l+6R}$. Each $j\in A_0$ is $\tildelta$-incorrect for $\sigma[\Ga]$
i.e. satisfies
\eq{\sum_{k:k\neq j}|w_{jk}^\pm(\mp,\sigma[\Ga]_k)|>\tildelta\,.}
Since $|w_{jk}^\pm(\mp,\sigma[\Ga]_k)|=2\phi_{jk}(\pm,\sigma[\Ga]_k)$, 
\eq{V_{\sigma[\Ga]}(j;\pm)=
\sum_{k:k\neq j}\phi_{jk}(\pm,\sigma[\Ga]_k)>\frac{\tildelta}{2}\,.}
We bound the surface energy from below as follows:
\al{\|\Ga\|&\geq \frac{1}{2}\sum_{j\in A_0}\sum_{\substack{k\in B_R(j)\cap \Ga
}}\sum_{l:l\neq k}\phi_{kl}(\sigma[\Ga]_k,\sigma[\Ga]_l)\nonumber\\
&=\frac{1}{2}\sum_{j\in A_0}\sum_{\substack{k\in B_R(j)\cap \Ga
}}V_{\sigma[\Ga]}(k;\sigma[\Ga]_k)
\geq \frac{1}{2}\sum_{j\in A_0}\sum_{\substack{k\in B_R(j)\cap C_j^{(l)}}
}V_{\sigma[\Ga]}(k;\sigma[\Ga]_k)\\
&\phantom{=\frac{1}{2}\sum_{j\in A_0}\sum_{\substack{k\in B_R(j)\cap \Ga
}}V_{\sigma[\Ga]}(k;\sigma[\Ga]_k)}
\geq \frac{1}{2}\sum_{j\in A_0}\sum_{\substack{k\in B_R(j)\cap C_j^{(l)}
\\ d(k,j)\leq \frac{\tildelta}{4c_2}R}}V_{\sigma[\Ga]}(k;\sigma[\Ga]_k)}
where $c_2$ was defined in Lemma \ref{KL7}. Moreover we have, using
\eqref{K32}, for each $k$ of the sum,
\al{V_{\sigma[\Ga]}(k;\sigma[\Ga]_k)&=V_{\sigma[\Ga]}(j;\sigma[\Ga]_k)
+\big(V_{\sigma[\Ga]}(k;\sigma[\Ga]_k)-V_{\sigma[\Ga]}(j;\sigma[\Ga]_k)\big)\\
&\geq \frac{\tildelta}{2}-c_2\frac{d(k,j)}{R}\geq
\frac{\tildelta}{2}-c_2\frac{\tildelta}{4c_2}=\frac{\tildelta}{4}\,.}
We have used the fundamental fact that the correctness of a point $j$ does not
depend on the value taken by the spin $\sigma_j$.
This gives the lower bound
\eq{\|\Ga\|\geq \frac{1}{2}|A_0|\frac{1}{2^d}|B_{\frac{\tildelta}{4c_2}R}(0)|
\frac{\tildelta}{4}\geq \frac{\tildelta}{2^{d+3}}
|B_{\frac{\tildelta}{4c_2}R}(0)||B_{l+6R}(0)|^{-1}|\Ga|\geq \rho|\Ga|\,.}
\ep
Since the Peierls constant is uniform in $\ga$, we will be able to 
study the van der Waals limit at fixed $\beta$. Proposition
\ref{KP1} allows to define, for $N=1,2,\dots$, the following
numbers called \grasA{isoperimetric constants}:
\eq{\label{K38}
K(N)\pardef \inf\Big\{\kappa>0:V(\Ga)^\dmusd\leq
\kappa\|\Ga\|,\,\text{ for all }\Ga,\,V(\Ga)\geq N \Big\}\,.}
These constants will play a crucial role in the construction of the phase
diagram and in the study of non-analyticity. Some of their properties are given
in the following lemma.
\begin{lem}\label{KL10}
The sequence $K(N)$ is decreasing and there exists positive
constants $c_-,c_+$ such that
\eq{\label{K39}c_-\ga\leq \inf_NK(N)\leq\sup_NK(N)\leq c_+\ga\,.}
As a consequence, the following limit exists
\eq{\label{K40}K(\infty)\pardef \lim_{N\to\infty}K(N)\,.}
Moreover, there exists for all $\epsilon>0$ a sequence $(\Ga_N)_{N\geq 1}$,
$\lim_{N\to\infty}V(\Ga_N)=+\infty$, such that for $N$
large enough, 
\eq{\label{K41}(1-\epsilon)K(\infty)\|\Ga_N\|\leq V(\Ga_N)^\dmusd\leq
(1+\epsilon)K(\infty)\|\Ga_N\|\,.}
\end{lem}
\bp
$K(N)$ is decreasing by definition. For the upper bound, use the Peierls
condition and Lemma \ref{KL9} hereafter: for all $\Ga$,
\eq{\frac{V(\Ga)^\dmusd}{\|\Ga\|}\leq \frac{V(\Ga)^\dmusd}{\rho|\Ga|}
\leq \frac{1}{\rho l}=\frac{1}{\rho \nu}\ga\equiv c_+\ga\,.}
For the lower bound, we explicitly construct a large 
contour of cubic shape. Fix $N$ and take $M\in \bN$ so that
$\Lambda_M=[-M;+M]^d\cap \bZ^d$, $\Lambda_N\in \cC^{(l)}$, 
$|\Lambda_M|\geq 2N$.
Consider the configuration
$\sigma$ defined by $\sigma_i=-1$ if $i\in \Lambda_M$, $\sigma_i=+1$ if $i\in
\Lambda_M^c$. Clearly, $I^*(\sigma)$ contains a single contour $\Ga_M$ (of type
$+$). Using
\eqref{K7}, $\|\Ga_M\|\leq |\Ga_M|\leq 2l|\partial_1^+\Lambda_M|=
2\nu R|\partial_1^+\Lambda_M|$.
Taking $M$ large enough guarantees $|\Lambda_M|\geq 
V(\Ga_M)\geq \frac{1}{2}|\Lambda_M|$. This gives, since 
$|\partial_1^+\Lambda_M|=2d|\Lambda_M|^\dmusd$,
\eq{\frac{V(\Ga_M)}{\|\Ga_M\|}\geq
\frac{1}{2}\frac{1}{2\nu R}\frac{|\Lambda_M|}{|\partial_1^+\Lambda_M|}
\geq\frac{\ga}{8d\nu}V(\Ga_M)^\usd\equiv c_-\ga V(\Ga_M)^\usd\,.}
The existence of the sequence $(\Ga_N)_{N\geq 1}$ follows 
from the definition of $K(N)$ and from the existence of the 
limit $K(\infty)$.
\ep
\begin{lem}\label{KL9}
Let $B\in \cC^{(l)}$, and let $A$ be 
a maximal $R$-connected component of $B^c$. Then 
\eq{\label{K37}|B|\geq |\partial_{l}^+A|\geq l|A|^{\frac{d-1}{d}}\,.}
\end{lem}
\optionel{
\bp
Consider the edge boundary $\delta^+A\pardef\{e=\langle i,j\rangle:i\in A,j\in
A^c\}$, where $\langle i,j\rangle$ means that $i,j$ are nearest neighbours.
Decompose $\delta^+A=E_1\cup\dots\cup E_d$, where $E_\alpha$ is the set of edges
of $\delta^+A$ that are parallel to the coordinate axis $\alpha$. Suppose 
$e=\langle i,j\rangle$, $i\in A$, $j\in A^c$. Since $A$ is maximal, 
$C_j^{(l)}\subset B$. Moreover, 
\eq{T_e\pardef\big\{
j,j+(j-i),j+2(j-i),\dots,j+(\frac{l}{2}-1)(j-i)\big\}\subset B\,.}
For all $e,e'\in E_\alpha$, $T_e\cap T_{e'}=\emptyset$. So for all $\alpha$,
\eq{|\partial_{l}^+A|\geq \Big|\bigcup_{e\in E_\alpha}T_e\Big|=\sum_{e\in
T_\alpha}|T_e|=\frac{l}{2}|E_\alpha|\,.}
Considering the inequality $|\delta^+A|\leq d\max_{\alpha}|E_\alpha|$ and the
standard isoperimetric inequality $|\delta^+A|\geq 2d |A|^{\dmusd}$ finishes the
proof.
\ep}
\nouvellepage
\section{Restricted Phases}\label{KSrestricted}
Restricted phases intervene when a set of contours $\{\Ga\}$ is fixed (with a
configuration $\sigma_\Ga$ on each of them) and when we re-sum
over all the configurations that have this same set of contours.
The set of configurations having the
same set of contours was completely characterized in Proposition \ref{KP0.1}. 
We are thus naturally led 
to consider systems living in a volume $\Lambda$ 
with a boundary
condition $\cb$, with the constraint that 
each point $i\in[\Lambda]_R$ must be $\delta$-correct.
Our aim is to obtain a polymer representation for the partition function
of such systems, and to show that the associated free energy 
behaves analytically at $h=0$.
As will be seen, the presence of the constraint will allow to treat the system
in a way very similar to a high temperature expansion. The study of restricted
phases we present was invented by Bovier and Zahradn\'\i k in \cite{BZ2}. 
At a few points our development
differs slightly from theirs, so we expose all the details.\\
A source of complication will be that the definition of polymers, as well as
their weights, will depend
on the boundary conditions specified outside $\Lambda$. Typically, the $\Lambda$
we want to consider is the volume between a given set of contours and the
boundary of a box. That is, the boundary condition is specified partly by the
spins on the contours and partly by the boundary condition outside the
box. To have an
idea of the objects we are going to construct, see Figure \ref{KF12}.\\

We will only treat the case $+$, the case $-$ being 
similar by symmetry. Fix
$0<\tildelta<\delta<2^{-d}$. Consider any finite set $\Lambda\in\cC^{(l)}$.
First of all, we must consider boundary conditions of the following type:
\begin{defin}\label{KD5}
A boundary condition $\cb\in\Omega_{\Lambda^c}$ is 
$+$-\emph{\gras{admissible}} if each $i\in [\Lambda]_R$ is
$(\tildelta,+)$-correct for the configuration $+_\Lambda\cb$.
\end{defin}
\noindent More intuitively, a $+$-admissible 
boundary condition means that when looked from any point $i$
inside of $\Lambda$, there is a majority of spins $+1$ on the boundary.
In our case (i.e. with the step function), this 
can be formulated as: for
each $i\in[\Lambda]_R$,
\eq{|B_R^\bigdot(i)\cap B|\leq 
{\textstyle\frac{{\tildelta}}{{2}}}
|B_R(i)|\,,} 
where the set $B$ is
defined by
\eq{\label{K44}B=B(\cb)\pardef \{i\in\Lambda^c:(\cb)_i=-1\}\,.}
In this
sense, these boundary conditions are ``good''; there is hope in being able to
control the
$+$-phase in the volume $\Lambda$. Notice that the 
boundary condition specified by a contour on its interior is always admissible.
This is the reason why the 
parameter $\tildelta$ was introduced in their definition.
\\
We define the function that allows to realize the constraint obtained after
Proposition \eqref{KP0.1}:
consider a $+$-admissible boundary condition $\cb\in\Omega_{\Lambda^c}$.
Let $i\in[\Lambda]_R$, $\sigma_\Lambda\in\Omega_\Lambda$, and define
\eq{\label{KS41.01}
\UN_i(\sigma_\Lambda)\pardef
\begin{cases}
1&\text{ if $i$ is $(\delta,+)$-correct for $\sigma_\Lambda\cb$}\\
0&\text{ otherwise }.
\end{cases}}
Then define
\eq{\label{K41.1}\UN(\sigma_\Lambda)
=\UN_{\cb}(\sigma_\Lambda)\pardef\prod_{i\in[\Lambda]_R}
\UN_i(\sigma_\Lambda)\,.}
Notice that $\UN(+_\Lambda)=1$ since $\cb$ is $+$-admissible.
The hamiltonian we use for the restricted 
system is the one obtained after the
re-formulation of Lemma \ref{KL5} in a region of $+$-correct points.
Set $\sigma\pardef\sigma_\Lambda\cb$. 
The \grasA{restricted partition function with boundary condition} $\cb$ is 
\eq{\label{K42}
\rZ^+(\Lambda;\cb)\pardef
\sum_{\sigma_\Lambda\in\Omega_\Lambda}
\UN(\sigma_\Lambda)\exp\Big(-\beta\sum_{\substack{\{i,j\}
\cap\Lambda\neq\emptyset\\i\neq j}}
w_{ij}^+(\sigma_i,\sigma_j)-\beta\sum_{i\in\Lambda}
U^+(\sigma_i)\Big)\,.}
We will show that $\rZ^+$ can be put in the form $\rZ^+=e^{\beta h|\Lambda|}\rcZ$,
where $\rcZ$ is the partition function of a polymer model, having a normally 
convergent cluster expansion in 
\eq{H_+=\big\{h\in\bC:\real
h>-{\textstyle\frac{1}{8}}\big\}\,.}
The reason for $\log \rZ^+$ to behave analytically at 
$h=0$ is that the presence
of contours is suppressed by $\UN(\sigma_\Lambda)$, and that 
on each spin $\sigma_i=-1$ acts an effective
magnetic field 
\eq{U^+(-1)=h+\sum_{j:j\neq i}\phi_{ij}=1+h\,,} 
which is close to $1$ when $h$ is in a neighbourhood of $h=0$.
\subsection{Representation with Polymers}
The influence of a boundary condition can always be interpreted as a magnetic
field acting on sites near the boundary. We thus rearrange the terms of the
hamiltonian as follows:
\eq{\sum_{\substack{\{i,j\}\subset\Lambda\\i\neq j}}
w_{ij}^+(\sigma_i,\sigma_j)+\sum_{i\in\Lambda}
\Big(U^+(\sigma_i)+\sum_{j\in \Lambda^c}w^+_{ij}(\sigma_i,(\cb)_j)\Big)\,.}
By defining an new effective non-homogeneous magnetic field 
\eq{\mu_i^+(\sigma_i)\pardef U^+(\sigma_i)+h
+\sum_{j\in \Lambda^c}w^+_{ij}(\sigma_i,(\cb)_j)\,,}
we can extract a volume term from $\rZ^+$ and get $\rZ^+=e^{\beta
h|\Lambda|}\rcZ$, where
\eq{\label{K43}
\rcZ^+\pardef \sum_{\sigma_\Lambda\in\Omega_\Lambda}
\UN(\sigma_\Lambda)\exp\Big(-\beta\sum_{\substack{\{i,j\}
\subset\Lambda\\i\neq j}}
w_{ij}^+(\sigma_i,\sigma_j)-\beta\sum_{i\in\Lambda}
\mu^+_i(\sigma_i)\Big)\,,}
Notice that the field $\mu^+_i(\sigma_i)$ becomes independent of $\cb$ when
$d(i,\Lambda^c)>R$. 
Since $w_{ij}^+(\sigma_i,\sigma_j)=0$ if $\sigma_i=+1$ or $\sigma_j=+1$ and
$\mu^+_i(+1)=0$, we need only consider points $i$ with $\sigma_i=-1$, which
will be identified with the vertices of a graph. Each vertex of this graph
will then get a factor $e^{-\beta\mu^+_i(-1)}$. When $h\in H_+$,
\eq{\label{K43.1}
\real\mu^+_i(-1)=1+2\real h+\sum_{j\in \Lambda^c}w^+_{ij}(-,(\cb)_j)\geq
1-2{\textstyle \frac{1}{8}}-\tildelta> {\textstyle \frac{1}{2}}\,.}
We used the fact that $\tildelta<2^{-d}$.\\
The formulation of $\rcZ$ in terms of polymers will be a three step procedure.
We first express $\rcZ$
as a sum over graphs, satisfying a certain constraint inherited from
$\UN(\sigma_\Lambda)$. Then, we associate to each
graph a spanning tree and re-sum over all graphs having the same spanning tree.
We will see that the weights of the trees obtained have good decreasing
properties. Finally, the constraint is expanded, yielding sets on which the
constraint is \emph{violated}. These sets are linked with trees. After a second
partial re-summation, this yields a sum over polymers, which are nothing but
particular graphs with vertices living on $\bZ^d$ and whose 
edges are of length at most $R$.
\subsubsection{A sum over graphs} 
Let $\cG_\Lambda$ be the family of simple non-oriented graphs
$G=(V,E)$ where $V\subset \Lambda$, each edge $e=\{i,j\}\in E$ has $d(i,j)\leq
R$. For $e=\{i,j\}$,
set $w_e^+\pardef w_{ij}^+(-,-)$. Notice that $\omega_e^+=-2\phi_{ij}\leq 0$.
Define also $\mu^+_i\pardef \mu^+_i(-1)$.
Expanding the product over edges leads to the following expression
\al{\label{K45}
\rcZ=\sum_{\substack{
G\in \cG_\Lambda}}\UN(V(G))\prod_{e\in E(G)}(e^{-\beta
w_e^+}-1)\prod_{i\in V(G)}e^{-\beta\mu^+_i}\,,}
where $\UN(V)\pardef\UN(\sigma_\Lambda(V))$, and
$\sigma_\Lambda(V)\in\Omega_\Lambda$ is defined by
$\sigma_\Lambda(V)_i=-1$ if $i\in V$, $+1$ otherwise.
With this formulation in terms of graphs, the constraint $\UN(V(G))=1$ is
satisfied if and only if for each $i\in[\Lambda]_R$,
\eq{\label{K45.1}
\sum_{\substack{e=\{i,j\}\\j\in V(G)\cup B}}|w_e^+|\leq \delta\,.}
Moreover, the fact that the boundary condition $\cb$ is $+$-admissible reduces
to
\eq{\label{K45.2}\sum_{\substack{e=\{i,j\}\\j\in B}}|w_e^+|\leq \tildelta\,.}

\subsubsection{A sum over trees.}
Suppose we are given an algorithm that assigns to each connected graph
$G_0$ a deterministic spanning tree $T(G_0)$, in a translation invariant way
(that is if $G_0'$ is obtained from $G_0$ by translation then $T(G_0')$ is
obtained from $T(G_0)$ by the same translation). 
To be precise, we consider the Penrose algorithm considered in
Chapter 3 of \cite{Pf}~\footnote{The Penrose algorithm requires the choice of an
origin among the vertices of the graph. We choose
this origin as the smallest vertex of the
graph with respect to the lexicographical order.}. 
We apply the Penrose algorithm to each component of
each graph $G$ appearing in the partition function.
Let $\cT_\Lambda\subset \cG_\Lambda$ denote the set of all
forests. We have
\al{\rcZ&=\sum_{\substack{T\in\cT_\Lambda}}
\UN(V(T))\prod_{\ttt\in T} \omega^+(\ttt)\,,\label{K48.1}}
where the product is over trees of $T$,
and the weight of each tree is defined by
\eq{\label{K49}
\omega^+(\ttt)\pardef\sum_{\substack{G\in\cG_\Lambda:\\
T(G)=\ttt}}\prod_{e\in E(G)}(e^{-\beta
w_e^+}-1)\prod_{i\in V(G)}e^{-\beta\mu^+_i}\,.}
Notice that isolated sites $\{i\}\subset \Lambda$
are also considered as trees. In this case,
$\omega^+(\{i\})=e^{-\beta \mu^+_i}$. The following lemma
shows how the re-formulation in terms of trees allows to take advantage of the
constraint.
\begin{lem}\label{KL11}
Let $T\in\cT_\Lambda$ be a forest such that
$\UN(V(T))=1$. Then for each tree $\ttt\in T$,
\eq{\label{K49.1}
\|\omega^+(\ttt)\|_{H_+}\leq \prod_{e\in E(\ttt)}(e^{-\beta
w^+_e}-1)\prod_{i\in V(\ttt)}
e^{-\frac{1}{4}\beta}\,.}
\end{lem}
\bp For each $\ttt\in T$, let $E^*(\ttt)$ denote the set of edges of the maximal
$\partial$-connected 
graph of $\{G\in\cG_\Lambda:T(G)=\ttt\}$ 
(see \cite{Pf}). We can express the weight as follows:
\al{\omega^+(\ttt)&=\prod_{e\in E(\ttt)}(e^{-\beta
w_e^+}-1)\prod_{i\in V(\ttt)}e^{-\beta\mu^+_i}
\sum_{\substack{G\in\cG_\Lambda:\\T(G)=\ttt}}
\prod_{e\in E(G)\backslash E(\ttt)}(e^{-\beta w_e^+}-1)\\
&=\prod_{e\in E(\ttt)}(e^{-\beta
w_e^+}-1)\prod_{i\in V(\ttt)}
e^{-\beta\mu^+_i}\prod_{e\in E^*(\ttt)\backslash
E(\ttt)}e^{-\beta w_e^+}\,.}
Since $\UN(V(T))=1$, the constraint \eqref{K45.1} is satisfied, and the last 
product can be bounded by:
\al{\prod_{e\in E^*(\ttt)\backslash
E(\ttt)}e^{\beta|w_e^+|}&\leq \prod_{i\in V(\ttt)}
\prod_{\substack{e=\{i,j\}\\j\in V(\ttt)}}e^{\beta|w_e^+|}\\
&=\prod_{i\in V(\ttt)}
\exp{\beta\sum_{\substack{e=\{i,j\}\\j\in V(\ttt)}}|w_e^+|}
\leq\prod_{i\in V(\ttt)}e^{\beta\delta}\,.\label{K49.2}}
This gives the result, since $\real \mu^+_i\geq \frac{1}{2}$ by \eqref{K43.1}, 
and $\delta\leq 2^{-d}\leq \frac{1}{4}$.
\ep
Notice that to obtain \eqref{K49.2}, we only needed that the bound
\eq{\label{K49.3}
\sum_{\substack{e=\{i,j\}\\j\in V(\ttt)}}|w_e^+|\leq \delta\,\quad\quad \forall
\,i\in V(\ttt)}
be satisfied. This is weaker than
\eqref{K45.1} and 
clearly $\UN(V(T))=1$ only if \eqref{K49.3} is satisfied for all
$\ttt\in T$.
In the sequel we can thus assume that the trees we consider always satisfy 
\eqref{K49.3}, independently of each other.
So the bound \eqref{K49.1} can always be used.
A direct consequence of the last lemma is the following result which shows that
trees and their weights satisfy the main condition ensuring convergence of 
cluster expansions.
\begin{cor} \label{KC11}
Let $0<c\leq \frac{1}{8}\beta$, $\epsilon>0$. There exists $\ga_0>0$ and 
$\beta_1=\beta_1(\epsilon)$ such that for all $\ga\in(0,\ga_0)$, $\beta\geq
\beta_1$, the following bound holds:
\eq{\label{K49.4}
\sum_{\ttt:V(\ttt)\ni 0}\|\omega^+(\ttt)\|_{H_+}e^{c|V(\ttt)|}\leq
\epsilon\,.}
\end{cor}
\bp
Using Lemma \ref{KL11},
\eq{\|\omega^+(\ttt)\|_{H_+}e^{c|V(\ttt)|}\leq \prod_{e\in E(\ttt)}
(e^{-\beta w^+_e}-1)\prod_{i\in V(\ttt)}
e^{-\frac{1}{8}\beta}\,.}
When $\ttt$ is a single isolated point (the origin), then we have a factor
$e^{-\frac{1}{8}\beta}$. When $V(\ttt)\ni 0$, $E(\ttt)\neq \emptyset$, we
define the \grasA{generation} of $\ttt$, $\mathrm{gen}(\ttt)$, as the number of
edges of the longest self avoiding path in $\ttt$ starting at the origin. 
The
sum in \eqref{K49.4} is bounded by
\al{e^{-\frac{1}{8}\beta}+\sum_{g\geq 1}&
\sum_{\substack{\ttt:V(\ttt)\ni 0\\ \mathrm{gen}(\ttt)=g}}
\prod_{e\in E(\ttt)}
(e^{-\beta w^+_e}-1)\prod_{i\in V(\ttt)}
e^{-\frac{1}{8}\beta}\\
&\leq e^{-\frac{1}{8}\beta}+\sum_{g\geq 1}e^{-\frac{1}{16}\beta g}
\sum_{\substack{\ttt:V(\ttt)\ni 0\\ \mathrm{gen}(\ttt)=g}}
\prod_{e\in E(\ttt)}
(e^{-\beta w^+_e}-1)\prod_{i\in V(\ttt)}
e^{-\frac{1}{16}\beta}\\
&\leq e^{-\frac{1}{8}\beta}+\sum_{g\geq 1}e^{-\frac{1}{16}\beta
g}\alpha_g\,,}
where we defined ($V_l(\ttt)$ is the set of leaves of the tree $\ttt$):
\eq{\alpha_g\pardef\sum_{\substack{\ttt:V(\ttt)\ni 0\\ \mathrm{gen}(\ttt)=g}}
\prod_{e\in E(\ttt)} (e^{-\beta w^+_e}-1)
\prod_{i\in V(\ttt)\backslash V_l(\ttt)}e^{-\frac{1}{16}\beta}
\prod_{i\in V_l(\ttt)}e^{-\frac{1}{32}\beta}}
Before going further, we define
\eq{\label{K49.5}\ga_0\pardef\sup
\Big\{\ga>0:2c_\ga\ga^d\sup_sJ(s)\leq{\textstyle \frac{1}{64}}\Big\}\,.}
Since $e^{-\beta w^+_e}-1\leq \beta|w_e^+|e^{\beta|w^+_e|}$ and
$|w_e^+|=2\phi_{ij}$ we can bound,
when $\ga\leq\ga_0$,
\eq{\sum_{e\ni 0}\big(e^{-\beta w^+_e}-1\big)e^{-\frac{1}{32}\beta}\leq\beta
e^{-\frac{1}{64}\beta}\sum_{e\ni 0}|\omega^+_e|\leq 2\beta
e^{-\frac{1}{64}\beta}\equiv\beta\zeta(\beta)\,.}
We are going to show
that $\alpha_{g+1}\leq \alpha_g$ for all $g\geq 1$.
Clearly, a tree $\ttt$ of generation $g+1$ can be obtained from a sub-tree 
$\ttt'\subset \ttt$
of generation $g$ by attaching edges to leaves of $\ttt'$. Let $x$ be a leaf of
$\ttt'$. The sum over all possible edges (if any) attached at $x$ is
bounded by
\eq{\nonumber 
1+\sum_{k\geq 1}\frac{1}{k!}\sum_{e_1\ni x}\dots\sum_{e_k\ni
x}\prod_{i=1}^k\big(e^{-\beta w^+_{e_i}}-1\big)e^{-\frac{1}{32}\beta}\leq 
1+\sum_{k\geq 1}\frac{1}{k!}(\beta\zeta(\beta))^k=e^{\beta\zeta(\beta)}\,.}
Assuming $\beta$ is large enough so that $\zeta(\beta)\leq \frac{1}{32}$, 
the weight of the leaf $x$ changes into
$e^{-\frac{1}{16}\beta}e^{\beta\zeta(\beta)}\leq e^{-\frac{1}{32}\beta}$, which
is exactly what appears in $\alpha_g$. 
This shows that $\alpha_{g+1}\leq\alpha_g$. We
then have $\alpha_{g+1}\leq\alpha_g\leq \dots\leq \alpha_1$. 
Like we just did, it is easy to see that $\alpha_1\leq e^{-\frac{1}{32}\beta}$.
This proves the result.
\ep
\subsubsection{A sum over polymers.}
\begin{figure}[tbp]
\begin{center}
\input{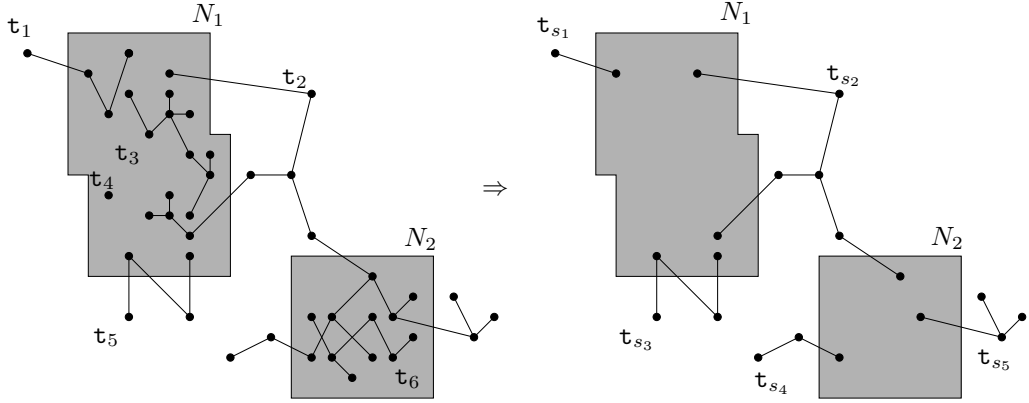}
\end{center}
\caption{The decimation 
procedure $P'_l\Rightarrow P_l$. 
The hatched polygons represent the body $\cB(P_l)$
and the legs are the trees
$\{\ttt_{s_1},\ttt_{s_2},\ttt_{s_3},
\ttt_{s_4},\ttt_{s_5}\}$. Each $\ttt_{s_j}$ is a sub-tree of some $\ttt_i$.}
\label{KF10}
\end{figure}

After the partial re-summation over the graphs having the same spanning tree,
the constraint $\UN(V(T))$ in
\eqref{K48.1} still depends on the relative positions of the trees.
This ``multi-body interaction'' can be worked out by expanding
\eq{\nonumber \UN(V(T))=\prod_{i\in [\Lambda]_R}\UN_i(V(T))
=\prod_{i\in [\Lambda]_R}(1+\UN_i^c(V(T)))
=\sum_{M\subset [\Lambda]_R}\prod_{i\in M} \UN_i^c(V(T))\,,}
where $\UN_i^c(V(T))\pardef \UN_i(V(T))-1$. This yields
\eq{\label{K55}\rcZ=\sum_{\substack{
T\in \cT_\Lambda}}\sum_{M\subset [\Lambda]_R}
\Big(\prod_{i\in M}\UN_i^c(V(T))\Big)\Big(
\prod_{\ttt\in T}\omega^+(\ttt)\Big)\,.}
Consider a pair $(T,M)$ in \eqref{K55}. Let $i\in M$.
The function $\UN_i^c(V(T))$ is non-zero only when $i$ is not
$(\delta,+)$-correct; it depends on the presence of
trees of $T$ in the $R$-neighbourhood of $i$ and eventually on the points of 
$B(\cb)$ if $B_R(i)\cap\Lambda^c\neq \emptyset$. 
To make these dependencies only local, we are going to
link the $R$-neighbourhood of points of $M$ with the trees of $T$.\\
Consider the graph $N=N(M)$ defined as follows: the vertices of $N$ are given by
\eq{V(N)\pardef\bigcup_{i\in M}B_R(i)\,.}
Then, $N$ has an edge between $x$ and $y$ if and only if
$\langle x,y\rangle$ is a pair of nearest neighbours of the same box $B_R(i)$
for some $i\in M$.
The graph $N$ decomposes naturally into connected components (in the
sense of graph theory) $N_1,N_2,\dots,N_K$. Some of these components can
intersect $\Lambda^c$.\\
We then link trees $\ttt_i\in T$ 
with components $N_j\in N$. 
To this end, we define an abstract graph $\hat{G}$: to each tree 
$\ttt_i\in T$,
associate an abstract vertex $w_i$ and to each component $N_j$ an abstract
vertex $z_j$. The edges of $\hat{G}$ are defined as follows: $\hat{G}$
has only edges between vertices $w_i$ and $z_j$, and this occurs if
and only if $V(\ttt_i)\cap V(N_j)\neq \emptyset$. 
Consider a connected component of $\hat{G}$, 
whose vertices 
$\{w_{i_1},\dots, w_{i_l},z_{j_1},\dots,z_{j_l}\}$ correspond to 
a set $P_l'=\{\ttt_{i_1},\dots, \ttt_{i_l},N_{j_1},\dots,N_{j_l}\}$.
We change $P_l'$ into a set $P_l$, using the following decimation
procedure: if $P_l'=\{\ttt_{i_1}\}$ is a single tree then $P_l\pardef P_l'$.
Otherwise, \\
1) delete from $P_l'$ all trees $\ttt_{i_k}$ that have no edges,\\
2) for all tree $\ttt_{i_k}$ containing at least one edge, 
delete all edges $e\in E(\ttt_{i_k})$
whose both end-points lie in the same component $N_{j_m}$.\\
The resulting set is of the form
$P_l=\{\ttt_{s_1},\dots,\ttt_{s_l},N_{j_1},\dots,N_{j_l}\}$, where each tree
$\ttt_{s_i}$ is a sub-tree of one of the trees $\{\ttt_{i_1},\dots,
\ttt_{i_l}\}$. $P_l$ is called a
\grasA{polymer}. The decimation 
procedure $P'_l\Rightarrow P_l$ is depicted on Figure \ref{KF10}.\\
The \grasA{body} of $P_l$ is $\cB(P_l)\pardef V(N_{j_1})\cup
\dots\cup V(N_{j_l})$. The \grasA{legs} of $P_l$, $\cL(P_l)$, are 
the trees $\{\ttt_{s_1},\dots,\ttt_{s_l}\}$.\\
A polymer can have no body (in which case it is a tree of $\cT_\Lambda$), or
no legs (in which case it is a single component $N_{j_1}$). We define the
\gras{support} $V(P)$ as the total set of sites:
\eq{V(P)\pardef \bigcup_{\ttt\in \cL(P)}V(\ttt)\cup \bigcup_{i}V(N_i)\,.}
Often we denote $V(P)$ also by $P$.
Two polymers are \grasA{compatible} if and only if $V(P_1)\cap
V(P_2)=\emptyset$, denoted $P_1\sim P_2$.
We have thus associated to each pair $(T,M)$ a family of pairwise compatible 
polymers $\{P\}\pardef\varphi(T,M)$. The set of all possible
polymers constructed in this way
is denoted $\cP_\Lambda^+ (\cb)$. The
representation of $\rcZ$ in terms of polymers is then
\eq{\label{K56}\rcZ= \sum_{\substack{\{P\}\subset
\cP^+_\Lambda(\cb)\\\text{compat.}}}\prod_{P\in\{P\}}\omega^+(P)\,,}
where the weight is defined by
\eq{\label{K57}
\omega^+(P)\pardef \sum_{\substack{(T,M):\\\varphi(T,M)=P}}\Big(\prod_{i\in M}
\UN_i^c(V(T))\Big)\Big(\prod_{\ttt\in T}\omega^+(\ttt)\Big)}
We should have in mind that $\omega^+(P)$ depends on the position of $P$ inside
the volume $\Lambda$, via the boundary condition $\cb$: more precisely if
$\cB(P)\cap \Lambda^c\neq\emptyset$ or if there
exists a leg $\ttt\in\cL(P)$ such that $d(\ttt,\Lambda^c)\leq R$.
Therefore, we define the family of  \grasA{free polymers} whose 
weight depends only on the
intrinsic structure of $P$, and not on the boundary condition.
The family $\cP^+$ is translation invariant, as well as the weight of each of
its polymers.
To any finite family $\cP$, we associate the partition function
\eq{\rcZ(\cP)\pardef\sum_{\substack{\{P\}\subset
\cP\\\text{compat.}}}\prod_{P\in\{P\}}\omega^+(P)\,,}
where the product equals $1$ when $\{P\}=\emptyset$.
For instance, we have obtained
\eq{\label{K64.0}
\rZ^+(\Lambda;\cb)=e^{\beta h|\Lambda|}\rcZ(\cP^+_\Lambda(\cb))\,.}
Everything we have done until now can be done for a $-$-admissible
boundary condition $\tau_{\Lambda^c}$, yielding a family of 
polymers $\cP^-_\Lambda(\tau_{\Lambda^c})$,
with weights $\omega^-(P)$. In this case, sites get a
factor $e^{-\beta\mu^-_i}$. In particular, if we 
consider the spin-flipped boundary condition
$-\cb$ defined by $(-\cb)_i\pardef -(\cb)_i$, which is $-$-admissible, we have
when $h$ is purely
imaginary~\footnote{Here, $\overline{z}$ denotes the complex conjugate of $z$.},
\eq{\label{K64.01}
\overline{\rcZ(\cP^+_\Lambda(\cb))}=\rcZ(\cP^-_\Lambda(-\cb))\,.}
This symmetry holds only because we consider free polymers.
\subsection{Analyticity of the Restricted Phases}\label{KSanalrestr}
\begin{figure}[b]
\begin{center}
\input{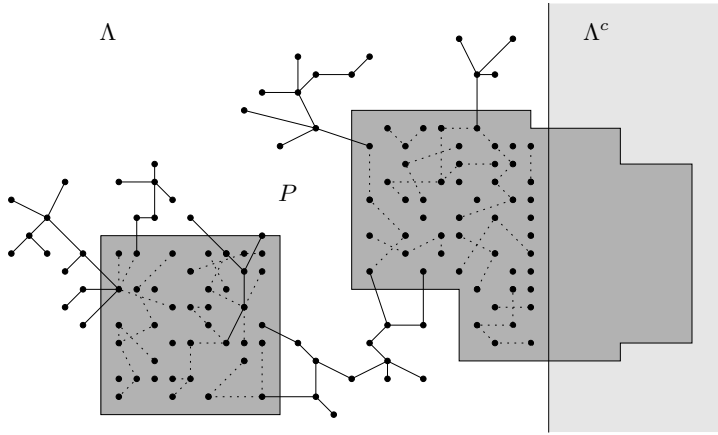}
\caption{{ The re-summation of 
Lemma \ref{KL11.1}. We emphasized the fact that the forest $T$
must have many points in $\cB(P)\cap \Lambda$, as was shown in \eqref{K58.4}.}}
\label{KF12}\end{center}
\end{figure}

Define the \grasA{restricted pressures} by
\eq{\label{K64.02}
p_{r,\ga}^\pm\pardef\lim_{\Lambda\nearrow\bZ^d}\frac{1}{\beta|\Lambda|}\log
\rZ^\pm(\Lambda;\pm_{\Lambda^c})\,\,,}
where the thermodynamic limit is taken along a sequence of cubes.
A result of the present section is that the restricted pressures, unlike
the total pressure $p_\ga$, behave analytically at $h=0$.\\
We study the weight $\omega^+(P)$ ($\omega^-(P)$ is similar by
symmetry).
The point is that we linked trees with the $R$-neighbourhood of points of
the set $M$, and we must now see that this thickening does not destroy, from the
point of view of entropy, the
uniformity we have been able to obtain with respect to the scaling parameter
$\gamma$. Moreover, the body of polymers can intersect $\Lambda^c$. At this
point we will see that $\delta-\tildelta>0$ is crucial.
\begin{lem}\label{KL11.1}
There exists $\beta_2$ and $\tau_0>0$ such that
for all $\beta\geq \beta_2$ and for all $\gamma\in (0,\ga_0)$, the following
holds: each polymer $P\in \cP_\Lambda^+(\cb)$ satisfies
\al{\label{K64.03}
\|\omega^+(P)\|_{H_+}
&\leq e^{-\tau_0\beta|\cB(P)|}\prod_{e\in \cL(P)}(e^{-\beta w_e^+}-1)
\prod_{i\in\cL(P)}e^{-\frac{1}{12}\beta}\,.}
\end{lem}
\bp Remember that the bound \eqref{K49.1} holds for
each tree under consideration. If $\cB(P)=\emptyset$, then $P$ is a tree and 
the result follows from Lemma \ref{KL11}. Otherwise,
\eq{\nonumber\|\omega^+(P)\|_{H_+}\leq 
\sum_{\substack{(T,M):\\\varphi(T,M)=P}}\Big(\prod_{i\in M}
|\UN_i^c(V(T))|\Big)
\prod_{\ttt\in T}\Big(\prod_{e\in E(\ttt)}(e^{-\beta w_e^+}-1)\prod_{i\in
V(\ttt)}e^{-\frac{1}{4}\beta}\Big)\,.}
Consider a pair $(T,M)$ such that $\varphi(T,M)=P$. Let 
$i_0\in M$, and assume 
$\UN_{i_0}^c(V(T))\neq 0$. This implies, with regard to \eqref{K45.1},
\eq{\label{K58.1}\sum_{\substack{e=\{i_0,j\}\\ j\in V(T)\cup B}}|w_e^+|
>\delta\,.}
But, according to \eqref{K45.2}, we have
\eq{\sum_{\substack{e=\{i_0,j\}\\ j\in B}}|w_e^+|\leq \tildelta\,.}
This implies the crucial lower bound
\eq{\sum_{\substack{e=\{i_0,j\}\\ j\in V(T)}}|w_e^+|\geq \delta-\tildelta
>0\,.}
Since $|w_e^+|=2\phi_{ij}\leq 2c_\gamma \gamma^d \sup_s J(s)$, we can 
find a constant $c_3$ such that 
\eq{\label{K58.8} |V(T)\cap B_R^\bigdot(i_0)|
>(\delta-\tildelta)c_3|B_R(i_0)|\,.}
In this sense, the forests that contribute to $\omega^+(P)$ accumulate
in the neighbourhood of each point $i_0\in M$. See Figure \ref{KF12}.
Let $M_0$ be any $2R$-approximant of $M$. Then we have $|\cB(P)|\leq
|M_0||B_{3R}(0)|$ and so
\eq{\label{K58.4} |V(T)\cap\cB(P)|\geq 
\sum_{i_0\in M_0}|V(T)\cap B_R(i_0)|\geq (\delta-\tildelta) c_4|\cB(P)|\,}
where $c_4$ is a constant.
Now, each $i\in V(T)$ gets a factor $e^{-\frac{1}{4}\beta}=
e^{-3\frac{1}{12}\beta}$. One factor
$e^{-\frac{1}{12}\beta}$ contributes to 
extract a term decreasing exponentially
fast with the size of $\cB(P)$, using \eqref{K58.4}:
\eq{e^{-\frac{1}{12}(\delta-\tildelta) c_4\beta |\cB(P)|}\,.}
A second factor $e^{-\frac{1}{12}\beta}$
contributes to the weight of the legs. Extracting this
contribution gives
\eq{\prod_{e\in \cL(P)}(e^{-\beta w_e^+}-1)\prod_{i\in
\cL(P)}e^{-\frac{1}{12}\beta}\,,}
The last factor $e^{-\frac{1}{12}\beta}$ is used to re-sum 
over all the possible configurations of $T$ inside
the body $\cB(P)$ (see Figure \ref{KF12}), that is over all forests $T'$,
$V(T')\subset \cB(P)$, where each tree $\ttt'\in T'$ gets a weight bounded by
\eq{\label{K58.88}
\omega_0(\ttt')\pardef\prod_{e\in E(\ttt')}(e^{-\beta
w_e^+}-1)\prod_{i\in V(\ttt')}e^{-\frac{1}{12}\beta}\,.} 
The remaining sum is thus bounded by
\al{\label{K58.89}\sum_{T':V(T')\subset \cB(P)}
&\prod_{\ttt'\in T'}\omega_0(\ttt')\equiv\Theta_0(\cB(P))\,.}
This partition function can be studied with a convergent cluster expansion.
Proceeding as we did in Corollary \ref{KC11}, we can take $\beta$ sufficiently
large so that the weight $\omega_0(\ttt')$ satisfies \eqref{K49.4}. We can then
guarantee that
\eq{|\log \Theta_0(\cB(P))|\leq |\cB(P)|\,.}
The sum over all possible sets $M$ such that $N(M)$ has a set of vertices given 
by $\cB(P)$ is bounded by $2^{|\cB(P)|}$.
Altogether these bounds give
\al{e^{-\frac{1}{12}(\delta-\tildelta) 
c_4\beta |\cB(P)|}2^{|\cB(P)|} e^{|\cB(P)|}
\equiv e^{-\tau_0\beta |\cB(P)|}\,,\nonumber}
which finishes the proof.
\ep
We now give the consequence of this lemma, namely that polymers satisfy the
main criterion needed for having a convergent cluster expansion.
\begin{cor}\label{KCO.1} 
Let $0<c\leq \min(\frac{\tau_0}{2},\frac{1}{24})\beta$, $\epsilon>0$.
There exists $\beta_3=\beta_3(\epsilon)$, such that
for all $\beta\geq \beta_3$ and for all $\gamma\in (0,\ga_0)$, the following
holds: 
\eq{\label{K59.1}\sum_{\substack{P:
V(P)\ni 0}}\|\omega^+(P)\|_{H_+}e^{c|{V}(P)|}\leq
\epsilon\,.}
\end{cor}
\bp 
Lemma \ref{KL11.1} allows to bound 
\al{\label{K59.20}
\|\omega^+(P)\|_{H_+}\leq \Big(\prod_{N\in P}\omega_0(N)\Big)
\Big(\prod_{\ttt\in \cL(P)}\omega_0(\ttt)\Big)\equiv \omega_0(P)\,,}
where the weight of each component of the body 
$N$ is $\omega_0(N)\pardef
e^{-\tau_0\beta|V(N)|}$ and the weight of each leg $\ttt$
was defined in \eqref{K58.88}. Fix $\epsilon>0$ small.
It is easy to show that when $\beta$ is large enough,
\eq{\label{K60.02}\sum_{\substack{N:V(N)\ni
0}}\omega_0(N)e^{(c+\epsilon)|V(N)|}\leq {\textstyle
\frac{1}{2}}\epsilon\,,}
and, proceeding like in Corollary \ref{KC11},
\eq{\label{K60.2}\sum_{\ttt:V(\ttt)\ni
0}\omega_0(\ttt)e^{(c+\epsilon)|V(\ttt)|}\leq{\textstyle
\frac{1}{2}}\epsilon\,.}
Let $n(P)$ denote the number of objects (components $N$ and trees $\ttt$)
contained in $P$. That is, if $P=\{\ttt_1,\dots,\ttt_L,N_1,\dots,N_K\}$, then
$n(P)=L+K$. We will show by induction on $N=1,2,\dots$\footnote{We thank
Daniel Ueltschi for suggesting
this method, which will be used
again later in the study of the phase diagram.} that
\eq{\lambda_N\pardef
\sum_{\substack{P:V(P)\ni 0\\ n(P)\leq N}}\omega_0(P)e^{c|V(P)|}\leq
\epsilon\,,} which will finish the proof.
If $N=1$ then $P$ can be either a single component $N$ or a tree $\ttt$. The
bound then follows from \eqref{K60.02} and \eqref{K60.2}. Suppose $\beta$ is
large and that the bound
holds for $N$. If $P$ satisfies $V(P)\ni 0$, $n(P)\leq N+1$, we choose an
object of $P$ that contains the origin (which can be a tree $\ttt_0$ or a
component $N_0$), and decompose $P$ as follows: either
$P=\{N_0\}\cup \{P_1,\dots,P_k\}$ with $V(N_0)\ni 0$, $V(P_i)\cap V(N_0)\neq
\emptyset$, $n(P_i)\leq N$, $P_i\sim P_j$ for $i\neq j$, or
$P=\{\ttt_0\}\cup \{P_1,\dots,P_k\}$ with $V(\ttt_0)\ni 0$, and $V(P_i)\cap
V(\ttt_0)\neq \emptyset$, $n(P_i)\leq N$, $P_i\sim P_j$ for $i\neq j$.
In the first case, we have, using the induction hypothesis and \eqref{K60.02},
\al{\sum_{N_0:V(N_0)\ni 0}&\omega_0(N_0)e^{c|V(N_0)|}
\sum_{k\geq 0}\frac{1}{k!}
\Big(\sum_{\substack{P:V(P)\cap V(N_0)\neq \emptyset\\ 
n(P)\leq N}}\omega_0(P)e^{c|V(P)|}\Big)^k\\
&\leq \sum_{N_0:V(N_0)\ni 0}\omega_0(N_0)e^{c|V(N_0)|}
\sum_{k\geq 0}\frac{1}{k!}
\big(|V(N_0)|\lambda_N\big)^k\\
&\leq \sum_{N_0:V(N_0)\ni 0}\omega_0(N_0)e^{c|V(N_0)|}
e^{\epsilon|V(N_0)|}
\leq {\textstyle\frac{1}{2}}\epsilon\,.}
In the second case the same computation yields, using \eqref{K60.2},
\al{\nonumber\sum_{\ttt_0:V(\ttt_0)\ni 0}&\omega_0(\ttt_0)e^{c|V(\ttt_0)|}
\sum_{k\geq 0}
\frac{1}{k!}
\Big(\sum_{\substack{P:V(P)\cap V(\ttt_0)\neq \emptyset\\ 
n(P)\leq N}}\omega_0(P)e^{c|V(P)|}\Big)^k\\
&\leq \sum_{\ttt_0:V(\ttt_0)\ni 0}
\omega_0(\ttt_0)e^{c|V(\ttt_0)|}e^{\epsilon|V(\ttt_0)|}\leq
{\textstyle\frac{1}{2}}\epsilon\,.}
This shows that $\lambda_{N+1}\leq \epsilon$ and finishes the proof.
\ep
\noindent  We now
state the main result concerning restricted phases and their analyticity
properties, again only for the case $\#=+$. We refer to Appendix
\ref{KSnewCLEXP} for notations. Here polymers play the role of \animals{}.
Clusters of polymers associated to $\cP^+_\Lambda(\cb)$ 
are denoted $\hatP\in \hatcP^+_\Lambda(\cb)$. By
Lemma \ref{KL50}, \eqref{K59.1} implies
\eq{\label{K70}
\sup_{x\in\Lambda}\sum_{\hatP \ni x}\|\omega^+(\hatP)\|_{H_+}\leq
\sup_{x\in\Lambda}\sum_{\hatP \ni x}|\omega_0(\hatP)|\leq \eta(\epsilon)\,,}
where the weights $\omega^+(\hatP)$ and $\omega_0(\hatP)$ are defined like in
\eqref{K601.1}.
Since $\epsilon$ can be made arbitrarily small by taking $\beta$ large enough,
we will replace $\eta(\epsilon)$ by a function $\epsilon_r(\beta)$, where the
subscript $r$ is to indicate that this function concerns the restricted phase.
We define $\tilde{H}_+\pardef\{\real h>
-{\textstyle\frac{1}{16}}\}\subset H_+$.
\begin{theo}\label{KT3} Let $\beta$ be large enough, $\ga\in(0,
\ga_0)$. Let $\Lambda\in\cC^{(l)}$ and $\cb$ be a $+$-admissible boundary
condition.
Then $\rcZ(\cP_\Lambda^+(\cb))$ has a
cluster expansion that converges normally in $H_+$, given by
\eq{\label{K65}
\log \rcZ(\cP_\Lambda^+(\cb))=\sum_{\hatP\in \hatcP^+_\Lambda(\cb)}
\omega^+(\hatP)\,.}
The maps $h\mapsto \log \rcZ(\cP_\Lambda^+(\cb))$, $h\mapsto p_{r,\ga}^+(h)$ 
are analytic in $H_+$.
Moreover there exists a function $\epsr(\beta)$,
$\lim_{\beta\nearrow\infty}\epsr(\beta)=0$, such that 
\al{\big\|\log \rcZ(\cP_\Lambda^+(\cb))&\big\|_{H_+}\leq \epsr(\beta)
|\Lambda|\,,\quad \sum_{\substack{\hatP\in \hatcP^+_\Lambda(\cb)\\ \hatP\ni 0}}
\|\omega^+(\hatP)\|_{H_+}\leq \epsr(\beta)\,,
\label{K65.1}\\
\big\|&\deriv{h}\log \rcZ(\cP_\Lambda^+(\cb))\big\|_{\tilde{H}_+}\leq 
\epsr(\beta)|\Lambda|\,.\quad \label{K65.2}}
\end{theo}
The proof of the theorem follows easily from Lemma \ref{KL50}. Analyticity
follows from the fact that the convergence is normal on $H_+$. 
The bound on the
first derivative is obtained by using the Cauchy formula:
any disc of radius $\frac{1}{16}$ centered at
$z\in\tilde{H}_+$ is contained in $H_+$. 
This also implies the existence of a constant $C_r>0$
such that for all integer $k\geq 2$,
\eq{\label{K65.4}\frac{1}{|\Lambda|}
\left|\derivk{h}{k}\log \rZ^+(\Lambda;\cb)\right|_{h=0}\leq
C_r^kk!\,,\,\quad\,|p_{r,\ga}^{+(k)}(0)|\leq C_r^kk!\,.}
\nouvellepage
\section{The Phase Diagram}\label{KSextended}
Throughout this section and until the end of the paper we assume
$\ga\in(0,\ga_0)$ is fixed, where $\ga_0$ was given in \eqref{K49.5}.
To start with, consider the partition function
\eq{\label{K300.1}
Z^+(\Lambda)\pardef\sum_{\substack{\sigma_\Lambda\in\Omega_\Lambda^+}}
e^{-\beta H_\Lambda(\sigma_\Lambda+_{\Lambda^c})}\,,}
where
\eq{\label{K300}\Omega_\Lambda^\newplus
\pardef\{\sigma_\Lambda\in\Omega_\Lambda:
d(I^*(\sigma_\Lambda+_{\Lambda^c}),\Lambda^c)>l\}\,.}
For each $\sigma_\Lambda\in\Omega_\Lambda^\newplus$, the decomposition of
$I^*(\sigma_\Lambda+_{\Lambda^c})$ into connected components yields an
admissible family
$\{\Ga\}$, such that
$\Ga\subset \Lambda$ and $d(\Ga,\Lambda^c)>l$ for each $\Ga\in\{\Ga\}$.
Then, $\Lambda$ is decomposed
into $\Lambda=\{\Ga\}\cup \Lambda^+\cup\Lambda^-$, where $\Lambda^\#$ are the
points of
$\Lambda\backslash\{\Ga\}$ that are $(\delta,\#)$-correct for
the configuration $\sigma_\Lambda+_{\Lambda^c}$.\\
In \eqref{K300.1}, we re-sum over the configurations $\sigma_{\Lambda^+}$ (resp.
$\sigma_{\Lambda^+}$) on $\Lambda^+$ (resp.
$\Lambda^-$) that yield the same set of contours $\{\Ga\}$.
In Proposition
\ref{KP0.1} we characterized explicitely the constraints satisfied by the
configurations $\sigma_{\Lambda^\pm}$: each point $i\in [\Lambda^+]_R$ 
must be $(\delta,+)$-correct for the configuration
$\sigma_{\Lambda^+}+_{\Lambda^c}\sigma_{\{\Ga\}}$, 
where $\sigma_{\{\Ga\}}$ is the
configuration specified by the contours on the union of their supports. 
Similarly, each point $i\in [\Lambda^-]_R$ 
must be $(\delta,-)$-correct for the configuration
$\sigma_{\Lambda^-}\sigma_{\{\Ga\}}$. 
Using the re-formulation of the hamiltonian given in Lemma \ref{KL5} we 
get:
\eq{\label{K303}Z^+(\Lambda)=\sum_{\{\Ga\}\subset
\Lambda}\Big(\prod_{\Ga\in\{\Ga\}}\rho(\Ga)\Big)
\rZ^+(\Lambda^+;+_{\Lambda^c}
\sigma_{\{\Ga\}})\rZ^-(\Lambda^-;\sigma_{\{\Ga\}})\,,}
where the sum is over admissible families of contours, and
\eq{\rho(\Ga)\pardef e^{-\beta H_\Ga(\sigma[\Ga])}\,.}
Notice that when $\{\Ga\}=\emptyset$, then $\Lambda\equiv \Lambda^+$ and
the summand of \eqref{K303}
equals $\rZ^+(\Lambda;+_{\Lambda^c})$.
Since they are subject to
boundary conditions that depend on the family of contours $\{\Ga\}$, the
restricted phases induce an interaction among the contours. 
Nevertheless, the boundary conditions
imposed by the contours and $+_{\Lambda^c}$ on
$\Lambda^+$ and $\Lambda^-$ are admissible (in the sense of Definition 
\ref{KD5}). This implies that the results of Section \ref{KSrestricted} can be
used for the restricted partition functions appearing in \eqref{K303}.\\

Since we need to represent the partition function with objects whose
compatibility is purely geometrical, we need to proceed by induction, and
consider systems living in the interior of external contours. Therefore, we must
study functions similar as \eqref{K303}, with an arbitrary $+$-admissible 
boundary condition $\cb$. We thus define
\eq{\label{K400bis} \Theta^+(\Lambda;\cb)\pardef
\sum_{\{\Ga\}\subset
\Lambda}\Big(\prod_{\Ga\in\{\Ga\}}\rho(\Ga)\Big)
\rZ^+(\Lambda^+;\cb \sigma_{\{\Ga\}})
\rZ^-(\Lambda^-;\sigma_{\{\Ga\}})\,,}
Contours always lie at least at distance $l$ from $\Lambda^c$.
The external contours contours of $\{\Ga\}$ can be subject to
particular constraints (as will appear, for example, in Section
\ref{KSderivatives}), but we omit it in the notation. Notice that the 
empty family $\{\Ga\}=\emptyset$, the summand 
corresponds to a pure restricted phase 
$\rZ^+(\Lambda;\cb)$.\\

The aim, in the study of $\Theta^+(\Lambda;\cb)$, is
to extract from \eqref{K400bis} a global
contribution of the restricted phase.
In the Ising model, the same operation 
amounts to extract
the trivial term $e^{\beta h|\Lambda|}$. Here we extract $\rZ^+(\Lambda,\cb)
=e^{\beta
h|\Lambda|}\rcZ(\cP^+_\Lambda(\cb))$, and our aim is to reach the representation
\eqref{K415}.
The deviations from the restricted phase
will be modelized by \emph{chains}, i.e. contours linked by clusters of polymers
(polymers describe the restricted phase).
In Section
\ref{KSlinking}, we expose this linking procedure. In 
Section \ref{KSentropy} we show how to handle
the entropy of chains, preserving the uniformity in the scaling parameter $\ga$.
In Section \ref{KSdomains} we study the weights of
chains and their dependence on the magnetic field near $\real h=0$, i.e. at
coexistence. In Section \ref{KSpure} we study {pure phases}, i.e. $\{\real
h>0\}$ and $\{\real h<0\}$. \\

\subsection{The Linking Procedure}\label{KSlinking}

We first express $\Theta^+(\Lambda;\cb)$ as a sum over external
contours. By Lemma \ref{KL4.1},
each external contour is of type $+$. 
Let $\{\Ga\}$ be a family of external contours.
Then, $\Lambda$ is decomposed into $$\Lambda=\ext_\Lambda{\{\Ga\}}\cup
\{\Ga\}\cup \bigcup_{\Ga\in\{\Ga\}}\inte\Ga\,,$$
where $\ext_\Lambda\{\Ga^{}\}\pardef
\Lambda\cap \bigcap_{\Ga\in\{\Ga\}}\ext \Ga^{}$.
For each family of admissible 
external contours $\{\Ga\}$, we
re-sum over the configurations whose external contours are given exactly by
$\{\Ga^{}\}$. This induces, for all $\Ga^{}$, a partition function
$\Theta^-(\inte\Ga;+\sigma_{\Ga})$, which can be expressed as in
\eqref{K400bis}. On $\ext_\Lambda\{\Ga\}$, we get a restricted partition
function $\rZ^+(\ext_\Lambda\{\Ga\};\cb\sigma_{\{\Ga\}})$.
We thus have
\al{\nonumber
\Theta^+&(\Lambda;\cb)=\\
&\rZ^+(\Lambda;\cb)
+\sum_{\substack{\{\Ga\}\subset \Lambda\\\text{ext.}}}
\rZ^+(\ext_\Lambda\{\Ga^{}\};\cb\sigma_{\{\Ga^{}\}})\prod_{\Ga^{}}
\rho(\Ga^{})\Theta^-(\inte\Ga^{};\sigma_{\Ga^{}})\,,}
where the sum is over non-empty families of external contours.
Consider the configuration $-\sigma_\Ga$ obtained by spin-flipping $\sigma_\Ga$,
i.e. $(-\sigma_\Ga)_i\pardef -(\sigma_\Ga)_i$ for all $i\in\Ga$. We introduce
the functions $\rZ^+(\inte\Ga;-\sigma_\Ga)$ and $\Theta^+(\inte\Ga;
-\sigma_\Ga)$ and consider, for a while,
\eq{\frac{\rZ^+(\ext_\Lambda\{\Ga^{}\};\cb\sigma_{\{\Ga^{}\}})
\prod_\Ga\rZ^+(\inte\Ga;-\sigma_\Ga)}{\rZ^+(\Lambda;\cb)}\,.}
Using the
polymer representation of Section \ref{KSrestricted}, we consider the family of
polymers
$\cP^+_\ext\pardef\cP^+_{\ext_\Lambda\{\Ga^{}\}}(\cb\sigma_{\{\Ga^{}\}})$
associated to 
$\rZ^+(\ext_\Lambda\{\Ga^{}\};\cb\sigma_{\{\Ga^{}\}})$, the families
$\cP_{\inte\Ga}^+\pardef \cP^+_{\inte\Ga}(-\sigma_\Ga)$ 
associated to each of the  
$\rZ^+(\inte\Ga;-\sigma_\Ga)$, as well as the
family $\cP^+_\Lambda\pardef\cP^+_\Lambda(\cb)$ associated to 
$\rZ^+(\Lambda;\cb)$. Since the expansions of these functions are
absolutely convergent, we can rearrange the terms. The 
volume contributions from $\ext_\Lambda\{\Ga\}$ and $\bigcup_\Ga\inte\Ga$
cancel, and we get
\al{\nonumber
\frac{\rcZ(\cP^+_{\ext})\prod_{\Ga}\rcZ(\cP^+_{\inte\Ga})}{\rcZ
(\cP^+_\Lambda)}&=
\exp\Big(\sum_{\hatP}\pm\omega^+(\hatP)
+\sum_\Ga E_\Ga^+\Big)\,,}
where we used the abbreviation 
\eq{\sum_{\hatP}\pm\omega^+(\hatP)\equiv 
\sum_{\substack{\hatP\in \hatcP^+_\ext\\ d(\hatP,\{\Ga^{}\})\leq
R}}\omega^+(\hatP)
-\sum_{\substack{\hatP\in \hatcP^+_\Lambda\\ d(\hatP,\{\Ga^{}\})\leq
R\\ \hatP\cap \ext_\Lambda\{\Ga^{}\}\neq \emptyset}}
\omega^+(\hatP)\,.}
The sign $\pm$ in front of $\omega^+(\hatP)$ is
chosen in function of the sum to which $\hatP$ belongs.
Define $\lambda^+(\hatP)\pardef e^{\pm
\omega^+(\hatP)}-1$ and expand
\eq{e^{\sum_{\hatP}\pm\omega^+(\hatP)}=\prod_{\hatP}(1+\lambda^+(\hatP))
=\sum_{\{\hatP_1,\dots,\hatP_n\}}\prod_{i=1}^n\lambda^+(\hatP_i)\,.}
The function $E_\Ga^+$ depends only on the structure of $\Ga$, and
is given by
\eq{\label{K202.00}
E_\Ga^+=\sum_{\substack{\hatP\in\hatcP^+_{\inte\Ga}\\ d(\hatP,\Ga)\leq
R}}\omega^+(\hatP)-\sum_{\substack{\hatP\in \hatcP^+\\\hatP\cap
\ext\Ga=\emptyset\\ d(\hatP,\Ga)\leq R}}\omega^+(\hatP)\,,}
where $\hatcP^+$ denotes the family of clusters associated to free polymers of
type $+$. Notice that $E_\Ga^+$ is analytic
in $H_+$. Since $|[\Ga]_R|\leq 3^d|\Ga|$ we have, 
if $\beta$ is large enough (see Theorem \ref{KT3})
\eq{\label{K202.01}
\|E^+_\Ga\|_{H_+}\leq\frac{1}{3}|\Ga|\,,\quad
\|\deriv{h}E^+_\Ga\|_{\tilde{H}_+}\leq \frac{1}{3}|\Ga|\,.}
If we define the weight (we denote $+\sigma_\Ga\equiv \sigma_\Ga$)
\eq{\omega^+(\Ga)\pardef \rho_1(\Ga)
\frac{\Theta^-(\inte\Ga;+\sigma_\Ga)}{\Theta^+(\inte\Ga;-\sigma_\Ga)}
\,,\label{K412}}
with $\rho_1(\Ga)\pardef \rho(\Ga)e^{-\beta h|\Ga|}e^{E_\Ga^+}$,
we have
\al{\frac{\Theta^+(\Lambda;\cb)}{\rZ^+(\Lambda;\cb)}&=
1+\sum_{\substack{\{\Ga^{}\}\subset \Lambda\\ \text{ext.}}}
\sum_{\{\hatP_1,\dots,\hatP_n\}}\Big(\prod_{i=1}^n\lambda^+(\hatP_i)\Big)
\Big(\prod_\Ga
\omega^+(\Ga)\frac{\Theta^+(\inte\Ga;-\sigma_\Ga)}{\rZ^+(\inte\Ga;
-\sigma_\Ga)}\Big)\,.\nonumber}
We can then repeat the same procedure of summing inside external contours of
$\Theta^+(\inte\Ga;-\sigma_\Ga)$, etc. 
This procedure continues until we reach
contours whose interior can't contain any contour. At the end,
\eq{\label{K411}\frac{\Theta^+(\Lambda;\cb)}{\rZ^+(\Lambda;\cb)}=
1+\sum_{\{\Ga\}\subset \Lambda}\sum_{\{\hatP\}}
\Big(\prod_{\hatP}\lambda^+(\hatP)\Big)\Big(\prod_{\Ga}\omega^+(\Ga)\Big)\,,}
where the sum $\{\Ga\}$ contains contours of type $+$, and 
each cluster $\hatP$ lies at distance at
most $R$ from one or several contours of $\{\Ga\}$. 
For this reason, the weight of some
polymers can depend on the configuration $\sigma_\Ga$ of the 
contours $\Ga$ that lie in their neighbourhood (or on $\cb$).\\
We get rid of these dependencies by linking polymers to contours.
Like we did in Section \ref{KSrestricted} 
(when linking trees with components of the graph $N$), we associate
to each pair $(\{\Ga\},\{\hatP\})$ an abstract 
graph $\hat{G}$ as follows: each contour
$\Ga_j\in\{\Ga\}$ is represented by an abstract vertex $z_j$, each cluster
$\hatP_k\in\{\hatP\}$ is represented by an abstract vertex $w_k$. This defines
$V(\hat{G})$. Then, we put an edge between $z_j$ and $w_k$ if and only if
$d(\Ga_j,\hatP_k)\leq R$. We also put an edge between $w_{k_1}$ and $w_{k_2}$ if
and only if $V(\hatP_{k_1})\cap V(\hatP_{k_2})\neq \emptyset$.\\
Each connected component of $\hat{G}$, with vertices, say,
$\{z_{j_1},\dots,z_{j_l},w_{k_1},\dots,w_{k_l}\}$, represents a subset of
$\{\Ga\}\cup\{\hatP\}$ given by
$\eGa=\{\Ga_{j_1},\dots,\Ga_{j_l},\hatP_{k_1},\dots,\hatP_{k_l}\}$. $\eGa$ is
called a \grasA{chain of contours}, or simply a \grasA{chain}.
We denote by $\{\eGa\}$ the family of chains associated to the 
pair $(\{\Ga\},\{\hatP\})$. The chains of $\{\eGa\}$ are of \grasA{type} $+$,
and pairwise \grasA{compatible} by definition. 
The \grasA{support} 
of $\eGa$, also written $\eGa$, denotes the union
$\bigcup_{\Ga\in\eGa}\Ga\cup\bigcup_{\hatP\in\eGa}\hatP$.
Notice that if two chains
$\eGa,\eGa'$ are not compatible, then $b(\eGa)\cap b(\eGa')\neq \emptyset$,
where
\eq{\label{K412.2}b(\eGa)\pardef
\bigcup_{\Ga\in\eGa}[\Ga]_l\cup\bigcup_{\hatP\in\eGa}\hatP\,.}
The weight of a
chain is defined by
\eq{\label{K413}
\omega^+(\eGa)\pardef\Big(\prod_{\hatP\in\eGa}\lambda^+
(\hatP)\Big)\Big(\prod_{\Ga\in\eGa}\omega^+(\Ga)\Big)\,,} and depends only 
on the intrinsic structure of the chain $\eGa$ (except, maybe, if
$d(\eGa,\Lambda^c)\leq R$).
The final representation of the partition function is thus
\al{\Theta^+(\Lambda;\cb)&=\rZ^+(\Lambda;\cb)\sum_{\{\eGa\}}
\prod_{\eGa\in\{\eGa\}}\omega^+(\eGa)\label{K414}\\
&\equiv\rZ^+(\Lambda;\cb)\eZ^+(\Lambda;\cb)\,.\label{K415}}
In \eqref{K414}, the product is defined to be equal to $1$ when
$\{\eGa\}=\emptyset$.
This last expression nicely expresses the fact that chains of contours describe
deviations from a restricted phase.
For the restricted phase, there corresponds a family $\cP^+_\Lambda(\cb)$
associated to $\rZ^+(\Lambda;\cb)$.
Similarly, there corresponds a family of chains $\cX^+_\Lambda(\cb)$ associated
to $\eZ^+(\Lambda;\cb)$. The
partition function can be written in terms of these families as
\eq{\label{K415.1}\Theta^+(\Lambda;\cb)=e^{\beta h|\Lambda|}
\rcZ(\cP^+_\Lambda(\cb))\eZ(\cX^+_\Lambda(\cb))\,.}
By definition,
$\eZ(\cX^+_{\Lambda}(\cb)):=1$ when 
$\cX^+_{\Lambda}(\cb)=\emptyset$.
Everything that was done until now can be applied also to the case
where $\cb$ is $-$-admissible, yielding chains of type $-$.
\subsection{The Entropy of Chains}\label{KSentropy}
Before starting the analysis of the weights, we show how a
priori bounds on the weights $\lambda^+(\hatP)$ and $\omega^+(\Ga)$ allow to
handle the summation of weights of chains. In this section we assume that
$|\lambda^+(\hatP)|\leq \lambda_0(\hatP)$, 
$|\omega^+(\Ga)|\leq \rho_0(\Ga)$, i.e.
\eq{\label{K416}|\omega^+(\eGa)|\leq 
\Big(\prod_{\hatP\in\eGa}\lambda_0
(\hatP)\Big)\Big(\prod_{\Ga\in\eGa}\rho_0(\Ga)\Big)\equiv
\omega_0(\eGa)\,.}
\emph{Convention:} Now 
and in the sequel we will always use a subscript ``$0$'' in the weight of an
object to specify that it depends only on the geometric structure of the
object (as we did in \eqref{K59.20}, Section \ref{KSanalrestr}). 
That is, such weights will always be translation invariant. When a
weight is defined for an object, we use the same letter for the weight of the
clusters of such objects (see Appendix \ref{KSnewCLEXP}).\\

The proof of the following lemma is essentially the same as the one 
of Corollary
\ref{KCO.1}. We use the notations $|\hatP|\pardef |\bigcup_{P\in\hatP}V(P)|$,
$|\eGa|\pardef\sum_{\Ga\in \eGa}|\Ga| +\sum_{\hatP\in\eGa}|\hatP|$.
\begin{lem}\label{KL13.05}
Let $c>0$, $\epsilon>0$, and assume the weights $\lambda_0(\hatP)$, 
$\rho_0(\Ga)$ satisfy the bounds
\eq{\label{K417}\sum_{\hatP\ni
0}\lambda_0(\hatP)e^{(c+\epsilon(2^d+1)|\hatP|}\leq
\frac{\epsilon}{2}\,,\quad
\sum_{\Ga:[\Ga]_l\ni 0}
\rho_0(\Ga)e^{(c+\epsilon)|[\Ga]_l|}\leq 
\frac{\epsilon}{2}\,.}
Then the weight $\omega_0(\eGa)$ satisfies the condition \eqref{K603} of 
Lemma \ref{KL50}. Namely,
\eq{\label{K418}\sum_{\eGa:b(\eGa)\ni 0}
\omega_0(\eGa)e^{c|b(\eGa)|}\leq \epsilon\,.}
\end{lem}
\bp
For a chain $\eGa=\{\Ga_1,\dots,\Ga_L,\hatP_1,\dots,\hatP_M\}$, let
$n(\eGa)\pardef L+M$ denote the number of objects composing $\eGa$ (a cluster is
considered as a single object). We show by induction on $N=1,2,\dots$ that 
\eq{\label{K419}\xi_N\pardef\sum_{\substack{\eGa:b(\eGa)\ni 0\\n(\eGa)\leq N}}
\omega_0(\eGa)e^{c|b(\eGa)|}\leq \epsilon}
If $n(\eGa)=1$ then $\eGa$ contains a single object, i.e. a contour. Then
$\xi_1\leq \epsilon$ follows from \eqref{K417}. So suppose \eqref{K419} holds
for $N$, and consider $\xi_{N+1}$; this sum can be bounded by a sum in which
each chain $\eGa$ is 
decomposed into $[\Ga_0]_l\ni 0$, $\eGa\ni \Ga_0$, or into
$\hatP_0\ni 0$, $\eGa\ni \hatP_0$. This means:\\
1) in the first case, $\eGa$ decomposes into 
$\eGa=\{\Ga_0\}\cup\{\eGa_1,\dots,\eGa_K\}$~\footnote{The chains $\eGa_i$ are
obtained as follows: consider the abstract connected graph $\hat{G}$ 
associated to the chain
$\eGa$. Then, remove all the edges of $\hat{G}$ that are 
adjacent to the vertex $z_0$ representing $\Ga_0$ and $z_0$ itself,
and consider the decomposition of the remaining graph into connected components.
These components are exactly the representatives of $\eGa_1,\dots,\eGa_K$.} with 
$[\Ga_0]_l\ni 0$, $d(\eGa_i,\Ga_0)\leq R$, $n(\eGa_i)\leq N$ for all 
$i=1,\dots,K$,
$\eGa_i\cap\eGa_j= \emptyset$ for all $i\neq j$. The contribution to $\xi_{N+1}$
is thus bounded by
\al{&\sum_{\Ga_0:[\Ga_0]_l\ni 0}
\rho_0(\Ga_0)e^{c|[\Ga_0]_l|}\sum_{K\geq 0}\frac{1}{K!}\prod_{i=1}^K
\sum_{\substack{\eGa_i:d(\eGa_i,\Ga_0)\leq R\\ n(\eGa_i)\leq N}}
\omega_0(\eGa_i)e^{c|b(\eGa_i)|}\label{K420}\\
&\leq \sum_{\Ga_0:[\Ga_0]_l\ni 0}
\rho_0(\Ga_0)e^{c|[\Ga_0]_l|}\sum_{K\geq 0}\frac{1}{K!}
\big(|[\Ga_0]_R|\xi_N\big)^K\leq \sum_{\Ga_0:[\Ga_0]_l\ni 0}
\rho_0(\Ga_0)e^{(c+\epsilon )|[\Ga_0]_l|}\leq 
\frac{\epsilon}{2}\nonumber\,,}
where we used the induction
hypothesis $\xi_N\leq \epsilon$.\\
2) in the second case, $\eGa=\{\hatP_0\}\cup\{\eGa_1,\dots,\eGa_K\}$ with 
$\hatP_0\ni 0$, $d(\eGa_i,\hatP_0)\leq R$, $n(\eGa_i)\leq N$ for all 
$i=1,\dots,K$, $\eGa_i\cap\eGa_j= \emptyset$ for all $i\neq j$. A chain $\eGa_i$
of this decomposition can be of two types: i) there exists a cluster
$\hatP\in\eGa_i$ such that $\hatP\cap\hatP_0\neq \emptyset$. Then the
contribution from these chains is at most
\eq{|\hatP_0|\sum_{\substack{\eGa_i:b(\eGa_i)\ni 0\\n(\eGa_i)\leq N}}
\omega_0(\eGa_i)e^{c|b(\eGa_i)|}= |\hatP_0|\xi_N\leq |\hatP_0|\epsilon.}
ii) there exists $\Ga\in\eGa_i$,
$\Ga\cap\{[\hatP_0]_R\}_l\neq \emptyset$, where the thickening $\{\cdot\}_l$ was
defined in \eqref{K26}. Notice that the set $\{[\hatP_0]_R\}_l\in\cC^{(l)}$
contains at most $2^d|\hatP_0|$ cubes $C^{(l)}$. Since contours are
composed of cubes $C^{(l)}$, the contribution from these chains can be bounded
by 
\eq{2^d|\hatP_0|\xi_N\leq 2^d\epsilon|\hatP_0|\,.}
We can then proceed like in \eqref{K420}, and get a 
contribution to $\xi_{N+1}$ bounded by
\eq{\sum_{\hatP_0\ni 0}\lambda_0(\hatP_0)e^{c|\hatP_0|}
e^{\epsilon(2^d+1)|\hatP_0|}\leq \frac{\epsilon}{2}\,}
Altogether, this shows that $\xi_{N+1}\leq \epsilon$.
\ep
\subsection{Domains of Analyticity}\label{KSdomains}
In this section we consider the dependence of the weights $\omega^+(\eGa)$
on the magnetic field $h\in \bC$, in a neighbourhood of $\{\real h=0\}$. For
obvious reasons, the domain in which $\omega^+(\eGa)$ can be shown to be
analytic depends on the contour $\Ga\in\eGa$ that has the largest interior. 
Everything we
say in this section holds for chains of both types, but for the
sake of simplicity, the statements will be given only for
chains of type $+$.\\
The domains of analyticity depend on the isoperimetric constants $K(N)$
defined in \eqref{K38}. Consider the reals
\eq{\label{K101}R(N)\pardef \frac{\theta}{2 K(N)N^{\usd}}\,,}
where $\theta\in (0,1)$ will play an important role later 
in the study of the
derivatives. We know from Lemma \ref{KL10} that $R(N)N^\usd$ is
increasing and that 
\eq{\label{K101.1}\lim_{N\to\infty}R(N)N^\usd=\frac{\theta}{2K(\infty)}\,.}
Since we want the domains of
analyticity to be decreasing with the size of the contours, we define
\eq{\label{K102}R^*(N)\pardef \min\,\{R(N'): 1\leq N'\leq N\}\,.}
The sequences $R^*(N)$ and $R(N)$ have the same asymptotic behaviour, as the
following lemma shows.
\begin{lem}\label{KL13.1}
\eq{\lim_{N\to\infty}R^*(N)N^\usd=\frac{\theta}{2K(\infty)}}
\end{lem}
\optionel{\bp
First  notice that there exists an unbounded increasing sequence
$N_1,N_2,\dots,$ such that $R^*(N_i)=R(N_i)$. This is a direct consequence of
the bounds
\eq{R^*(N)\leq R(N)\leq \frac{\theta}{2K(\infty)N^\usd}\,.}
Since $R(N)N^\usd$ increases, it is sufficient to show that $R^*(N)N^\usd$ is
increasing. Consider the interval $[N,N+1]$. We have two possibilities: 1)
$R(N+1)\geq R^*(N)$. In this case, $R^*(N+1)(N+1)^\usd=R^*(N)(N+1)^\usd\geq
R^*(N)N^\usd$. 2) $R(N+1)\leq R^*(N)$. In this case, $R^*(N+1)(N+1)^\usd
=R(N+1)(N+1)^\usd\geq R(N)N^\usd\geq R^*(N)N^\usd$.
\ep}
For $r>0$, consider the strip
\eq{\label{K103}U(r)\pardef\{z\in\bC:|\real z|<r\}\,.}
Generally, we will 
restrict our attention to small magnetic fields, that is $h\in U_0\pardef
U(h_0)$ where $h_0$ will be taken small enough. For instance, $h_0<\frac{1}{16}$
so that the results on the restricted phases can be used in $U_0$.\\
We define the \grasA{domain of analyticity} for a contour:
\eq{\label{K430}U_\Ga\pardef U(R^*(V(\Ga)))\cap U_0\,,}
and for a chain $\eGa$:
\eq{\label{K431}U_\eGa\pardef \bigcap_{\Ga\in\eGa}U_\Ga\,.}
That is, $U_\eGa=U_{\Gamax}$, where $\Gamax\in\eGa$ has the
largest interior. Notice that the domains $U_\Ga,U_\eGa$ depend on $\theta$.
Set $V(\eGa)\pardef V(\Gamax)=\max\{V(\Ga):\Ga\in \eGa\}$.
The main result of this section is the following.
\begin{pro}\label{KP5}
Let $\theta\in(0,1)$, $\epsilon>0$, $c>0$ small enough. 
There exists $\beta_1=\beta_1(\theta, \epsilon)$ 
such that for all $\beta\geq \beta_1$, the following holds. For each chain
$\eGa$,
$h\mapsto \omega^+(\eGa)$ is analytic in $U_\eGa$. Moreover,
\eq{\label{K108}
\|\omega^+(\eGa)\|_{U_\eGa}< \omega_0(\eGa)\,,\quad 
\big\|\deriv{h}\omega^+(\eGa)\big\|_{U_\eGa}< \omega_0(\eGa)\,,}
where $\omega_0(\eGa)$ is defined via the weights $\lambda_0(\hatP)$ and
$\rho_0(\Ga)$ given in \eqref{K432.1}-\eqref{K432.2} hereafter, and 
satisfies \eqref{K418}.
\end{pro}
Before starting the proof of Proposition \ref{KP5}, we give explicitly the
weights $\lambda_0(\hatP)$ and $\rho_0(\Ga)$.
These weights are defined such that they can be used throughout the section,
also when bounding the first derivative of $\omega^+(\eGa)$. 
As will be seen, the
non-trivial part of $\omega^+(\Ga)$ will be bounded by:
\eq{\label{K432}\Big\|\frac{
\Theta^-(\inte\Ga^{};+\sigma_{\Ga^{}})}
{\Theta^+(\inte\Ga^{};-\sigma_{\Ga^{}})}\Big\|_{U_\Ga}\leq
e^{\beta\theta\|\Ga\|}e^{\frac{2}{3}|\Ga|}\,.} 
Using \eqref{K202.01},
$\|\rho_1(\Ga)\|_{U_0}\leq e^{-\beta\|\Ga\|}e^{2\beta
h_0|\Ga|}e^{\frac{1}{3}|\Ga|}$, this suggests to
define the weight $\rho_0(\Ga)$ in the following way:
\eq{\label{K432.1}\rho_0(\Ga)\pardef D_1\beta
|\Ga|^{\ddmu}e^{-(1-\theta)\beta\|\Ga\|}e^{2\beta h_0|\Ga|}e^{|\Ga|}\,.}
The term $D_1\beta |\Ga|^{\ddmu}$ has been added to take into account
other contributions, especially when studying the first derivative.
For clusters we get, using the definition of $\lambda^+(\hatP)$ and \eqref{K70},
\al{\|\lambda^+(\hatP)\|_{H_+}&\leq\|\omega^+(\hatP)\|_{H_+}
e^{\|\omega^+(\hatP)\|_{H_+}}\nonumber\\
&\leq |\omega_0(\hatP)|
e^{|\omega_0(\hatP)|}\leq
|\omega_0(\hatP)|e^{\epsilon_r}<D_2|\omega_0(\hatP)|\equiv
\lambda_0(\hatP)\label{K432.2}\,.}
The numerical constants $D_1,D_2$ are assumed to be fixed and 
sufficiently large, in order to cover all the cases that will appear in the
sequel.
\begin{lem}\label{KL15}
Let $\theta\in(0,1)$, $c>0$, 
and $\epsilon>0$ be small enough. Assume
$2h_0\leq\frac{1}{2}(1-\theta)\rho$ ($\rho$ is the Peierls constant). There
exists $\beta_1=\beta_1(\theta,\epsilon)$ such that for all $\beta\geq
\beta_1$, the hypothesis \eqref{K417} of Lemma \ref{KL13.05} are satisfied.
\end{lem}
\bp
Define a new weight for polymers (see \eqref{K59.20}):
\eq{\tildomega_0(P)\pardef \omega_0(P)e^{(c+\epsilon(2^d+1))|P|}\,.}
If $\beta$ is large enough, we can proceed as in \eqref{K70} and get
\eq{\nonumber\sum_{\hatP\ni 0}\lambda_0(\hatP)e^{(c+\epsilon(2^d+1))|\hatP|}=
D_2\sum_{\hatP\ni 0}|\omega_0(\hatP)|e^{(c+\epsilon(2^d+1))|\hatP|}
\leq D_2
\sum_{\hatP\ni 0}|\tildomega_0(\hatP)|\leq \frac{\epsilon}{2}\,.}
This shows the first inequality of \eqref{K417}. For the second, we use the
Peierls condition $\|\Ga\|\geq \rho |\Ga|$ (Proposition \ref{KP1}). This
gives
\al{\sum_{\Ga:[\Ga]_l\ni 0}\rho_0(\Ga)e^{(c+\epsilon )|[\Ga]_l|}&\leq 
D_1 \beta \sum_{\Ga:[\Ga]_l\ni 0}|\Ga|^\ddmu e^{-(1-\theta)\beta\rho|\Ga|}
e^{2\beta
h_0|\Ga|}e^{|\Ga|}e^{(c+\epsilon )|[\Ga]_l|}\\
&\leq D_1 \beta \sum_{\Ga:[\Ga]_l\ni 0}|\Ga|^\ddmu e^{-\frac{1}{2}
(1-\theta)\beta\rho|\Ga|}e^{|\Ga|}e^{(c+\epsilon )|[\Ga]_l|}\,.}
Since $|[\Ga]_l|\leq 3^d|\Ga|$, a standard 
Peierls estimate allows to bound this sum by $\frac{\epsilon}{2}$ as
soon as $\beta$ is large enough. 
\ep
Until now we have denoted by $\epsilon_r=\epsilon_r(\beta)$ the small function 
appearing in the
study of the restricted phases. Similarly, we denote by
$\epsilon_c=\epsilon_c(\beta)$ 
the small
function appearing in the study of chains. These two parameters are assumed to
have a common bound $\max\{\epsilon_r,\epsilon_c\}\leq\epsilon$, which is
small.\\
Consider the weight $\omega^+(\Ga)$ given \eqref{K412}.
We can use the linking procedure for the partition functions
$\Theta^\pm(\inte\Ga;\mp\sigma_\Ga)$, yielding
\eq{\label{K433.3}\omega^+(\Ga)=\rho_1(\Ga)\frac{
e^{-\beta h V(\Ga)}\rcZ(\cP^-_{\inte\Ga}(+\sigma_\Ga))
\eZ(\cX^-_{\inte\Ga}(+\sigma_\Ga))}{e^{+\beta h
V(\Ga)}\rcZ(\cP^+_{\inte\Ga}(-\sigma_\Ga))
\eZ(\cX^+_{\inte\Ga}(-\sigma_\Ga))}\,.}
\bp[Proof of Proposition \ref{KP5}:]
The proof will be done by induction. We say a contour $\Ga$ is of \grasA{class} 
$n$ if $V(\Ga)=n$. A chain
is of \grasA{class} $n$ if $V(\eGa)=n$.\\
Consider a contour $\Ga$ of small class (say, of class smaller than $l^{d}$).
Then the last ratio appearing in \eqref{K433.3} equals $1$.
We bound $\omega^+(\Ga)$ at $h=x+iy\in U_\Ga$. First,
\eq{\label{K435.3}
|e^{-2\beta hV(\Ga)}|\leq e^{2\beta |x| V(\Ga)}\leq 
e^{2\beta R^*(V(\Ga))V(\Ga)}\leq
e^{2\beta R(V(\Ga))V(\Ga)}\leq e^{\theta \beta\|\Ga\|}\,,}
where we used the definition of the isoperimetric constants $K(\cdot)$ 
given in \eqref{K38}. Then, write
\eq{\label{K436}\frac{\rcZ(\cP^-_{\inte\Ga}(+\sigma_\Ga))_{h}}{\rcZ
(\cP^+_{\inte\Ga}(-\sigma_\Ga))_{h}}=
\frac{\rcZ(\cP^-_{\inte\Ga}(+\sigma_\Ga))_{h}}{\rcZ
(\cP^-_{\inte\Ga}(+\sigma_\Ga))_{iy}}
\frac{\rcZ(\cP^-_{\inte\Ga}(+\sigma_\Ga))_{iy}}{\rcZ
(\cP^+_{\inte\Ga}(-\sigma_\Ga))_{iy}}
\frac{\rcZ(\cP^+_{\inte\Ga}(-\sigma_\Ga))_{iy}}{\rcZ
(\cP^+_{\inte\Ga}(-\sigma_\Ga))_{h}}}
The middle term has modulus $1$ by symmetry (see \eqref{K64.01}). 
The two other terms can be treated as follows:
\al{\label{K436.1}
\Big|\log \frac{\rcZ(\cP^-_{\inte\Ga}(+\sigma_\Ga))_{h}}{\rcZ
(\cP^-_{\inte\Ga}(+\sigma_\Ga))_{iy}}\Big|
=\Big|\int_{0}^x\diff s\deriv{s}\log
\rcZ(\cP^-_{\inte\Ga}(+\sigma_\Ga))_{s+iy}\Big|
\leq |x|\epsilon_rV(\Ga)\,,}
We used Theorem \ref{KT3}. Proceeding as in \eqref{K435.3}, we get
\eq{\Big\|\label{K437}\frac{\rcZ(\cP^-_{\inte\Ga}(+\sigma_\Ga))}{\rcZ
(\cP^+_{\inte\Ga}(-\sigma_\Ga))}
\Big\|_{U_\Ga}\leq e^{\theta\epsilon_r\|\Ga\|}\leq
e^{\frac{1}{3}|\Ga|}\,,}
when $\beta$ is large enough.
Altogether this gives
\eq{\label{K438}\|\omega^+(\Ga)\|_{U_\Ga}\leq
\|\rho_1(\Ga)\|_{U_\Ga}e^{\theta\beta\|\Ga\|}e^{\frac{1}{3}|\Ga|}\leq
e^{-(1-\theta)\beta\|\Ga\|}e^{2\beta h_0|\Ga|}e^{2\frac{1}{3}|\Ga|}
< \rho_0(\Ga)\,,}
Since $\|\lambda^+(\hatP)\|_{U_0}<\lambda_0(\hatP)$, 
we have shown the first inequality of \eqref{K108} 
for chains of small class. For the derivative, a Cauchy
estimate (any disc centered at $h\in U_0$ with radius
$\frac{1}{16}$ is contained in $H_+$) gives
\eq{\label{K439}\big\|\deriv{h}\lambda^+(\hatP)\big\|_{U_0}\leq
16\|\lambda^+(\hatP)\|_{H_+}\,.}
For contours, 
\al{\deriv{h}\omega^+(\Ga)&=\omega^+(\Ga)\deriv{h}\log \omega^+(\Ga)\nonumber\\
=\omega^+(\Ga)\Big(-&\beta \deriv{h}H_\Ga(\sigma[\Ga])
-\beta|\Ga|+\deriv{h}E^+_\Ga
-2\beta V(\Ga)
+\deriv{h}\log\frac{\rcZ(\cP^-_{\inte\Ga}(+\sigma_\Ga))}{\rcZ
(\cP^+_{\inte\Ga}(-\sigma_\Ga))}\Big)\nonumber}
Using $V(\Ga)\leq |\Ga|^{\ddmu}$ (this is a consequence of Lemma \ref{KL9}) 
and
\eq{\label{K440}\Big\|\deriv{h}\log\frac{\rcZ(\cP^-_{\inte\Ga}(+\sigma_\Ga))}{
\rcZ
(\cP^+_{\inte\Ga}(-\sigma_\Ga))}\Big\|_{U_\Ga}\leq 2\epsilon_rV(\Ga)\,,}
this gives the upper bound
\eq{\label{K441}\Big\|\deriv{h}\omega^+(\Ga)\Big\|_{U_\Ga}\leq
6\beta|\Ga|^\ddmu \|\omega^+(\Ga)\|_{U_\Ga}\,,}
which implies, as can be seen easily, that
\eq{\label{K442}\big\|\deriv{h}\omega^+(\eGa)\big\|_{U_\eGa}< 
\omega_0(\eGa)\,,}
With Lemma \ref{KL13.05}, this shows the proposition for chains of small class.
Suppose it has been shown for chains of class $\leq n$. 
By this induction hypothesis, \eqref{K418} and
Lemma \ref{KL50}, a cluster expansion can be used for
the partition functions containing chains.
Let $\eGa$ be a chain
of class $n+1$, and consider $\Ga\in\eGa$.
The treatment of the restricted phases is the same, and
we must study the ratio
\eq{\label{K443}\frac{\eZ(\cX^-_{\inte\Ga}(+\sigma_\Ga))_h}{\eZ(\cX^+_{\inte\Ga}
(-\sigma_\Ga))_h}=
\frac{\eZ(\cX^-_{\inte\Ga}(+\sigma_\Ga))_h}{\eZ(\cX^-_{\inte\Ga}
(+\sigma_\Ga))_{iy}}
\frac{\eZ(\cX^-_{\inte\Ga}(+\sigma_\Ga))_{iy}}{\eZ(\cX^+_{\inte\Ga}
(-\sigma_\Ga))_{iy}}
\frac{\eZ(\cX^+_{\inte\Ga}(-\sigma_\Ga))_{iy}}{\eZ(\cX^+_{\inte\Ga}
(-\sigma_\Ga))_{h}}\,.}
Again the middle term has modulus $1$ and the rest is treated using
the induction hypothesis.
\eq{\label{K444}\Big|\log\frac{\eZ(\cX^-_{\inte\Ga}(+\sigma_\Ga))_h}{
\eZ(\cX^-_{\inte\Ga}(+\sigma_\Ga))_{iy}}\Big|=\Big|\int_0^x\diff
s\deriv{s}\log\eZ(\cX^-_{\inte\Ga}(+\sigma_\Ga))_{s+iy}\Big|\leq
|x|\epsilon_cV(\Ga)\,.} 
This implies
\eq{\label{K445}\Big\|\frac{\eZ(\cX^-_{\inte\Ga}(+\sigma_\Ga))}{\eZ(\cX^+_{\inte\Ga}
(-\sigma_\Ga))}\Big\|_{U_\Ga}\leq e^{\theta\epsilon_c\|\Ga\|}
\leq e^{\frac{1}{3}|\Ga|}\,.}
For the weight of $\Ga$, we thus have (compare with \eqref{K438}):
\eq{\label{K445.1}\|\omega^+(\Ga)\|_{U_\Ga}\leq
e^{-(1-\theta)\beta\|\Ga\|}e^{2\beta h_0|\Ga|}e^{3\frac{1}{3}|\Ga|}
< \rho_0(\Ga)\,.}
For the derivative, use again the induction hypothesis, and bound
\eq{\label{K446}
\Big\|\deriv{h}\log\frac{\eZ(\cX^-_{\inte\Ga}(+\sigma_\Ga))}{
\eZ(\cX^+_{\inte\Ga}(-\sigma_\Ga))}\Big\|_{U_\Ga}\leq 2\epsilon_cV(\Ga)\,.}
It is easy to check that \eqref{K441} still holds which, in turn, implies
\eqref{K442}. This shows the proposition.
\ep

\subsection{Pure Phases}\label{KSpure}
In the last section we gave for each chain $\eGa$ a
domain $U_\eGa$ in which the weight
$\omega^+(\eGa)$ behaves analytically. The size of the domain $U_\eGa$ shrinks
to $\{\real h=0\}$ when the size of the largest contour of $\eGa$ increases.
In the present section we show that the weights $\omega^+(\eGa)$ can actually be
controlled when $0<\real h<h_+$ where $h_+$ is fixed, independently of the size
of $\eGa$. This treatment is standard and was first introduced by
Zahradn\'\i k \cite{Z1}.\\
We consider only chains of type $+$, the case $-$ being similar by symmetry.
Define
\eq{\label{K450}
U_+\pardef \{z\in \bC:0<\real h<h_+\}\,,}
where $0<h_+\leq\min\{\frac{1}{16},\frac{\rho}{2}\}$ is fixed ($\rho$
is the Peierls constant).
In Section \ref{KSderivatives}, domains will have to be made
optimal, with $\theta$ close to $1$, but here we choose
$\theta\pardef\frac{1}{2}$. The main result of this section is the following
\begin{pro}\label{KP6}
Let $\epsilon,c>0$ be small enough. There exists $\beta_2=\beta_2(\epsilon)$
such that for all $\beta\geq \beta_2$, the following holds. For each chain
$\eGa$ of type $+$, $h\mapsto \omega^+(\eGa)$ is analytic in $U_+$, and 
\eq{\label{K451}
\|\omega^+(\eGa)\|_{U_+}\leq \omega_0(\eGa)\,,}
where $\omega_0(\eGa)$ satisfies \eqref{K418}.
\end{pro}
\bp
Since $U_+\subset H_+$, clusters $\hatP$ and restricted phases are under
control. For each $\Ga$, we use the representation \eqref{K412} (rather than
\eqref{K433.3}). The main ingredient of the proof is the following lemma, whose
proof is standard and can be found, e.g. 
in \cite{Z1} or \cite{FP} (with minor
modifications due to the fact that we are working with analytic 
restricted phases rather than ground states).
\begin{lem}\label{KL16}
Let $\beta$ be large enough. Then for each contour $\Ga$ of type $+$,
we have $\Theta^+(\inte\Ga;-\sigma_\Ga)\neq 0$ on $U_+$ and 
\eq{\label{K452}\Big\|\frac{\Theta^-(\inte\Ga;+\sigma_\Ga)}{
\Theta^+(\inte\Ga;-\sigma_\Ga)}\Big\|_{U_+}\leq e^{\frac{2}{3}|\Ga|}\,.}
\end{lem}
\noindent The 
proof of Proposition \ref{KP6} finishes by using Lemma \ref{KL13.05}.
\ep

\nouvellepage
\section{Derivatives of the Pressure}\label{KSderivatives}

In this section we prove Theorem \ref{KT2}, adapting the mechanism used by
S.N. Isakov for the Ising model. Although estimates of Theorem \ref{KT2} hold
for the pressure density $p_\ga$, we will always work in a finite volume
$\Lambda$, and obtain bounds on the derivatives of the pressure 
that are uniform in
the volume. As in the preceding section, we assume $\ga\in(0,\ga_0)$ 
is fixed.\\

We consider a 
box $\Lambda=[-M,+M]^d\cap\bZ^d$, with $M$ large, chosen so
that $\Lambda\in\cC^{(l)}$. Outside $\Lambda$ we fix the spins to the
value $+1$, i.e. we consider the set $\Omega_\Lambda^+$, 
defined in \eqref{K300} and the 
associated partition function $Z^+(\Lambda)$ defined in \eqref{K300.1}.
The finite volume pressure $p_{\ga,\Lambda}^+$ is defined by
\eq{\label{K303.1}p_{\ga,\Lambda}^\newplus
\pardef \frac{1}{\beta|\Lambda|}\log Z^\newplus(\Lambda)\,,}
Clearly, this function equals the density pressure of \eqref{K17} in the
thermodynamic limit.
Consider the set $\cC^+(\Lambda)$ of all external contours
of type $+$ associated to the set $\Omega_\Lambda^+$. 
Remember that
$V(\Ga)=|\inte\Ga|$, where $\inte\Ga$ denotes the union of all components of
$\Ga^c$ with label $-$. The family $\cC^+(\Lambda)$ can be totally
ordered, with
an order relation denoted $\ppetit$, such that $V(\Ga')\leq V(\Ga)$ when
$\Ga'\ppetit \Ga$. When $\Ga$ is not the smallest contour we denote its
predecessor (w.r.t. $\ppetit$) by $i(\Ga)$.\\
For a given external contour $\Ga\in \cC^+(\Lambda)$, consider the set 
\eq{\label{K305}\Omega^\newplus_\Lambda(\Ga)\pardef
\{\sigma_\Lambda\in\Omega_\Lambda^\newplus:
\Ga'\ppetit \Ga\text{ for all external
contour }\Ga'\text{ of }\sigma_\Lambda+_{\Lambda^c}\}\,,}  
and define the partition function
\eq{\label{K306}\Theta_\Ga^\newplus(\Lambda)\pardef
\sum_{\sigma_\Lambda\in\Omega^\newplus_\Lambda(\Ga)}
\exp\big(-\beta H_\Lambda(\sigma_\Lambda+_{\Lambda^c})\big)\,.}
When $\Ga$ is the largest contour then clearly 
$\Theta_\Ga^\newplus(\Lambda)=Z^\newplus(\Lambda)$ and when
$\Ga$ is the smallest contour, we define 
$\Theta^\newplus_{i(\Ga)}(\Lambda)\pardef Z_r^\newplus(\Lambda)$.
We also introduce the following set in which the presence of
$\Ga$ is \emph{forced}:
\al{\label{K307}
\Omega^\newplus_\Lambda[\Ga]\pardef
\{\sigma_\Lambda\in\Omega_\Lambda^\newplus:&\,\Ga'
\ppetit \Ga\text{ for all external
contour }\Ga'\text{ of }\sigma_\Lambda+_{\Lambda^c}\\
&\text{and }\Ga\text{ is a contour of }\sigma_\Lambda+_{\Lambda^c}\}\,.}
The partition function $\Theta^\newplus_{[\Ga]}
(\Lambda)$ is defined as \eqref{K306},
with $\Omega^\newplus_\Lambda[\Ga]$ in place of $\Omega^\newplus_\Lambda(\Ga)$.
We have the following fundamental identity:
\eq{\label{K308}\Theta_\Ga^\newplus(\Lambda)
=\Theta_{i(\Ga)}^\newplus(\Lambda)+\Theta^\newplus_{[\Ga]}(\Lambda)\,.}
A crucial idea of Isakov is to consider the following identity.
\eq{\label{K309}
Z^\newplus(\Lambda)=Z_r^\newplus(\Lambda)\prod_{\Ga\in\cC^+(\Lambda)}
\frac{\Theta^\newplus_{\Ga}(\Lambda)}{
\Theta^\newplus_{i(\Ga)}(\Lambda)}\,.}
Then, the logarithm is written as a \emph{finite} sum:
\eq{\label{K310}
\log Z^\newplus(\Lambda)=\log Z_r^\newplus(\Lambda)
+\sum_{\Ga\in\cC^+(\Lambda)}
u_\Lambda^+(\Ga)\,,}
where 
\eq{u_\Lambda^+(\Ga)\pardef \log 
\frac{\Theta^\newplus_{\Ga}(\Lambda)}{
\Theta^\newplus_{i(\Ga)}(\Lambda)}\,.}
Using \eqref{K308} we can write $u_\Lambda^+(\Ga)=\log(1+\monphi)$, where
\eq{\label{K311}\monphi\pardef \frac{\Theta^\newplus_{[\Ga]}
(\Lambda)}{\Theta^\newplus_{i(\Ga)}(\Lambda)}}
Non-analyticity of the pressure is examined by studying high order derivatives
of the functions $\monphi$ at $h=0$, using the Cauchy formula
\eq{\monphi^{(k)}(0)=\frac{k!}{2\pi i}\int_C\frac{\monphi(z)}{z^{k+1}}
\diff z\,.}
To obtain bounds on $\monphi^{(k)}(0)$, we exponentiate $\monphi$ and
use a stationary phase analysis to estimate the integral.
The contour $C$ will be chosen in a $k$-dependent way. If the domain
$U_\Ga\ni 0$ in which $\monphi$ is analytic is too small, then no information
(not even the sign!) can be given about $\monphi^{(k)}(0)$.\\

\begin{figure}[htbp]
\begin{center}
\input{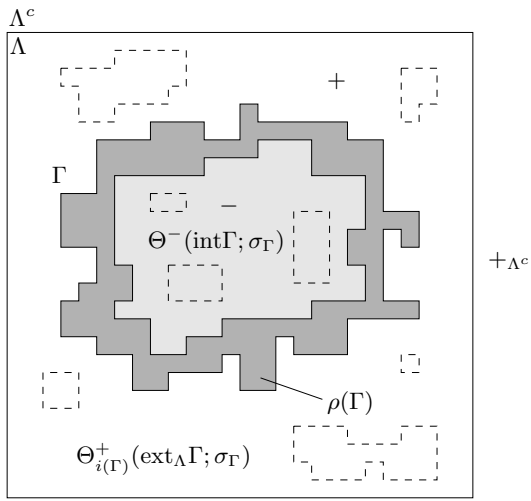}
\end{center}
\caption{{ The decomposition 
\eqref{K313} of the partition function
$\Theta^+_{[\Ga]}(\Lambda)$.}}
\label{KF13}
\end{figure}

For a while, consider the structure of the partition function
$\Theta^+_{[\Ga]}(\Lambda)$. We write
$\Lambda=\ext_\Lambda\Ga\cup\Ga\cup \inte\Ga$,
where $\ext_\Lambda\Ga\pardef \ext\Ga\cap \Lambda$.
By construction, $\ext_\Lambda\Ga$ and 
$\inte\Ga$ are at distance at least $l>2R$. We will 
therefore consider $\ext_\Lambda\Ga$ and $\inte\Ga$ as independent
systems (see Figure \ref{KF13}).
The sums over configurations on $\ext_\Lambda\Ga$ and $\inte\Ga$ can be done
separately, yielding
\eq{\label{K313}
\Theta^\newplus_{[\Ga]}(\Lambda)
=\rho(\Ga)\Theta^+_{i(\Ga)}(\ext_\Lambda\Ga;+_{\Lambda^c}\sigma_\Ga)
\Theta^-_{}(\inte\Ga;\sigma_\Ga)\,,} 
All the contours of these partition functions are at distance larger than $l$
from $\Ga$, and 
have an interior smaller than $V(\Ga)$. The point is that we control
these functions for $h\in U_\Ga$, where $U_\Ga\subset \bC$ 
is a domain that
depends \emph{only on the volume of }$\Ga$.\\

The program for the rest of the section is the following.
In Section \ref{KSSexpon} we show that $\monphi$ can be exponentiated, using the
results of Section \ref{KSextended}. We then use
a stationary phase analysis and obtain upper and lower bounds
on some derivatives of $\monphi$ and $u_\Lambda^+(\Ga)$ at $h=0$.
In Section \ref{KSSmechanism} we fix $k$ and take the box 
$\Lambda$ large enough. For a class of contours called $k$-large and thin, the
$k$-th derivative of $u_\Lambda^+(\Ga)$ can be estimated from below, using the
results of Section \ref{KSSexpon}.
This gives a lower bound on $p_{\ga,\Lambda}^{+(k)}(0)$. 
In Section \ref{KSlimithermo} we show that for $p^+_{\ga,\Lambda}$, the 
operations
$\lim_{\Lambda}$ and $(\cdot)^{(k),\leftarrow}(0)$ commute, leading to the proof of our
main results.
\subsection{Study of the Functions $\monphi$}
\label{KSSexpon}
The following lemma requires the
main results of Sections \ref{KSrestricted} and \ref{KSextended}. After that,
the proof of non-analyticity of the pressure will essentially follow 
the argument of Isakov (see \cite{I1}, \cite{I2} or \cite{FP}).
\begin{lem}\label{KL30}
Let $\theta\in(0,1)$, $\beta$ large enough. 
Then the following holds. For all
contour $\Ga\in\cC^+(\Lambda)$ with $V(\Ga)\neq 0$ there exists a map $h\mapsto
g_{\Lambda}^+(\Ga)(h)$ analytic in the strip 
$U_\Ga$, such
that for all $h\in {U_\Ga}$, $\monphi$ can be {exponentiated}:
\eq{\label{K209}
\monphi=\exp\big(-\beta \|\Ga\|-2\beta hV(\Ga)
+2\beta V(\Ga)g_{\Lambda}^+(\Ga)\big)\,.}
Moreover, we have the following local estimate
\eq{\label{K210}2\beta V(\Ga)|g_{\Lambda}^+(\Ga)(0)|\leq
\delta_1(\beta)\beta\|\Ga\|\,,}
and a uniform bound on the first derivative
\eq{\label{K211}
\Big\|\deriv{h}g_{\Lambda}^+(\Ga)\Big\|_{U_\Ga}\leq
\delta_2(\beta)+2\frac{|\Ga|}{V(\Ga)}\,.}
The functions $\delta_i$ are such that $\lim_{\beta\nearrow\infty}\delta_i=0$.
\end{lem}
\bp
Consider 
$\Theta_{[\Ga]}^+(\Lambda)$. We have seen
how to re-sum over configurations on $\ext_\Lambda\Ga$ and $\inte\Ga$. 
We write
\eq{\label{K212}
\monphi=\rho(\Ga)\frac{\Theta^+_{i(\Ga)}(\ext_\Lambda\Ga;+_{\Lambda^c}
\sigma_\Ga)
\Theta^+(\inte\Ga;-\sigma_\Ga)}{\Theta^+_{i(\Ga)}(\Lambda)}
\frac{\Theta^-(\inte\Ga;+\sigma_\Ga)}{\Theta^+(\inte\Ga;-\sigma_\Ga)}\,.}
All the volume contributions coming from the
first quotient will be shown to vanish.
The partition functions $\Theta_{i(\Ga)}^+(\ext_\Lambda;\sigma_{\Ga})$ 
and $\Theta^\pm(\inte\Ga;\mp\sigma_\Ga)$ are of the type \eqref{K400bis}. 
We can therefore apply the linking procedure and obtain a representation of the
form \eqref{K415.1} for each of them:
\al{\Theta_{i(\Ga)}^+(\ext_\Lambda\Ga;+_{\Lambda^c}\sigma_{\Ga})&=
e^{\beta h|\ext_\Lambda\Ga|}
\rcZ(\cP^+_{\ext_\Lambda\Ga})
\eZ(\cX^+_{\ext_{\Lambda}\Ga})\label{K212.1}\\
\Theta^\pm(\inte\Ga;\mp\sigma_\Ga)&=e^{\pm\beta hV(\Ga)}
\rcZ(\cP^\pm_{\inte\Ga}) \eZ(\cX^\pm_{\inte\Ga})\,,\label{K213}}
where we omitted, in the notation, to mention that the 
families of polymers and chains always
depend on the boundary conditions specified by $+_{\Lambda^c}$ 
and $\sigma_\Ga$. Moreover, the family $\cX^+_{\ext_\Lambda\Ga}$ 
contains  chains $\eGa$ that satisfy $V(\eGa)\leq V(\Ga)$.
In the same way:
\eq{\label{K214}
\Theta_{i(\Ga)}^+(\Lambda)=e^{\beta h|\Lambda|}
\rcZ(\cP^+_\Lambda)\eZ(\cX^+_\Lambda)\,,}
where the 
families $\cP_\Lambda^+$ and $\cX_\Lambda^+$ depend only on the
boundary condition $+_{\Lambda^c}$. Using the definition of $\rho(\Ga)$, it is 
easy to see that $\monphi$ has the form
\eqref{K209}, where $g_\Lambda^+(\Ga)$ is defined by
\al{\label{K217}
2\beta V(\Ga)g_\Lambda^+(\Ga)\pardef-\beta
\sum_{i\in\Ga}u((\sigma_\Ga)_i)-\beta h|\Ga|+\log Q_r+\log Q_c\,,}
where $u(\sigma_i)=-h\sigma_i$, and the quotients $Q_r,Q_c$ are defined by
\al{Q_r(h)&\pardef
\frac{\rcZ(\cP^+_{\ext_\Lambda\Ga})\rcZ(\cP^+_{\inte\Ga})}{\rcZ(\cP^+_\Lambda)}
\frac{\rcZ(\cP^-_{\inte\Ga})}{\rcZ(\cP^+_{\inte\Ga})}
\label{K215.1}\\
Q_c(h)&\pardef
\frac{\eZ(\cX^+_{\ext_\Lambda\Ga})\eZ(\cX^+_{\inte\Ga})}{\eZ(\cX^+_\Lambda)}
\frac{\eZ(\cX^-_{\inte\Ga})}{\eZ(\cX^+_{\inte\Ga})}\,,
\label{K215.2}}
Since all the families of chains involved contain contours with an interior
smaller than $\Ga$, 
$h\mapsto g_\Lambda^+(\Ga)$ is analytic in the strip $U_\Ga$
(by Proposition \ref{KP5}). Rearranging the
terms of the cluster expansions for $Q_r$ leads to
\al{\log Q_r=&\log\frac{\rcZ(\cP^-_{\inte\Ga})}{\rcZ(\cP^+_{\inte\Ga})}
+\sum_{\substack{\hatP\in
\hatcP^+_{\ext_\Lambda\Ga}\\ 
\hatP\cap [\Ga]_R\neq\emptyset}}\omega^+(\hatP)
+\sum_{\substack{\hatP\in 
\hatcP^+_{\inte\Ga}\\ 
\hatP\cap [\Ga]_R\neq\emptyset}}\omega^+(\hatP)
-\sum_{\substack{\hatP\in 
\hatcP^+_{\Lambda}\\ 
\hatP\cap [\Ga]_R\neq\emptyset}}\omega^+(\hatP)\nonumber}
Notice that the volume contributions from $\ext_\Lambda\Ga$ cancel, and that the
three
sums are boundary terms. By symmetry, the quotient equals $1$ at
$h=0$, and so
\eq{\label{K220}|\log Q_r(0)|\leq 3\epsilon_r|[\Ga]_R|\,.}
For the derivative, using \eqref{K65.2} gives
\eq{\label{K221}
\Big\|\deriv{h}\log Q_r\Big\|_{\tilde{H}_+}
\leq 2\epsilon_rV(\Ga)+3\epsilon_r|[\Ga]_R|\,.}
The same computations can be done for $Q_c$. Clusters of chains are denoted
$\hateGa$. The contributions from $\ext_\Lambda\Ga$ also cancel. Indeed,
consider the difference
\eq{\sum_{\substack{\hateGa\in
\hatcX^+_{\ext_\Lambda\Ga}}}\omega^+(\hateGa)
-\sum_{\substack{\hateGa\in \hatcX^+_{\Lambda}}}\omega^+(\hateGa)}
Using Lemma \ref{KL4.1}, there exists
for all $\hateGa_1\in\hatcX^+_{\ext_\Lambda\Ga}$ with
$d(\hateGa_1,\Ga)>R$, a cluster $\hateGa_2\in\hatcX^+_{\Lambda}$, $\hateGa_2\cap
\ext_\Lambda\Ga\neq \emptyset$,
$d(\hateGa_2,\Ga)>R$,
such that $\omega^+(\hateGa_1)=\omega^+(\hateGa_2)$. We are thus left with
\al{\log Q_c=&\log\frac{\eZ(\cX^-_{\inte\Ga})}{\eZ
(\cX^+_{\inte\Ga})}
+\sum_{\substack{\hateGa\in
\hatcX^+_{\ext_\Lambda\Ga}\\
\hateGa\cap [\Ga]_R\neq\emptyset}}\omega^+(\hateGa)
+\sum_{\substack{\hateGa\in
\hatcX^+_{\inte\Ga}\\ 
\hateGa\cap [\Ga]_R\neq\emptyset}}\omega^+(\hateGa)
-\sum_{\substack{\hateGa\in \hatcX^+_{\Lambda}\\ 
\hateGa\cap [\Ga]_R
\neq\emptyset}}\omega^+(\hateGa)\,,\nonumber}
Using symmetry,
\eq{\label{K224}|\log Q_c(0)|\leq 3\epsilon_c|[\Ga]_R|\,.}
For the derivative, a similar treatment gives
\eq{\label{K225}
\Big\|\deriv{h}\log Q_c\Big\|_{U_\Ga}
\leq  2\epsilon_c V(\Ga)+3\epsilon_c|[\Ga]_R|\,.}
Estimates \eqref{K220} and \eqref{K224} yield 
\al{2\beta V(\Ga)|g_\Lambda^+(\Ga)(0)|\leq 3(\epsilon_r+\epsilon_c)|[\Ga]_R|
\leq \delta_1(\beta)\beta\|\Ga\|}
where $\delta_1(\beta)\pardef
3^{d+1}\beta^{-1}(\epsilon_r+\epsilon_e)\rho^{-1}$ ($\rho$ is the
Peierls constant). We get \eqref{K211}
by setting $\delta_2(\beta)\pardef
\beta^{-1}(\epsilon_r+\epsilon_e)$.
\ep
We are now in position of computing derivatives of the functions $\monphi$.
The main ingredient is the following theorem, which is a consequence of Corollary
\ref{LSPC0}
of Appendix \ref{KSLSP}. 
The first proof of this result was
given by Isakov; it is nothing but a stationary
phase analysis applied to the Cauchy integral giving the $k$-th derivative at
$h=0$ of a function of the type $e^{-cz+bf(z)}$.
\begin{theo}\label{KT4}
Let $r>0$, $F(z)=
\exp(-cz+bf(z))$ where $1\leq b\leq c$, and $f$ is analytic in a disc
$\{|z|<r\}$, taking real values on the real line, 
with a uniformly bounded derivative:
\eq{\label{K228}
\sup_{|z|<r}|f'(z)|\leq A<\frac{1}{25}\,.}
There exists $k_0=k_0(A)$ such that the following holds:
define $k_+=r(c-2b\sqrt{A})$. 
For all integer $k\in [k_0,k_+]$ there
exists $r_k\in(0,r)$ and $c_k>0$ satisfying
\eq{\label{K229}
\frac{ k}{c+bA}\leq r_k\leq \frac{k}{c-bA}\,,\quad
\frac{3}{10}\,\frac{1}{\sqrt{2\pi cr_k}}<
\,c_k < \frac{1}{\sqrt{cr_k}}\,,}
such that
\eq{\label{K230} F^{(k)}(0)=\frac{k!}{2\pi
i}\int_{|z|=r_k}\frac{F(z)}{z^{k+1}}\diff z=
k!\frac{c_k}{(-r_k)^k}F(-r_k)\,.}
In particular, $(-1)^kF^{(k)}(0)>0$.
Moreover, if $f$ satisfies the local condition
\eq{\label{K231} bf(0)\leq -\alpha r c\,,}
with $\alpha\in(\log 2,1)$, then for all $k\in[k_0,k_+]$ and $A$ sufficiently
small,
\eq{\label{K232}\big(\log(1+F)\big)^{(k)}(0)=
(1+a\cdot e^{-\frac{1}{2}\zeta k})F^{(k)}(0)\,,}
where $a$ is a bounded function of $k,c,b$ and $\zeta
=\zeta(\alpha)>0$.
\end{theo}
\noindent In 
Lemma \ref{KL30}, we have put $\monphi$ in the form $e^{-cz+bf(z)}$. In order
to satisfy \eqref{K228}, we must introduce a distinction among the contours.
Consider the function $\delta_2(\beta)$ of \eqref{K211}.
\begin{defin}\label{KDEF5}
A contour $\Ga\in\cC^+(\Lambda)$ is \emph{\grasA{thin}} if $|\Ga|\leq
\frac{\delta_2(\beta)}{2}V(\Ga)$, and \emph{\grasA{fat}} if it is not thin.
\end{defin}
\noindent Now, any 
thin contour $\Ga$ satisfies, when $\beta$ is large enough,
\eq{\label{K231.1}\Big\|\deriv{h}g_\Lambda^+(\Ga)\Big\|_{{U_\Ga}}\leq
2\delta_2(\beta)\equiv A(\beta)<\frac{1}{25}\,.}
\begin{lem}\label{KL31}
There exists $k_0$ such that when $\beta$ is sufficiently large, the following
holds. For all thin contour $\Ga$, define
\eq{\label{K232.1}
k_+(\Ga)\pardef 2\beta V(\Ga)R^*(V(\Ga))(1-2\sqrt{A})\,.}
Then for all integer $k\in[k_0,k_+(\Ga)]$, we have 
\al{
(-1)^ku_\Lambda^+(\Ga)^{(k)}(0)&\geq
\frac{1}{10}\big(2\beta V(\Ga)D_-\big)^k
e^{-(1+\delta_1(\beta))
\|\Ga\|}\label{K233.1}\\
(-1)^ku_\Lambda^+(\Ga)^{(k)}(0)&\leq 20
\big(2\beta V(\Ga)D_+\big)^ke^{-(1-\delta_1(\beta))
\|\Ga\|}\,,\label{K233.2}}
where $\lim_{\beta\to\infty}D_\pm=1$.
\end{lem}
\bp
Let $\Ga$ be a thin contour. Consider $\monphi$ in its exponentiated form
\eqref{K209}. We apply Theorem \ref{KT4} with $c=b=2\beta V(\Ga)$,
$f=g_\Lambda^+(\Ga)-\frac{1}{2}\frac{\|\Ga\|}{V(\Ga)}$, $r=R^*(V(\Ga))$,
and $A=A(\beta)$. \eqref{K231.1} guarantees \eqref{K228}.
There exists $r_k=r_k(\Ga)$ and $c_k=c_k(\Ga)$ such that
\eq{(-1)^k\monphi^{(k)}(0)=k!\frac{c_k}{(r_k)^k}
\monphi(-r_k)\,.}
Using the analyticity of $g_\Lambda^+(\Ga)$ in ${U_\Ga}$, we have with 
\eqref{K229}
\al{\monphi(-r_k)&=e^{-\beta \|\Ga\|}e^{cr_k}e^{cg_\Lambda^+(\Ga)(0)}
e^{c(g_\Lambda^+(\Ga)(-r_k)-g_\Lambda^+(\Ga)(0))}\\
&\geq e^{-\beta \|\Ga\|}e^{\frac{k}{1+A}}e^{-\delta_1\beta\|\Ga\|}
e^{-\frac{A}{1-A}k}\\
&=e^{-(1+\delta_1)\beta \|\Ga\|}e^ke^{-\frac{2A}{1-A^2}k}\,.}
Using the Stirling formula and the estimates for $r_k,c_k$, we get
\eq{(-1)^k\monphi^{(k)}(0)\geq
\frac{1}{5}\big(2\beta V(\Ga)D_-\big)^ke^{-(1+\delta_1)\beta \|\Ga\|}\,,}
where
\eq{D_-(\beta)=(1-A)e^{-\frac{2A}{1-A^2}}\,.}
Using \eqref{K210} we can satisfy \eqref{K231}:
\al{bf(0)=2\beta V(\Ga)g_\Lambda^+(\Ga)(0)-\beta\|\Ga\|&\leq
-(1-\delta_1)\beta\|\Ga\|\\
&\leq -(1-\delta_1)2\beta V(\Ga)R^*(V(\Ga))\label{K238}\\
&=-(1-\delta_1)rc\,.}
In \eqref{K238} we used
\al{\|\Ga\|\geq \frac{1}{K(V(\Ga))}V(\Ga)^\ddmu&\geq 2V(\Ga)
\frac{\theta}{2 K(V(\Ga))V(\Ga)^\usd}\geq 2V(\Ga)R^*(V(\Ga))\label{K239}}
We can thus use \eqref{K232} once $\beta$ is large enough. This gives the lower
bound \eqref{K233.1}. The upper bound is obtained similarly.
\ep
\subsection{Derivatives of the Pressure in a 
Finite Volume}\label{KSSmechanism}
In this section, we
\emph{fix} $k$ large enough. When 
a thin contour satisfies $[k_0,k_+(\Ga)]\ni k$ then 
$u_\Lambda^+(\Ga)^{(k)}(0)$ can be estimated with
Lemma \ref{KL31}. To characterize this class of contours, we introduce a
$k$-dependent notion of size.
\begin{defin}\label{KDEF6}
Let $k\in\bN$, $\epsilon'>0$ small enough.
A contour $\Ga$ is \emph{\grasA{$k$-large}} if
$V(\Ga)\geq V_0(k)$ where
\eq{\label{K241}V_0(k)\pardef
\Big(\frac{K(\infty)(1+\epsilon')}{\theta \beta(1-2\sqrt{A})}
k\Big)^\ddmu\,,}
where $K(\infty)$ was defined in Lemma \ref{KL10}. $\Ga$ is \grasA{$k$-small} if
$V(\Ga)< V_0(k)$.
\end{defin}
\noindent Let $N_0(\epsilon')$ be such that for all $N\geq N_0(\epsilon')$ (see
Lemma \ref{KL13.1}),
\eq{\frac{1}{(1+\epsilon')}
\frac{\theta}{2K(\infty)N^\usd}\leq R^*(N)\leq \frac{\theta}{2K(\infty)
N^\usd}\,.}
Let $k_-=k_-(\epsilon',\ga)$ be such that when $k\geq k_-$ then $V_0(k)\geq
N_0(\epsilon')$. This definition implies that when $k\geq
k_-$, we have for all $k$-large contour $\Ga$
\al{k_+(\Ga)=2\beta V(\Ga)(1-2\sqrt{A})R^*(V(\Ga))&\geq
\frac{\theta\beta(1-2\sqrt{A})}{K(\infty)(1+\epsilon')}V(\Ga)^\dmusd\geq k\,.}
That is, the $k$-th 
derivative of a $k$-large thin contour can be studied with Lemma
\ref{KL31}.
The dependence of $k_-$ on $\ga$ comes from the bound $K(\infty)\geq c_-\ga$. We
therefore have $\lim_{\ga\searrow 0}k_-=+\infty$.
\begin{pro}\label{KP8} Let $\theta$ be close to $1$, $\beta$ large enough.
There exists a constant $C_1>0$ and an unbounded increasing sequence of integers
$k_1,k_2,\dots$ such that for large $N$, we have
whenever $\Lambda$ is sufficiently large,
\eq{\frac{(-1)^{k_N}}{|\Lambda|}\derivk{h}{k_N}\sum_{\Ga\in\cC^+(\Lambda)}
u_\Lambda^+(\Ga)\Big|_{h=0}\geq \big(C_1 
K(\infty)^{\ddmu}\beta^{-\frac{1}{d-1}}
\big)^{k_N}k_N!^{\ddmu}}
\end{pro}
\bp Fix $\epsilon>0$ small and consider the sequence
$(\Ga_N)_{N\geq 1}$ of Lemma \ref{KL10}. We have
$\lim_{N\to\infty}V(\Ga_N)=+\infty$ and when 
$N$ is large enough,
\eq{\label{K247.1}
(1-\epsilon)K(\infty)\leq\frac{V(\Ga_N)^\dmusd}{\|\Ga_N\|}
\leq (1+\epsilon)K(\infty)\,.}
The sequence $(k_N)_{N\geq 1}$ is defined such that the contribution from the 
contour $\Ga_N$ to $p_{\ga,\Lambda}^{+(k_N)}(0)$ is close to maximal. Let
\eq{\label{K248}k_N\pardef\Big\lfloor \dmusd
\beta\|\Ga_N\|\Big\rfloor\,.}
Since $\lim_{N\to\infty}V(\Ga_N)=+\infty$, we have
$\lim_{N\to\infty}k_N=+\infty$. 
From now on we consider $N$ large enough so 
that \eqref{K247.1} and \eqref{K247} hold 
and $k_N\geq\max\{k_0,k_-\}$.
When considering the $k_N$-th derivative, we use the following decomposition:
\eq{\label{K246}\sum_{\substack{\Ga\in\cC^+(\Lambda)}}=
\sum_{\substack{\Ga\in\cC^+(\Lambda)\\k_N-\text{large, thin}}}
+\sum_{\substack{\Ga\in\cC^+(\Lambda)\\k_N-\text{small, thin}}}
+\sum_{\substack{\Ga\in\cC^+(\Lambda)\\\text{fat}}}}
We show that the dominant term comes from $\Ga_N$, which belongs to the first
sum, and that the two other sums are negligible. To see that $\Ga_N$
appears in the first sum, we first show that $\Ga_N$ is $k_N$-large. Indeed, if
$\theta$ is close to $1$ and $\epsilon,\epsilon', A(\beta)$ are small,
\al{V_0(k_N)&\leq \Big(\frac{K(\infty)(1+\epsilon')}{
\theta(1-2\sqrt{A})}\frac{d-1}{d}\|\Ga_N\|\Big)^{\ddmu}\\
&\leq \Big(\frac{1}{\theta(1-2\sqrt{A})}\frac{1+\epsilon'}{1-\epsilon}
\frac{d-1}{d}\Big)^{\ddmu}V(\Ga_N)\leq V(\Ga_N)\,.}
Then we show that $\Ga_N$ is thin:
\eq{\frac{|\Ga_N|}{V(\Ga_N)}\leq\frac{1}{\rho}\frac{\|\Ga_N\|}{V(\Ga_N)}\leq
\frac{1}{\rho K(\infty)(1-\epsilon)}
\frac{1}{V_0(k_N)^\usd}\leq\frac{1}{2}\delta_2(\beta)\,\label{K247}}
Finally, we assume $\Lambda$ is large enough in
order to contain at least $\frac{1}{2}|\Lambda|$ translates of $\Ga_N$.
Then we apply Lemma \ref{KL31} to $u_\Lambda^+(\Ga_N)$. Using \eqref{K247.1},
\al{V(\Ga_N)^{k_N}&e^{-(1+\delta_1)\beta\|\Ga_N\|}\geq
\big((1-\epsilon)K(\infty)\|\Ga_N\|\big)^{\ddmu
k_N}e^{-(1+\delta_1)\beta\|\Ga_N\|}\\
&\geq\Big((1-\epsilon)K(\infty){\ddmu}\frac{1}{\beta}k_N\Big)^{\ddmu
k_N}e^{-(1+\delta_1)\ddmu(k_N+1)}\\
&\geq c(k_N)K(\infty)^{\ddmu k_N}\beta^{-\ddmu k_N}
\Big[{\ddmu} 
(1-\epsilon)e^{-\delta_1}\Big]^{\ddmu k_N}k_N!^{\ddmu}\,,}
where $c(k_N)\geq C_3 k_N^{-\frac{1}{2}}$ and we used the Stirling formula.
Since 
\eq{(-1)^{k_N}u_\Lambda^+(\Ga)^{(k_N)}(0)\geq 0} 
for all $k_N$-large thin
contour, we can bound the first sum from below using only the contributions
coming from the translates of $\Ga_N$. We get
\al{&\frac{(-1)^{k_N}}{|\Lambda|}\derivk{h}{k_N}
\sum_{\substack{\Ga\in\cC^+(\Lambda)\\k_N-\text{large, thin}}}
u_\Lambda^+(\Ga)\Big|_{h=0}\geq\nonumber\\
&\frac{c(k_N)}{20}2^{k_N}K(\infty)^{\ddmu k_N}\beta^{-\usdmu k_N}
\Big[{\ddmu}(1-\epsilon)e^{-\delta_1}D_-\Big]^{\ddmu k_N}
k_N!^{\ddmu}\,,\label{K249}}
Consider now a $k_N$-small thin contour, i.e. $R^*(V(\Ga))\geq R^*(V_0(k_N))$. 
Using
the Cauchy formula with a disc of radius $R^*(V_0(k_N))$ centered at $h=0$,
\eq{|u_\Lambda^+(\Ga)^{(k_N)}(0)|\leq k_N!\Big(\frac{1}{R^*(V_0(k_N))}
\Big)^{k_N}
\|u_\Lambda^+(\Ga)\|_{{U_\Ga}}\,.\label{K250}}
\begin{lem}\label{KL32}
Setting $\alpha_1=\alpha_1(\theta,\beta)\pardef
\rho^{-1}(1-\theta(1+A(\beta))-\delta_1(\beta))$. If $\beta$ is large enough, 
we have $\alpha_1>0$ and the bound 
\eq{\label{K251}
\|u_\Lambda^+(\Ga)\|_{{U_\Ga}}\leq
\frac{e^{-\beta\alpha_1|\Ga|}}{1-e^{-\beta\alpha_1|\Ga|}}}
\end{lem}
\bp
Using \eqref{K209}, \eqref{K210} and \eqref{K231.1},
\eq{\label{K252.1}\|\monphi\|_{{U_\Ga}}
\leq \sup_{h\in U_\Ga}
e^{-\beta(1-\delta_1)\|\Ga\|}e^{2\beta(1+A)|\real h|V(\Ga)}\leq 
e^{-\alpha_1\beta|\Ga|}<1\,,}
where we used the definition of the radius of analyticity:
\eq{\sup_{h\in U_\Ga}|h|
V(\Ga)\leq R^*(V(\Ga))V(\Ga)\leq R(V(\Ga))V(\Ga)\leq 
\frac{\theta}{2}\|\Ga\|\,.}
The proof finishes by using the Taylor expansion of $\log(1+x)$.
\ep
A standard Peierls estimate implies, when $\beta$ is large,
the existence of a constant $C_4$ such that
\eq{\sum_{\Ga\in\cC^+(\Lambda)}e^{-\beta\alpha_1|\Ga|}\leq C_4|\Lambda|\,.}
Using the Stirling formula, it easy to see that $k_N!k_N^{\frac{1}{d-1}k_N}
\leq k_N!^\ddmu {e}^{{\usdmu}k_N}$.
The contribution from the $k_N$-small 
contours is then bounded by
\al{&\frac{1}{|\Lambda|}\Big|\derivk{h}{k_N}
\sum_{\substack{\Ga\in\cC^+(\Lambda)\\k_N-\text{small, thin}}}
u_\Lambda^+(\Ga)\Big|_{h=0}\leq\nonumber\\
&C_5 2^{k_N}K(\infty)^{\ddmu k_N}\beta^{-\usdmu k_N}
\Big[{e^{\usdmu}}
\Big(\frac{1+\epsilon'}{\theta}\Big)^\ddmu
\Big(\frac{1}{1-2\sqrt{A}}
\Big)^\usdmu\Big]^{k_N} k_N!^\ddmu\label{K252}}
Since ${\frac{d}{d-1}}>
{e}^{{\frac{1}{d-1}}}$, 
the comparison of the square brackets of 
\eqref{K252} with those of 
\eqref{K249} shows that if $\theta$ is close to
$1$, if $\epsilon,\epsilon'$ are small, and if $\beta$ is  large enough, then
the contribution from the $k_N$-small contours is
negligible in comparison to the $k_N$-large ones.\\
We are then left with the contribution of the fat contours. We can
use a Cauchy bound
\al{\Big|\derivk{h}{k}u_\Lambda^+(\Ga)\Big|_{h=0}&\leq k!
\Big(\frac{1}{R^*(V(\Ga))}\Big)^{k}
\|u_\Lambda^+(\Ga)\|_{{U_\Ga}}\nonumber\\
&\leq k!\Big(\frac{2K(1)}{\theta}\Big)^{k}V(\Ga)^{\frac{k}{d}}
\frac{e^{-\beta\alpha_1|\Ga|}}{1-e^{-\beta\alpha_1|\Ga|}}\nonumber\\
&\leq k!\Big(\frac{2K(1)}{\theta}\Big(\frac{2}{\delta_2}\Big)^\usd
\Big)^{k}|\Ga|^{\frac{k}{d}}
\frac{e^{-\beta\alpha_1|\Ga|}}{1-e^{-\beta\alpha_1|\Ga|}}\nonumber}
Then a Peierls estimate leads to
\eq{\sum_{\Ga\in\cC^+(\Lambda)}
|\Ga|^{\frac{k}{d}}e^{-\alpha_1\beta|\Ga|}\leq
|\Lambda|\sum_{L\geq 1}L^{\frac{k}{d}}e^{-\alpha_1'\beta L}\leq |\Lambda|
(\alpha_1'\beta)^{-\frac{k}{d}}\Ga\big(\frac{k}{d}+1\big)\,,}
where $\Ga(x)$ is the Gamma-function. Using the Stirling formula,
it is then easy to show that the contribution from the fat contours is bounded by
\eq{\frac{1}{|\Lambda|}\Big|\derivk{h}{k}
\sum_{\substack{\Ga\in\cC^+(\Lambda)\\\text{fat}}}
u_\Lambda^+(\Ga)\Big|_{h=0}\leq \big(K(1)\beta^{-\frac{1}{d}}
D(k)\big)^kk!^\ddmu\,,}
where $\lim_{k\to\infty}D(k)=0$. The fat contours can thus always be ignored.
This finishes the proof of Proposition \ref{KP8}
\ep
With \eqref{K65.4}, we get the lower bound, for a large enough box $\Lambda$,
\al{|p_{\ga,\Lambda}^{+(k_N)}(0)|&\geq \big(C_1K(\infty)^\ddmu 
\beta^{-\usdmu}\big)^{k_N}k_N!^\ddmu-C_r^{k_N}k_N!\\
&\geq \big(C_-\ga^\ddmu 
\beta^{-\usdmu}\big)^{k_N}k_N!^\ddmu-C_r^{k_N}k_N!\,.\label{K254}}
We used the lower bound $K(\infty)\geq c_-\ga$ from Lemma \ref{KL10}.
Notice that we could extract the contribution of the translates of $\Ga_N$ 
to $p_{\ga,\Lambda}^{+(k_N)}(0)$
\emph{without} knowing its explicit
shape. This is where our formulation of the isoperimetric problems differs 
from
the one of Isakov. Notice also that the lower bound \eqref{K254} shows how
non-analyticity is detected in \emph{finite} volumes.
\subsection{Thermodynamic Limit; Proofs of Theorems \ref{KT1} and 
\ref{KT2}}\label{KSlimithermo}
To extend the bounds we have on $p_{\ga,\Lambda}^{+(k_N)}(0)$ to 
the infinite volume limit, we first
show that in the strip $U_+$ the derivatives of
the pressure are uniformly bounded. 
\begin{lem}\label{KL33}
Let 
$\beta$ be large enough. There exists $C_+>0$ 
such that for all $k\geq 2$, 
\eq{\sup_{\Lambda}\|p_{\ga,\Lambda}^{+(k)}\|_{U_+}\leq 
\big(C_+\ga^{\ddmu}\beta^{-\usdmu}\big)^kk!^\ddmu+C_r^kk!\,.}
\end{lem}
\bp Like in Section \ref{KSpure}, we can fix
$\theta\pardef \frac{1}{2}$.
The term $C_r^kk!$ comes from \eqref{K65.4}. Consider $u_\Lambda^+(\Ga)$ and the
representation \eqref{K212} of $\monphi$. From Lemma \ref{KL30}, $\monphi$ is
analytic in $U_\Ga$. From Proposition \ref{KP6} and Lemma \ref{KL16}, 
it is also analytic in $U_+$, i.e. in $U_+\cup U_\Ga$. Proceeding like in the
proof of Lemma \ref{KL30}, we get
\al{\Big\|\frac{\Theta^+_{i(\Ga)}(\ext_\Lambda\Ga;\sigma_\Ga)
\Theta^+(\inte\Ga;-\sigma_\Ga)}{\Theta^+_{i(\Ga)}(\Lambda)}
\Big\|_{U_+}&\leq \sup_{h\in U_+}e^{-\beta \real h|\Ga|}
e^{3(\epsilon_r+\epsilon_c)|[\Ga]_R|}\\
&=e^{3(\epsilon_r+\epsilon_c)|[\Ga]_R|}\,.}
Assume $3^{d+1}(\epsilon_r+\epsilon_c)\leq \frac{1}{3}$. Using \eqref{K452},
\eq{\|\monphi\|_{U_+}\leq e^{-\beta\|\Ga\|}e^{\beta h_+|\Ga|}e^{|\Ga|}\leq
e^{-\alpha_2\beta|\Ga|}<1\,.}
Notice that unlike in \eqref{K252.1}, $\alpha_2$ in independent of $\theta$.
This implies that $u_\Lambda^+(\Ga)$ is also analytic in $U_+\cup U_\Ga$. 
Set $\alpha_3=\min\{\alpha_1,\alpha_2\}$. Using
a disc of radius $R^*(V(\Ga))$ around each $h\in U_+$, we have
\al{\|u_\Lambda^+(\Ga)^{(k)}\|_{U_+}
&\leq k!\Big(\frac{1}{R^*(V(\Ga))}\Big)^k
\|u_\Lambda^+(\Ga)\|_{U_+\cup U_\Ga}\\
&\leq k!\Big(\frac{2K(1)}{\theta}\Big)^{k}V(\Ga)^{\frac{k}{d}}
\frac{e^{-\beta\alpha_3|\Ga|}}{1-e^{-\beta\alpha_3|\Ga|}}\\
&\leq k!\Big(\frac{2K(1)}{\theta l^\usdmu}\Big)^{k}|\Ga|^{\frac{k}{d-1}}
\frac{e^{-\beta\alpha_3|\Ga|}}{1-e^{-\beta\alpha_3|\Ga|}}}
We used the isoperimetric inequality of Lemma \ref{KL9}. Remember that $K(1)\leq
c_+\ga$ (Lemma \ref{KL10}), and that
$l=\nu\ga^{-1}$.
The proof finishes like for the upper bound on fat contours.
\ep
\begin{cor}\label{KC3}
For all $h'\in U_+\cup\{\real h=0\}$ and for all $k\in\bN$,
\eq{\label{K266}p_\ga^{(k),\leftarrow}(h')=
\lim_{\Lambda\nearrow \bZ^{d}}p_{\ga,\Lambda}^{+(k)}(h')=\lim_{h\searrow h'}
p_{\ga}^{(k)}(h)\,.}
\end{cor}
\bp
We show \eqref{K266} for $k=1$. By definition,
\al{p_\ga^{(1),\leftarrow}(h')&=\lim_{\delta\searrow 0}\frac{p_{\ga}(h'+
\delta)-p_\ga(h')}{\delta}\\
&= \lim_{\delta\searrow 0}\lim_{\Lambda\nearrow \bZ^d}
\frac{p_{\ga,\Lambda}^+(h'+\delta)-p^+_{\ga,\Lambda}(h')}{\delta}\\
&=\lim_{\delta\searrow 0}\lim_{\Lambda\nearrow\bZ^d}
\big(p_{\ga,\Lambda}^{+(1)}(h')+\frac{1}{2!}p_{\ga,\Lambda}^{+(2)}
(h(\delta))\delta\big)\,,}
where $\lim_{\delta\searrow 0}h(\delta)=h'$. The following lemma will allow to
permute 
the limits $\lim_{\delta\searrow 0}$ and $\lim_{\Lambda\nearrow\bZ^d}$.
\begin{lem}\label{KL34}
Let, for all $N\in \bN$, $\delta>0$, $b_N(\delta)=a_N+c_N(\delta)$, such that
$|c_N(\delta)|\leq D\delta$ uniformly in $N$, and $\lim_{N\to \infty}
b_N(\delta)=b(\delta)$ exists. Then $\lim_{N\to\infty}a_N$ and
$\lim_{\delta\searrow 0} b(\delta)$ exist and are equal. 
\end{lem}
\bp
We first show that $\lim_{\delta\searrow 0} b(\delta)$ exists. Let
$(\delta_k)$ be any sequence $\delta_k>0$ such that $\lim_{k\to\infty}
\delta_k=0$. Then we have
\eq{|b(\delta_{k})-b(\delta_{k'})|=|\lim_{N\to\infty}(c_N(\delta_k)
-c_N(\delta_{k'}))|\leq D(\delta_k+\delta_{k'})\,}
and so $\lim_{k\to\infty}b(\delta_k)$ exists. Fix
$\epsilon>0$. There exists $N_{\epsilon,\delta}$
such that if $N\geq N_{\epsilon,\delta}$ then
$|b_N(\delta)-b(\delta)|\leq \epsilon$. We then have
\eq{b(\delta)-\epsilon -D\delta\leq \liminf_{N\to\infty}
a_N\leq \limsup_{N\to\infty}a_N\leq
b(\delta)+\epsilon+D\delta\,,} which
finishes the proof, once we take $\epsilon\to 0$, $\delta\to 0$.
\ep 
Using the fact that the second
derivative is uniformly bounded on $U_+$ (Lemma \ref{KL33}),
this shows the first equality in \eqref{K266}. 
For the second, we only need to
consider the case where $h'=0$. 
\al{p_\ga^{(1),\leftarrow}(0)&=\lim_{\delta\searrow
0}\frac{p_\ga(\delta)-p_\ga(0)}{\delta}\\
&=\lim_{\delta\searrow
0}\Big[\frac{p_\ga(\delta)-p_\ga(\frac{\delta}{2})}{\delta}
+\frac{p_\ga(\frac{\delta}{2})-p_\ga(0)}{\delta}\Big]\\
&=\big(\lim_{\delta\searrow 0}\frac{1}{2}p_\ga^{(1)}(h(\delta))\big)
+\frac{1}{2}p_\ga^{(1),\leftarrow}(0)\,,}
where $h(\delta)\in[\frac{\delta}{2},\delta]$ and 
$\lim_{\delta\searrow 0}h(\delta)=0$. This shows 
\eq{p_\ga^{(1),\leftarrow}(0)=
\lim_{\delta\searrow 0}p_\ga^{(1)}(h(\delta))\,,} which extends easily
to any sequence $h\searrow 0$, since derivatives of any order are uniformly
bounded on $U_+$.
\ep
We can then complete the proofs of our main results.
\bp [Proof of Theorem \ref{KT2}:]
The bounds on $p_{\ga,\Lambda}^{(k)}(0)$ of \eqref{K254} 
and Lemma \ref{KL33} extend to the thermodynamic limit using Corollary
\ref{KC3}. 
\ep
\bp [Proof of Theorem \ref{KT1}:]
Using the symmetry $p_\ga(h)=p_\ga(-h)$, we can write, for $m\geq 0$,
\eq{\label{K267}f_\ga(m)=\sup_{h\geq 0}\big(hm-p_\ga(h)\big)\,.}
By the Theorem of Lee and Yang, $h\mapsto p_\ga(h)$ is analytic in $\{\real
h>0\}$. If $m^*\pardef p_\ga^{(1),\leftarrow}(0)$, then
$p_\ga:(0,+\infty)\to(m^*,1)$ is one-to-one, and for all $m\in (m^*,1)$,
\eq{f_\ga(m)=h(m)m-p_\ga(h(m))\,,}
where $h(m)$ is the unique solution of the equation $p_\ga^{(1)}(h)=m$. 
The GHS inequality (see \cite{GHS})
states that for all $h> 0$,
\eq{p_\ga^{(2),\leftarrow}(0)\geq p_\ga^{(2)}(h)>0\,.}
Since $p^{(2)}_\ga(h)\neq 0$ for all $h>0$; the
biholomorphic mapping theorem~\footnote{
Let $g:D\to\bC$ be an analytic function, $z_0\in D$ be a point such that
$g'(z_0)\neq 0$. Then there exists a domain $V\subset D$ containing 
$z_0$, such that the following holds: $V'=g(V)$ is a domain, and
the map $g:V\to V'$ has an inverse
$g^{-1}:V'\to V$ which is
analytic, and which satisfies, for all $\omega\in V'$,
${{g^{-1}}'(\omega)=\big(g'(g^{-1}(\omega))\big)^{-1}\,.}$
The proof of this result can be found in \cite{Rem}, pp. 281-282.}
implies that $m\mapsto h(m)$ is analytic in a complex neighbourhood of each
$m\in(m^*,1)$. So $f_\ga$, which is a composition of analytic maps, is
analytic on $(m^*,1)$.\\
Since $p_\ga^{(1),\leftarrow}(0)=\lim_{h\searrow 0}p_\ga^{(1)}(h)$ exists
(Corollary \ref{KC3}), we can extend $p_\ga^{(1)}$ to $[0,+\infty)$ by setting
 $p_\ga^{(1)}(0)\pardef p_\ga^{(1),\leftarrow}(0)$ and we can extend
 $h(\cdot)$ to $[m^*,1)$ by setting
 $h(m^*)\pardef 0$. \\
 We now show that $f_\ga$ has no analytic continuation at $m^*$. Assume this is
 wrong, and denote by $\tilde{h}(\cdot)$ the analytic continuation of $h(\cdot)$
 at $m^*$. Using the definition of $h(m)$, we compute
 \al{\tilde{h}^{(1)}({m^*})&=\lim_{\substack{m\searrow m^*\\ m\in \bR}}
\frac{h(m)-{h}(m^*)}{m-m^*}\\
&=\lim_{\substack{m\searrow m^*\\ m\in
\bR}}\frac{h(m)-h(m^*)}{p_\ga^{(1)}(h(m))-p_\ga^{(1)}(0)}\\
&=\Big(\lim_{h\searrow 0}\frac{p_\ga^{(1)}(h)-p_\ga^{(1)}(0)}{h}
\Big)^{-1}=\big(p_\ga^{(2),\leftarrow}(0)\big)^{-1}\neq 0\,,}
So $\tilde{h}(\cdot)$ maps a neighbourhood of $m^*$ on a neighbourhood of $0$,
and $\tilde{h}^{(1)}({m^*})\neq 0$.
The biholomorphic mapping theorem implies that $\tilde{h}(\cdot)$ can be inverted
in an open
neighbourhood of $h=0$ and that this inverse, which coincides with 
$p_\ga^{(1)}$ on the positive real line, is analytic,
 a contradiction with the non-analyticity of $p_\ga$ at $h=0$ 
 (Theorem \ref{KT2}).
\ep
\section{Conclusion}
Our analysis has lead to the following representation of the pressure for $h\geq
0$: 
\eq{\label{K700}p_\ga(h)=p^+_{r,\ga}(h)+s_\ga^+(h)\,,}
where $p_{r,\ga}^+$ is the restricted pressure. As we
have seen in Section \ref{KSrestricted}, $p_{r,\ga}^+$, which describes a
homogeneous phase with positive magnetization, behaves analytically at $h=0$. On
the other side, $s_\ga^+$ contains the contributions from droplets (contours)
of any
possible sizes, and is responsible for the non-analytic 
behaviour of the pressure at
$h=0$. Non-analyticity can be detected only in the very high
order derivatives of $s_\ga^+$, although $s_\ga^+$ contributes essentially
nothing to the pressure when $\ga$ is small.
 Indeed, $s_\ga^+$ can be expressed as a
sum over clusters of chains, and each chain contains at least one contour. Since
the length $|\Ga|$ of a contour is bounded below by the size of a cube
$C^{(l)}$, we have
\eq{\label{K701}
\|s_\ga^+\|_{U_{+}}\leq ae^{-b\beta \ga^{-d}}\,,}
where $a,b>0$ are constants.\\
For the pressure, the Lebowitz-Penrose Theorem takes the form (see \cite{Pr}):
\eq{\label{K702}
p_0(h)\pardef \lim_{\ga\searrow 0}p_\ga(h)=\inf_{m\in[-1,+1]}
\phi_h(m)\,,}
where
\eq{\label{K703}
\phi_h(m)\pardef -hm-\frac{1}{2}m^2-\frac{1}{\beta}I(m)\,.} 
The bound \eqref{K701} implies, for $h\geq 0$,
\eq{\label{K704}p_0(h)=\lim_{\ga\searrow 0}p_{r,\ga}^+(h)=
\inf_{m\geq 0}\phi_h(m)\,.}
\begin{figure}[htbp]
\begin{center}
\input{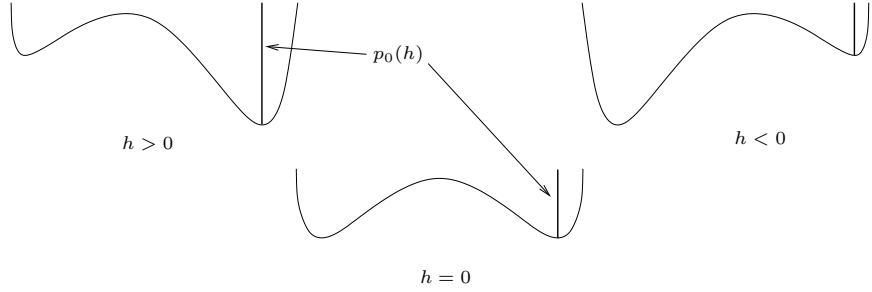}
\end{center}
\caption{{ The 
function $\phi_h(m)$ for different values of $h$. For $h\geq 0$, the 
vertical line
represents the value taken by $p_0(h)$. When $h<0$ it represents 
its analytic continuation, given by $\lim_{\ga\searrow 0}p_{r,\ga}^+$.}}
\label{KF30}
\end{figure}
From this last expression, the analytic continuation of the pressure, in the van
der Waals limit, at $h=0$, can be
understood easily: for $h>0$, $\phi_h(m)$ has a unique global minima at 
$m^*(h,\beta)>0$. When $h<0$ this minima is only local, but provides the analytic
continuation at $h=0$. The identity \eqref{K704} shows that the constraint on
the local magnetization, in $p_{r,\ga}^+$, has the effect of selecting the
minima $m^*(h,\beta)$, which is global when $h>0$ and local when $h<0$. This
mechanism is illustrated on Figure \ref{KF30}. When $\ga>0$, this scenario
breaks down: droplets are well defined, and they are all
stable at $h=0$, creating arbitrarily large fractions of the
$-$ phase. As we saw, this gives 
a contribution $k!^{\ddmu}$ to the $k$-th derivative of the
pressure.
\nouvellepage
\appendix
\section{A Stationary Phase Analysis}\label{KSLSP}
The following theorem is a generalization of a result due to Isakov \cite{I1}.
Let $\cD_\rho(t)\pardef\{z\in\bC:|z-t|<\rho\}$.
\begin{theo}\label{LSPT0}
Let $\rho>0$, 
$F(z)=\exp(-cz+bf(z))$ where $1\leq b\leq c$, and $f$ is analytic in a disc
$\cD_\rho(0)$, with a uniformly bounded derivative:
\eq{\label{LSP0}\sup_{z\in \cD_\rho(0)}|f'(z)|\leq A<\frac{1}{25}\,.}
There exists $k_0=k_0(A)$ such that the following holds: let $t\in \cD_\rho(0)$
and define $k_+=(\rho-|t|)(c-2b\sqrt{A})$. For all integer $k\in [k_0,k_+]$ there
exists $z_k=r_ke^{i\varphi_k}\in\cD_\rho(0)$ and $c_k\in \bC$ such that 
\eq{\label{LSP1}
F^{(k)}(t)=k!\frac{c_k}{(-z_k)^k}F(-z_k+t)\,.}
When $\imag t=0$ and $f(z)$ takes real values for real $z$,
then $\varphi_k=0$ and $\imag c_k=0$, and we have the estimates
\eq{\label{LSP2}\frac{3}{10}\,\frac{1}{\sqrt{2\pi cr_k}}<
\mathrm{Re}\,c_k < \frac{1}{\sqrt{cr_k}}\quad
\text{and}\quad
\big|\mathrm{Im}\,c_k\big|\leq \,\frac{1}{\sqrt{cr_k}}\,,}
\eq{\label{LSP3}\big|\tan \varphi_k\big|\leq \frac{bA}{c-bA}\quad
\text{and} \quad \frac{ k\cos\varphi_k}{c+bA}\leq
r_k\leq \frac{ k\cos\varphi_k}{c-bA}\,.}
\end{theo}
\noindent We 
have not indicated, for notational convenience, the dependence of $r_k,
\varphi_k, c_k$ on $t$.
A consequence of this Theorem, for $t\in (-\rho,+\rho)$, is
given after the proof in Corollary \ref{LSPC0}.
Our theorem improves significantly the original
result of \cite{I1}, since 
we show that derivatives of the function can be
obtained anywhere in the disc of analyticity $\cD_\rho(0)$, although we don't
use it in our proof. In the course of the
proof, we make clear the fact that the stationary point $z_k=r_ke^{i\varphi_k}$
is solution of a \emph{system}
of equations (see \eqref{eqphi}-\eqref{eqpsi}), whereas
Isakov considered only the point $t=0$, and there a single equation suffices to
find $z_k$ since $\varphi_k=0$. Since this result is at the core of the proof of
non-analyticity, we have explicited every step of the proof.
\begin{proof}[Proof of Theorem \ref{LSPT0}:] We use the
Cauchy formula. Define $\kappa\in(0,1)$ by $\kappa\rho=|t|$.
For all $r\in(0,\rho(1-\kappa))$ we have
\begin{equation}\nonumber
\frac{1}{k!}F^{(k)}(t)=
\frac{1}{2\pi i}\int_{\partial \cD_r(t)}\frac{F(z)}{(z-t)^{k+1}}\,\diff z=
\frac{1}{2\pi r^k}\int_0^{2\pi} 
\frac{F(re^{i\varphi}+t)}{e^{ik\varphi}}\diff\varphi\,,
\end{equation}
where we have used the parametrization $z:=t+re^{i\varphi}$
for $\partial \cD_r(t)$.
Our aim is to extract the main contribution to this last integral. The
integrand, because of the form of $F$, has
a maximal value for $\varphi$ close to $\pi$. We thus make a
change of variable, $\varphi':=\varphi-\pi$, to obtain
\begin{equation}\label{bkak}
\frac{1}{k!}F^{(k)}(t)=\frac{(-1)^ke^{-ct}}{r^k}\frac{1}{2\pi}
\int_{-\pi}^{+\pi}e^{\phi(r,\varphi)+i\psi(r,\varphi)}\diff \varphi\,,
\end{equation}
where
\begin{align}
\phi(r,\varphi)&:=cr\cos\varphi+b\mathrm{Re}f(-re^{i\varphi}+t)\nonumber\\
\psi(r,\varphi)&:=cr\sin\varphi+b\mathrm{Im}f(-re^{i\varphi}+t)
-k\varphi\,.\nonumber
\nonumber
\end{align}
If $t\in\bR$ and $f$ is real at real points, $\overline{f(z)}=
f(\overline{z})$ and
therefore $\mathrm{Im}f(\overline{z}+t)=-\mathrm{Im}f(z+t)$. By symmetry
we get $F^{(k)}(t)\in\bR$.
The core of the proof is to choose $r$ in a specific manner. This is a
standard stationary phase analysis. To this end, we will need an estimate on
the second derivative of $f$. Using Cauchy allows to obtain, for all $z_0\in
\cD_\rho(0)$:
\eq{f''(z_0)=\frac{1}{2\pi
i}\int_{\partial\cD_{r'}(z_0)}\frac{f'(\omega)}{(\omega-z_0)^{2}}\diff \omega=
\frac{1}{2\pi}\int_{0}^{2\pi}\frac{f'(z_0+r'e^{i\theta})}{r'e^{i\theta}}\diff
\theta\,,} where $r'>0$ is such that $|z_0|+r'<\rho$. Using the uniform bound
$|f'|<A$ we obtain (we optimize taking the largest possible $r'$, namely
$\rho-|z_0|$)
\eq{\label{bornefseconde}|f''(z_0)|\leq\frac{A}{\rho-|z_0|}\,.}
Now, set $t\in\cD_\rho(0)$, $|t|=\kappa\rho$, and consider the
map $\varphi\mapsto f(-re^{i\varphi}+t)$. A direct computation yields 
\begin{align}
\frac{\diff}{\diff \varphi}f(-re^{i\varphi}+t)&=-ire^{i\varphi}f'(
-re^{i\varphi}+t)\nonumber\\
\frac{\diff^2}{\diff \varphi^2}f(-re^{i\varphi}+t)&=re^{i\varphi}
f'(-re^{i\varphi}+t)-(re^{i\varphi})^2f''(-re^{i\varphi}+t)\,.\nonumber
\end{align}
Using \eqref{bornefseconde} gives the bound
\eq{\label{bornefseconde2}\sup_{\varphi}\left|
\frac{\diff^2}{\diff \varphi^2}f(-re^{i\varphi}+t)\right|\leq rA +r^2
\frac{A}{\rho(1-\kappa)-r}=\frac{rA}{1-\frac{r}{\rho(1-\kappa)}}\,.}
We now turn to the existence of a saddle point.
\begin{lem}\label{LSPL1}
Let $t\in \cD_\rho(0)$, $|t|=\kappa\rho$. Then for all
$k\in[0,\rho(1-\kappa)(c-2b\sqrt{A})]$, the system 
\begin{align}
\derivp{\varphi}\phi(r,\varphi)&=0\label{eqphi}\\
\derivp{\varphi}\psi(r,\varphi)&=0\label{eqpsi}
\end{align}
has a solution $(r_k,\varphi_k)$ with $r_k$ and $\varphi_k$ satisfying the
following estimates:
\begin{equation}\label{varphi*}
\big|\tan \varphi_k\big|\leq \frac{bA}{c-bA}\quad
\text{and}
\quad
\frac{ k\cos\varphi_k}{c+bA}\leq
r_k\leq \frac{ k\cos\varphi_k}{c-bA}\,.
\end{equation}
\end{lem}
\begin{proof}
We explicit \eqref{eqphi}, \eqref{eqpsi}:
\begin{align}
\sin\varphi\big(c-b\mathrm{Re}f^\prime(-z+t)\big)-
\cos\varphi \,b\mathrm{Im}f^\prime(-z+t)&=0 \,; \nonumber\\
r\cos\varphi\big(c-b\mathrm{Re}f^\prime(-z+t)\big)+
r\sin\varphi\,b\mathrm{Im}f^\prime(-z+t)&=k\,.\nonumber
\end{align}
These two equations are equivalent to 
\begin{align}
k\sin\varphi&=
rb\mathrm{Im}f^\prime(-z+t))\,;\label{e3}\\
k\cos\varphi&=r(c-b\mathrm{Re}f^\prime(-z+t))\label{e4}\,.
\end{align}
Then, we see that any solution of the system \eqref{eqphi}, \eqref{eqpsi},
satisfies \eqref{varphi*}. To show that there exists a solution, we first
solve \eqref{e4} locally for some fixed $\varphi\in(-\frac{\pi}{2},
+\frac{\pi}{2})$ (so that $\cos\varphi>0$). Define the map
$r\mapsto \xi(r,\varphi):=r(c-b\mathrm{Re} f^\prime(-re^{i\varphi}+t))$.
Since $f$ is analytic, its real and imaginary parts are $C^\infty$
with respect to $r>0$ and $\varphi$ (see \cite{Rem}), so
$\xi$ is $C^\infty$. We have $\xi(0,\varphi)=0$, and
\begin{align}
\xi(\rho(1-\kappa),\varphi)&=\rho(1-\kappa)\big(c-b\mathrm{Re}f^\prime\big(-
\rho(1-\kappa)e^{i\varphi}+t\big)\big)\nonumber\\
&\geq \rho(1-\kappa)(c-bA)\nonumber\\
&\geq\rho(1-\kappa)(c-2b\sqrt{A})\geq k\geq k\cos
\varphi\nonumber
\end{align}
which proves the existence of some $r_\varphi\in (0,\rho(1-\kappa)]$ 
such that
$\xi(r_\varphi,\varphi)=k\cos \varphi$. Notice that we also have
that~\footnote{The definition of $k_+$, with $2\sqrt{A}$ instead of $\sqrt{A}$,
ensures the strict inequality $<1$.}
\begin{equation}\label{124421}
\frac{r_\varphi}{\rho(1-\kappa)}=
\frac{k\cos\varphi}{\rho(1-\kappa)(c-b{\mathrm Re}f^\prime)}
\leq\frac{c-2b\sqrt{A}}{c-bA}<1\,.
\end{equation}
We can then show that the solution $r_\varphi$ is unique, by verifying that 
$\frac{\partial}{\partial r}\xi(r,\varphi)$ is strictly
 positive at $r_\varphi$. First,
\begin{align*}
\frac{\partial}{\partial r} \xi(r,\varphi)
&= 
c-b\mathrm{Re}f^\prime(-re^{i\varphi}+t)-r
b\frac{\partial}{\partial r} \mathrm{Re}f^{\prime}(-re^{i\varphi}+t)\\ 
&=c-b\mathrm{Re}f^\prime(-re^{i\varphi}+t)
+rb\mathrm{Re}\big({\rm e}^{i\varphi}f^{\prime\prime}(-re^{i\varphi}+t)\big)
\nonumber\\
(\text{see} \;\eqref{bornefseconde})\quad&\geq 
c-\frac{bA}{1-\frac{r}{\rho(1-\kappa)}}\,.\nonumber
\end{align*}
At $r=r_\varphi$ we get (see \eqref{124421})
$$\left.\frac{\partial}{\partial r} \xi(r,\varphi)\right|_{r=r_\varphi}\geq
c-\frac{\sqrt{A}(c-bA)}{2-\sqrt{A}}\geq
c-\frac{c\sqrt{A}}{2-\sqrt{A}}\geq\frac{8c}{9}> 0\,,
$$ 
which proves uniqueness of $r_\varphi$.
The continuity of $\varphi\mapsto r_\varphi$ is a consequence of the 
implicit function theorem. We turn
to the second equation, and set $r=r_\varphi$. Using again
equations \eqref{e3}, \eqref{e4}, we have
\begin{equation}\nonumber
\tan\varphi=\frac{b\mathrm{Im}f^\prime\big(r_\varphi{\rm e}^{i\varphi}\big)}
{c-b\mathrm{Re}f^\prime\big(r_\varphi{\rm e}^{i\varphi}\big)}\,.
\end{equation}
On $(-\frac{\pi}{2},\frac{\pi}{2})$ the function
\begin{equation}\nonumber
\varphi\mapsto\frac{b\mathrm{Im}f^\prime\big(r_\varphi{\rm e}^{i\varphi}\big)}
{c-b\mathrm{Re}f^\prime\big(r_\varphi{\rm e}^{i\varphi}\big)}\,,
\end{equation}
is continuous and takes its values
in the interval $\big(\frac{-bA}{c-bA},\frac{bA}{c-bA}\big)$.
Therefore there exists a solution~\footnote{We have not yet shown that this
solution is unique.} $(r_k,\varphi_k)$,
$r_k:=r_{\varphi_k}$, of \eqref{e3} and
\eqref{e4}.
\end{proof}
Notice that we have explicit bounds on $\varphi_k$, such as
\eq{\label{bornesvarphik}|\sin\varphi_k|\leq|\tan \varphi_k|\leq
\frac{1}{24}\,\,\,,\,\,\,|\cos\varphi_k|\geq
\frac{4}{5}\,\,\,,\,\,\,|\varphi_k|\leq \frac{\pi}{8}\,,} and that we can
estimate, at $r=r_k$ (see \eqref{bornefseconde2} and \eqref{124421}),
\eq{\label{bornefseconde2rk}\sup_{\varphi}\left|
\frac{\diff^2}{\diff \varphi^2}f(-r_ke^{i\varphi}+t)\right|\leq
\frac{r_kA}{1-\frac{r_k}{\rho(1-\kappa)}}\leq r_kA 
\frac{c-bA}{2b\sqrt{A}-bA}\leq \frac{5}{9}\frac{c}{b}r_k \sqrt{A}\,.}
We now examine \eqref{bkak} when $r=r_k$. Defining
$z_k=r_ke^{i\varphi_k}$, we extract the value taken by the integrand at 
$z_k$:
\eq{\frac{1}{k!}F^{(k)}(t)
=\frac{F(-z_k+t)}{(-z_k)^k}\underbrace{\frac{1}{2\pi}
\int_{-\pi}^{+\pi}e^{\phi(r_k,\varphi)-\phi(r_k,\varphi_k)
+i\big(\psi(r_k,\varphi)-\psi(r_k,\varphi_k)\big)}\diff \varphi}_{\equiv c_k}}
We will estimate the integral in $c_k$ by decomposing $[-\pi,+\pi]$ in two
parts. The
first is $[-\pi,+\pi]\backslash [-\frac{\pi}{4},+\frac{\pi}{4}]$.
\begin{lem}\label{LSPL2}
For all $\delta>0$, there exists $k_1=k_1(\delta)$ such that for all $k\geq k_1$
we have
\eq{\left|\frac{1}{2\pi}\left(\int_{\frac{\pi}{4}}^{\pi}+\int_{-\pi}^{
-\frac{\pi}{4}}\right)e^{\phi(r_k,\varphi)-\phi(r_k,\varphi_k)
+i\big(\psi(r_k,\varphi)-\psi(r_k,\varphi_k)\big)}\diff \varphi\right|
\leq \delta \frac{1}{\sqrt{2\pi c r_k}}\,.}
\end{lem}
\begin{proof}
We have $|e^{i(\psi(r_k,\varphi)-\psi(r_k,\varphi_k))}|=1$.
First, consider the interval $[\frac{\pi}{4},\pi]$. On this interval,
$\cos\varphi\leq y(\varphi)$ where
$\varphi\mapsto y(\varphi):=\cos\frac{\pi}{4}-\sin\frac{\pi}{4}(\varphi
-\frac{\pi}{4})$ (we have $y(\pi)=-0,95\dots>-1$). We can thus compute
\begin{align}
cr_k(\cos\varphi-\cos\varphi_k)&=cr_k(\cos\varphi-\cos\frac{\pi}{4}
+\cos\frac{\pi}{4}-\cos\varphi_k)\nonumber\\
&\leq
-\frac{\sqrt{2}}{2}cr_k(\varphi-\frac{\pi}{4})+cr_k(\cos\frac{\pi}{4}
-\cos\varphi_k)\nonumber\\
&\leq
-\frac{\sqrt{2}}{2}cr_k(\varphi-\frac{\pi}{4})-\frac{2.3}{25}cr_k\nonumber\,, 
\end{align}
where we used \eqref{bornesvarphik} in the last step. For the other part
containing $f$,
\begin{align}
b\big(\real f(-re^{i\varphi}+t)-\real f(-re^{i\varphi_k}+t)\big)&\leq
b(\varphi-\varphi_k)\sup_\varphi\left|\deriv{\varphi}\real
f(-r_ke^{i\varphi_k}+t)\right|\nonumber\\
&\leq br_k A(\varphi-\varphi_k)\nonumber\\
&=br_k A(\varphi-\frac{\pi}{4}+\frac{\pi}{4}-\varphi_k)\nonumber\\
&\leq cr_k A(\varphi-\frac{\pi}{4})+\frac{2}{25}cr_k\nonumber
\end{align}
The first part of the integral can thus be bounded by:
\begin{align}
\frac{e^{-\frac{0.3}{25}cr_k}}{2\pi}
\int_{\frac{\pi}{4}}^\pi e^{-\frac{\sqrt{2}}{2}cr_k(\varphi-\frac{\pi}{4})
+cr_k
A(\varphi-\frac{\pi}{4})}\diff \varphi
&\leq\frac{e^{-\frac{0.3}{25}cr_k}}{2\pi}\int_{0}^\infty
e^{-(\frac{\sqrt{2}}{2}-A)x}\diff x\nonumber\\
&\leq
\frac{e^{-\frac{0.3}{25}cr_k}}{\sqrt{2\pi cr_k}(\frac{\sqrt{2}}{2}
-\frac{1}{25})}\frac{1}{\sqrt{2\pi cr_k}}\nonumber\\
&\leq \frac{\delta}{2}\frac{1}{\sqrt{2\pi cr_k}}\,,\nonumber
\end{align}
once $k$ is large enough, since $cr_k\geq  \frac{1}{1+A}\frac{4}{5}k$ (see
\eqref{varphi*}). The same can be done on $[-\pi,-\frac{\pi}{4}]$,
on which we use the function
$y(\varphi):=\cos\frac{\pi}{4}+\sin\frac{\pi}{4}(\varphi-\frac{\pi}{4})$.
\end{proof}
On the interval $[-\frac{\pi}{4},+\frac{\pi}{4}]$, we use Taylor expansions for
$\phi$ and $\psi$, around $\varphi=\varphi_k$. We have ($r=r_k$ is fixed)
\begin{align}
\phi(\varphi)&=\phi(\varphi_k)+0+\frac{1}{2!}(\varphi-\varphi_k)^2
\left.\frac{\diff^2}{\diff\varphi^2}\phi
\right|_{\tilde{\varphi}}\nonumber\\
\psi(\varphi)&=\psi(\varphi_k)+0+\frac{1}{2!}(\varphi-\varphi_k)^2
\left.\frac{\diff^2}{\diff\varphi^2}\psi
\right|_{\tilde{\tilde{\varphi}}}\,,\nonumber
\end{align}
where $\tilde{\varphi}$ and $\tilde{\tilde{\varphi}}$ are both functions of
$\varphi$. On the interval $[-\frac{\pi}{4},+\frac{\pi}{4}]$, we
have the estimates
\eq{\label{estimdersecpsi}
-\frac{10}{9}cr_k\leq \frac{\diff^2}{\diff\varphi^2}\phi
\leq -\frac{5}{9}cr_k
}
Indeed, since 
\eq{\frac{\diff^2}{\diff\varphi^2}\phi=-cr_k\cos\varphi
+b\frac{\diff^2}{\diff\varphi^2}\real f(-r_ke^{i\varphi}+t)\,,}
we use \eqref{bornefseconde2rk} and find
\eq{\frac{\diff^2}{\diff\varphi^2}\phi\geq -cr_k-\frac{5}{9}\sqrt{A}cr_k\geq
-\frac{10}{9}cr_k\,,} and the upper bound
\eq{\frac{\diff^2}{\diff\varphi^2}\phi\leq -cr_k\cos\frac{\pi}{4}
+\frac{5}{9}\sqrt{A}cr_k\leq-\frac{5}{9}cr_k\,.}
We can thus compute some upper bound on the integral over
$[-\frac{\pi}{4},+\frac{\pi}{4}]$ in $c_k$ as follows:
\begin{align}
\frac{1}{2\pi}\int_{-\frac{\pi}{4}}^{+\frac{\pi}{4}}
e^{\phi(r_k,\varphi)-\phi(r_k,\varphi_k)}\diff \varphi&\leq 
\frac{1}{2\pi}\int_{-\infty}^{+\infty}\exp
\big(-\frac{1}{2}\cdot\frac{5}{9}cr_k x^2
\big)\diff x\nonumber\\
&=\frac{3}{\sqrt{5}}\frac{1}{\sqrt{2\pi cr_k}}\nonumber
\end{align}
The upper bounds on $\real c_k$ and $\imag c_k$ can be obtained by taking, 
say, $\delta=\frac{1}{3}$ in Lemma \ref{LSPL2}, which gives \eq{|c_k|\leq
\delta\frac{1}{\sqrt{cr_k}}+\frac{3}{\sqrt{5}}\frac{1}{\sqrt{2\pi cr_k}}\leq 
\frac{1}{\sqrt{cr_k}}\,.}
The lower bound on $\real c_k$ is obtained by dividing
$[-\frac{\pi}{4},+\frac{\pi}{4}]=I_1\cup I_2$, where
$I_1=[\varphi_k-\ga,\varphi_k+\ga]$, and  
$\ga=\ga_k\in[-\frac{\pi}{8},+\frac{\pi}{8}]$ is defined by 
two conditions: first, fix $\ga$ small enough such that 
\eq{\sup_{\varphi\in I_1}|\sin \varphi|\leq
\frac{2A}{1-A}\,.} The existence of such a $\ga$ is guaranteed by
\eqref{varphi*}. This first choice implies that 
\begin{align}
\sup_{\varphi\in I_1}|\psi(\varphi)-\psi(\varphi_k)|&\leq \frac{1}{2}\ga^2
\sup_{\varphi\in I_1}\big(cr_k|\sin \varphi|+b\big|\frac{\diff^2}{\diff
\varphi^f}
\imag f(-r_ke^{i\varphi}+t)\big|\big)\nonumber\\
(\text{ see } \eqref{bornefseconde2rk})&\leq \frac{1}{2}\ga^2
\big(cr_k\frac{2A}{1-A}+\frac{5}{9}cr_k\sqrt{A}\big)\nonumber\\
&=\frac{1}{2}\ga^2cr_k\sqrt{A}\big(\frac{2\sqrt{A}}{1-A}+\frac{5}{9}
\big)\nonumber\\
&\leq \frac{1}{2}cr_k\sqrt{A}\ga^2\leq \frac{1}{10}cr_k\ga^2\nonumber
\end{align}
Then, the second condition on $\ga$ is the following:
\eq{\frac{1}{10}cr_k\ga^2\equiv\frac{\pi}{3}\,.}
Here, we might have to take $k$ large enough to make sure that 
$\ga\in[-\frac{\pi}{8},+\frac{\pi}{8}]$. Then, we have a lower bound:
\begin{align}
\real
\frac{1}{2\pi}\int_{I_1}e^{\phi(\varphi)-\phi(\varphi_k)+i\big(\psi(\varphi)
-\psi(\varphi_k)\big)}\diff\varphi&\geq\frac{\cos\frac{\pi}{3}}{2\pi}
\int_{I_1}e^{\phi(\varphi)-\phi(\varphi_k)}\diff \varphi\nonumber\\
(\text{ see }\eqref{estimdersecpsi})&\geq 
\frac{\cos\frac{\pi}{3}}{2\pi}\int_{I_1}\exp\big(-\frac{1}{2}\cdot
\frac{10}{9}cr_k(\varphi-\varphi_k)^2\big)\diff \varphi\nonumber\\
&=\frac{3}{\sqrt{80\pi}}\big(2\Phi(\sqrt{\frac{100}{27}\pi})-1\big)
\frac{1}{\sqrt{2\pi cr_k}}\nonumber\\
&\geq\frac{47}{100}\frac{1}{\sqrt{2\pi cr_k}}\label{borneck1}
\end{align}
The upper bound on
$I_2=[-\frac{\pi}{4},\varphi_k-\ga]
\cup[\varphi_k+\ga,+\frac{\pi}{4}]$ is obtained easily:
\begin{align}
\frac{1}{2\pi}\big(\int_{-\frac{\pi}{4}}^{\varphi_k-\ga}+
\int_{\varphi_k+\ga}^{\frac{\pi}{4}}\big)e^{\phi(\varphi)-
\phi(\varphi_k)}\diff \varphi &\leq 2\cdot\frac{1}{2\pi}
\int_{\ga}^{\frac{\pi}{4}+|\varphi_k|}\exp\big(
-\frac{1}{2}\cdot\frac{5}{9}cr_k x^2\big)\diff x\nonumber\\
&\leq 2\cdot\frac{1}{2\pi}\int_{\ga}^{+\infty}
\exp\big(
-\frac{1}{2}\cdot\frac{5}{9}cr_k x^2\big)\diff x\nonumber\\
&=\frac{6}{\sqrt{5}}\big(1-\Phi(\sqrt{\frac{50}{27}\pi})
\big)\frac{1}{\sqrt{2\pi cr_k}}\nonumber\\
&\leq \frac{3}{100}\frac{1}{\sqrt{2\pi cr_k}}\label{borneck2}
\end{align}
Taking $\delta:=\frac{14}{100}$ in Lemma \ref{LSPL2} and using
\eqref{borneck1}, \eqref{borneck2} gives
\eq{\real \,c_k\geq \big(\frac{47}{100}-\delta-\frac{3}{100}\big)
\frac{1}{\sqrt{2\pi cr_k}}=\frac{3}{10}\frac{1}{\sqrt{2\pi cr_k}}\,,} which
completes the proof.
\end{proof}
\begin{cor}\label{LSPC0}
Let $\rho>0$, $F(z)=
\exp(-cz+bf(z))$ where $1\leq b\leq c$, and $f$ is analytic in a disc
$\cD_\rho(0)$, taking real values on the real line, 
with a uniformly bounded derivative:
\eq{\label{conditfprime}
\sup_{z\in \cD_\rho(0)}|f'(z)|\leq A<\frac{1}{25}\,.}
There exists $k_0=k_0(A)$ such that the following holds: let $t\in
(-\rho,+\rho)$
and define $k_+=(\rho-|t|)(c-2b\sqrt{A})$. 
For all integer $k\in [k_0,k_+]$ there
exists $r_k>0$ and $c_k>0$ such that
\eq{\label{LSP1'}
F^{(k)}(t)=k!\frac{c_k}{(-r_k)^k}F(-r_k+t)\,.}
We have the estimates
\eq{\label{LSP23'}\frac{3}{10}\,\frac{1}{\sqrt{2\pi cr_k}}<
\,c_k < \frac{1}{\sqrt{cr_k}}\,,\quad
\frac{ k}{c+bA}\leq r_k\leq \frac{k}{c-bA}\,.}
In particular, $(-1)^kF^{(k)}(t)>0$.
Moreover, if $f$ satisfies the local condition
\eq{\label{conditf} -ct+bf(t)\leq -\alpha \rho c\,,}
with $\alpha\in(\log 2,1)$, then there exists function
$a=a(k,c,b)$, $\sup|a|<\infty$, and $\ga=\ga(\alpha)>0$ such that for all
$k\in[k_0,k_+]$ and $A$ sufficiently small:
\eq{\label{LSP25}\big(\log(1+F)\big)^{(k)}(t)=
(1+a\cdot e^{-\frac{\ga}{2}k})F^{(k)}(t)}
\end{cor}
\begin{proof}
The first part of the proof follows from Theorem \ref{LSPT0}.
For the second part, notice that \eqref{conditf} implies $|F(t)|\leq
e^{-\alpha\rho c}<1$. The
continuity of $f$
implies that in some neighborhood $V\subset \cD_\rho(0)$, 
$V\ni t$, we have $\sup_{z\in V}|F(z)|\leq e^{-\frac{1}{2}\alpha\rho c}<1$. 
In $V$ we can thus use the Taylor series for $\log(1+F)$:
\begin{align}
\log\big(1+F\big)=\sum_{n\geq 1}\frac{(-1)^{n+1}}{n}F^n=F
+\sum_{n\geq 2}
\frac{(-1)^{n+1}}{n}F^n\nonumber
\end{align}
Since the series converges absolutely and uniformly in $V$, we can derive it
term-wise with respect to $k$.
We then need to show that the following holds:
\begin{lem}\label{LSPL3} Let $\alpha\in(\log 2,1)$.
There exists a positive constant $K_0<\infty$ such that
for all $\lambda\in(\frac{\log 2}{\alpha},1)$ 
and for all $n\geq 2$, $k\in [k_0,k_+]$,
\eq{\left|(F^n)^{(k)}(t)\right|\leq
K_0e^{-\frac{\ga}{2}k}e^{-\alpha(1-\lambda)(n-1)k}
\left|F^{(k)}(t)\right|\,,}
where $\ga$ is given by $\ga:=\alpha\lambda-\log 2$. The constant $A$ in
\eqref{conditfprime} has to be taken small enough (depending on the value of
$\alpha$).
\end{lem}
\noindent Suppose for a while that the Lemma has been shown; we have
the following estimate:
\begin{align}
\left|\sum_{n\geq 2}\frac{(-1)^{n+1}}{n}
(F^n)^{(k)}(t)\right|
&\leq
\sum_{n\geq 2}\left|
(F^n)^{(k)}(t)\right|\nonumber\\
&\leq K_0 e^{-\frac{\ga}{2}k}\left|F^{(k)}(t)\right|
\sum_{n\geq 2}e^{-\alpha(1-\lambda)(n-1)k}\nonumber\\
&\leq K_0 \frac{e^{-\alpha(1-\lambda)k_0}}{1-e^{-\alpha(1-\lambda)k_0}}
e^{-\frac{\ga}{2}k}\left|F^{(k)}(t)\right|\,,\nonumber
\end{align}
which proves \eqref{LSP25}.
\end{proof}
\begin{proof}[Proof of Lemma \ref{LSPL3}:]
The point is that $F$ is of exponential type. We have 
\begin{align}
F(z)^n&=e^{-cnz+bnf(z)}\equiv e^{-c_nz+b_nf(z)}\,,
\end{align}
with $c_n=cn$, $b_n=bn$. For each $n=1,2,\dots$, we can apply 
Theorem \ref{LSPT0}:
there exists for all $k\in[k_0,k_{n,+}]$, where 
$k_{n,+}=nk_+$, some $r_{n,k}$ and $c_{n,k}$ such that 
\eq{\frac{1}{k!}(F^n)^{(k)}(t)=
\frac{c_{n,k}}{(-r_{n,k})^{k}}F(-r_{n,k}+t)^n}
Notice that $[k_0,k_+]\supset[k_0,k_{n,+}]$ for all $n$.
The constant $r_{n,k}$ is a solution of the equation $k=r(c_n-b_n \real 
f'(-r+t))$ and satisfies
\eq{\label{B39}\frac{k}{c_n+b_nA}\leq r_{n,k}\leq \frac{k}{c_n-b_nA}\,.}
The constant $c_{n,k}$ satisfies 
\eq{\label{LSP26}
\frac{3}{10}\frac{1}{\sqrt{2\pi c_nr_{n,k}}}\leq c_{n,k}\leq
\frac{1}{\sqrt{c_nr_{n,k}}}\,.}
We can then consider, for all $k\in[k_0,k_+]$:
\eq{\label{LSP26.5}\frac{(F^n)^{(k)}(t)}{F^{(k)}(t)}=\frac{c_{n,k}}{c_{1,k}}
\left(\frac{r_{1,k}}{r_{n,k}}\right)^k
\frac{F(-r_{n,k}+t)^n}{F(-r_{1,k}+t)}}
Notice that when $n$ increases, $r_{n,k}\searrow 0$ and $k_{n,+}\nearrow 
\infty$. Using \eqref{B39} and \eqref{LSP26}, we find
\eq{\frac{r_{1,k}}{r_{n,k}}\leq n\frac{1+A}{1-A}\,,}
and
\eq{\frac{c_{n,k}}{c_{1,k}}\leq
\frac{10}{3}\sqrt{2\pi}\sqrt{\frac{1+A}{1-A}}\equiv K_0\,.}
We must estimate
\al{\frac{F(-r_{n,k}+t)^n}{F(-r_{1,k}+t)}=\exp\big(
c(nr_{n,k}-&r_{1,k})-ct(n-1)\nonumber\\
&+b(nf(-r_{n,k}+t)-f(-r_{1,k}+t))\big)\label{LSP27}}
Using the definition of $r_{n,k}$ gives
\begin{align}
nr_{n,k}-r_{1,k}&=k\left[\frac{1}{c-b\real f'(-r_{n,k}+t)}-
\frac{1}{c-b\real f'(-r_{1,k}+t)}\right]\nonumber\\
&=k\frac{b\big(\real f'(-r_{n,k}+t)-\real f'(-r_{1,k}+t)\big)}{
(c-b\real f'(-r_{n,k}+t))(c-b\real f'(-r_{1,k}+t))}\nonumber\\
&\leq k \frac{2bA}{(c-bA)^2}\leq 
\frac{k}{c}\frac{2A}{(1-A)^2}\nonumber
\end{align}
We then compute the term involving $f$ (we use twice 
$|f(x)-f(x')|\leq A|x-x'|$):
\begin{align}
nf(-r_{n,k}+t)-&f(-r_{1,k}+t)\\
&=(n-1)f(t)+n(f(-r_{n,k}+t)-f(t))
-(f(-r_{1,k}+t)-f(t))\nonumber\\
&\leq (n-1)f(t)+ nAr_{n,k}+A r_{1,k}\nonumber\\
&\leq (n-1)f(t)+\frac{k}{c}\frac{2A}{1-A}\nonumber
\end{align}
We thus have 
\eq{\frac{F(-r_{n,k}+t)^n}{F(-r_{1,k}+t)}
\leq e^{\epsilon(A)k}e^{(-ct+bf(t))(n-1)}\,,}
where $\epsilon(A)=2A(2-A)(1-A)^{-2}$. Since $\rho c> k_+\geq k$, we can
use assumption
\eqref{conditf} and get, for all $\lambda\in(0,1)$, 
\eq{e^{(-ct+bf(t))(n-1)}
\leq e^{-\alpha k(n-1)}=e^{-\alpha\lambda k(n-1)}
e^{-\alpha(1-\lambda)k(n-1)}}
Since $\log n-\log 2\leq\frac{1}{2}(n-2)$ for all $n\geq 1$ we can compute the
following bound
\al{\sup_{n\geq 2}
n^ke^{-\alpha k\lambda(n-1)}&=\sup_{n\geq 2}
e^{k(\log n-\alpha \lambda (n-1))}\\
&\leq \sup_{n\geq 2}e^{k(\log
2-1+\alpha\lambda+n(\frac{1}{2}-\alpha\lambda))}\\
&\leq 
e^{k(\log2-\alpha\lambda)}\equiv e^{-\zeta k}\,,}
where we used the fact that $\lambda$ is chosen such that
$\zeta=\zeta(\alpha)>0$. Putting our bounds together we bound \eqref{LSP26.5}, 
when $A$ is small, by
\begin{align}
K_0\left(
\frac{1+A}{1-A}e^{\epsilon(A)}\right)^ke^{-\frac{1}{2}\zeta k}
e^{-\frac{1}{2}\zeta k}
e^{-\alpha(1-\lambda)k(n-1)}
\leq K_0e^{-\frac{1}{2}\zeta k}e^{-\alpha(1-\lambda)k(n-1)}\,.\nonumber
\end{align}
\end{proof}
\section{Cluster Expansion}\label{KSnewCLEXP}
Consider a countable set $\cD$ whose elements are
called \grasA{\animals}, and denoted $\ga\in\cD$.
To each \animal{} $\ga$ is associated a finite subset of $\bZ^d$, 
called the \grasA{support of} $\ga$. 
Usually we also denote the support by
$\ga$. In the cases we consider, the support is always an $R$-connected set.
Assume we are given a symmetric binary relation on $\cD$, denoted
$\sim$. We say two \animals{} $\ga,\ga'$ are \grasA{compatible} if
$\ga\sim\ga'$.
When $\ga$ and $\ga'$ are not compatible we write
$\ga\not\sim\ga'$.
We assume that the following condition is sufficient to characterize
incompatibility: for each each \animal{} $\ga$, there exists a set
$b(\ga)\subset \bZ^d$ such that if $\ga\not \sim\ga'$, then
$b(\ga)\cap b(\ga')\neq \emptyset$.\\
To each \animal{} $\ga\in\cD$ we associate a complex weight $\omega(\ga)\in\bC$.
The \grasA{partition function} is defined by
\eq{\label{K600}\Xi({\cD})\pardef\sum_{\substack{\{\ga\}\subset
\cD\\\text{compat.}}}\prod_{\ga\in\{\ga\}}\omega(\ga)\,,}
where the sum extends over all sub-families of $\cD$ of pairwise compatible
\animals{} (we assume this sum exists, which is the case in every
concrete situation). When $\{\ga\}=\emptyset$, we define the product over $\ga$
as equal to $1$. We are interested in studying the logarithm of the
partition function. To this end, we define the family $\hatcD$ of all maps
$\hatga:\cD\to\{0,1,2,\dots\}$. The \grasA{support} of $\hatga$ is the set
$\{\ga\in\cD:\hatga(\ga)\geq 1\}$. Usually we also denote the support of
$\hatga$ by $\hatga$. We will also write $\hatga\ni x$ if the support of
$\hatga$ contains an \animal{} whose support contains $x$.
A map $\hatga\in\hatcD$ is a \grasA{cluster of \animals{}} if its support
can't be decomposed into a disjoint union 
$S_1\cup S_2$ such that each $\ga_1\in S_1$ is
compatible with each $\ga_2\in S_2$.
Formally, the logarithm of the partition function has the form (see e.g 
\cite{Pf})
\eq{\label{K601}
\log \Xi({\cD})=\sum_{\hatga\in\hatcD}\omega(\hatga)\,,}
where the weight of $\hatga$ equals 
\eq{\label{K601.1}
\omega(\hatga)=a^T(\hatga)\prod_{\ga\in\cD}\omega(\ga)^{\hatga(\ga)}\,}
The functions
$a^T(\hatga)$ are purely combinatorial factors.
They equal zero if $\hatga$ is not a cluster.
The following is the technical lemma that
gives explicit conditions for the convergence of the development
\eqref{K601}. 
The proof is standard and can be adapted from \cite{Pf}.
\begin{lem}\label{KL50}
Let $\omega_0(\ga)$ be a positive weight such that
\eq{\label{K603}\sup_{x\in\bZ^d}\sum_{\ga:b(\ga)\ni x}
\omega_0(\ga)e^{|b(\ga)|}\leq \epsilon\,,}
where $0<\epsilon<1$.
Define $\omega_0(\hatga)$ as in \eqref{K601.1} with $\omega_0(\ga)$ in place of
$\omega(\ga)$. Then there
exists a function $\eta(\epsilon)$, $\lim_{\epsilon\to 0}\eta(\epsilon)=0$ such
that 
\eq{\sup_{x\in\bZ^d}
\sum_{\hatga\ni x}|\omega_0(\hatga)|\leq \eta(\epsilon)\,.}
\end{lem}
Typically, in the cases we consider, the 
weights are maps $z\mapsto \omega(\ga;z)$, analytic in a
domain $A\subset \bC$, and there exists a positive weight $\omega_0(\ga)$ such
that $\|\omega(\ga;\cdot)\|_A\leq \omega_0(\ga)$ for all $\ga$. Lemma \ref{KL50}
thus implies that the series \eqref{K601} 
is normally convergent on $A$. This guarantees analyticity of the logarithm of
$\Xi(\cD)$, by a standard Theorem of Weierstrass (see e.g. \cite{Rem}).


\begin{thebibliography}{xxxxx}

\bibitem[BKL]{BKL} Bricmont J., Kuroda K., Lebowitz J.L., \emph{First Order Phase
Transitions in Lattice and Continuous Systems: Extension of Pirogov-Sinai
Theory}, Commun. Math. Phys. \textbf{101}, 501-538, 1985.
\bibitem[BS]{BS} Bricmont J., Slawny J., \emph{Phase Transitions in Systems with a
Finite Number of Dominant Ground States}, J. Stat Phys. {\bf 54},
89-161, 1989.
\bibitem[BZ1]{BZ1} Bovier A., Zahradn\'\i k M., \emph{The Low-Temperature 
Phases of Kac-Ising Models}, J. Stat. Phys. {\bf 87}, 311-332, 1997.
\bibitem[BZ2]{BZ2} Bovier A., Zahradn\'\i k M., \emph{Pirogov-Sinai Theory 
for Long Range Spin Systems}, Mark. 
Proc. Rel. Fields, {\bf 8}, 443-478, 2002.
\bibitem[CP]{CP} Cassandro M., Presutti E., \emph{Phase Transitions in 
Ising Systems
with Long but Finite Range Interactions}, Markov Processes and Related
Fields, {\bf 2}, 241-262, 1996.
\bibitem[DS]{DS} Dinaburg E.I., Sinai Y.G, \emph{Contour Models with
Interactions and Applications}, Selecta Mathematica Sovietica, {\bf 7}, 3,
291-315, 1988.
\bibitem[F]{F} Fisher M.E., \emph{The Theory of Condensation and  
the Critical Point}, Physics {\bf 3}, 255-283 (1967). 

\bibitem[FP]{FP} Friedli S., Pfister C.-E., \emph{On the Singularity of
the Free Energy at First Order Phase Transition}, submitted to Commun. Math.
Phys., 2002 (available on cond-mat/0212015).
\bibitem[GHS]{GHS} Griffiths R.B., Hurst C.A., Sherman S., \emph{Concavity 
of
Magnetization of an Ising Ferromagnet in a Positive External Field}, 
J. Math. Phys. \textbf{11}, \textbf{3}, 790-795, 1970.
\bibitem[I1]{I1} Isakov S.N., \emph{Nonanalytic Features of the First 
Order Phase  
Transition in the Ising Model}, Commun. Math. Phys. {\bf 95}, 427-443, 1984. 
\bibitem[I2]{I2} Isakov S.N., \emph{Phase Diagrams and Singularity at the
 Point of a
Phase Transition of the First Kind in Lattice Gas Models},
Teoreticheskaya i Matematicheskaya 
Fizika, {\bf71}, 426-440, 1987.

\bibitem[Ku]{Ku} Kuratowski K., \emph{Topology}, vol. 1 and 2, Academic Press,
1968.
\bibitem[KUH]{KUH} Kac M., Uhlenbeck G.E., 
Hemmer P.C., \emph{On the van der Waals Theory of the
Vapor-Liquid Equilibrium}, J. Math. Phys. \textbf{4}, 216-228, 1963.
\bibitem[L1]{L1} Langer J.S., Annals of Physics {\bf 41}, 108-157, 1967.
\bibitem[LP]{LP} Lebowitz J.L., Penrose O., \emph{Rigorous Treatment of
 the Van Der
Waals-Maxwell Theory of the Liquid-Vapor Transition}, J. Math. Phys.
{\bf7}, 98-113, 1966.
\bibitem[LY]{LY} Lee T.D., Yang,C.N., \emph{Statistical Theory of State 
and Phase  
Transition I \& II}, Phys. Rev. {\bf 87}, 
404-419, 1952.
\bibitem[Pf]{Pf} Pfister, C.-E., \emph{Large Deviations and Phase 
Separation in  
the Two Dimensional Ising Model}, Helvetica Physica Acta,
{\bf 64}, 953-1054, 1991.
\bibitem[Pr]{Pr} Presutti E., \emph{From Statistical Mechanics to Continuum
Mechanics}, Lecture Notes, Max Planck Institute, Leipzig, 1999.
\bibitem[PS]{PS} Pirogov S.A., Sinai Ya. G.,\emph{ Phase Diagrams of 
Classical
 Lattice
Systems}, Teoreticheskaya i Matematicheskaya Fizika, \textbf{26, 1},
61-76, 1976.
\bibitem[Rem]{Rem} Remmert, \emph{Theory of Complex Functions}, 
Springer Verlag, 1991.
\bibitem[vdW]{vdW} van der Waals J.D., \emph{De Continuiteit van den Gas en
Vloeistoftoestand}, Academic Thesis, Leiden, 1873.
\bibitem[Z1]{Z1} Zahradn\'\i k, M., \emph{An alternate version of Pirogov-Sinai
 theory}, Commun. Math. Phys. {\bf 93}, 559-581 (1984).

\end{thebibliography}
\end{document}